\documentclass[11pt]{amsart}
\usepackage{fullpage}
\usepackage{amsmath}
\usepackage{graphicx} 
\usepackage{color,url}
\usepackage[normalem]{ulem}

\newcommand{\jcm}[1]{}
\newcommand{\pp}{\ }
\newcommand{\boundary}{\mathrm{boundary}}
\newcommand{\pop}{\mathrm{pop}}
\newcommand{\con}{\mathrm{conflicted}}

\newcommand{\area}{\mathrm{area}}
\newcommand{\mino}{\mathrm{AA}}
\newcommand{\capSize}{\small}

\renewcommand{\d}{\partial}
\newcommand{\J}{J}
\newcommand{\JJ}{J}
\newcommand{\dist}{\mathcal{R}}

\newcommand{\pr}{\mathcal{P}}
\renewcommand{\Pr}{\pr_{\beta}}

\title{Redistricting: Drawing the Line} 
\author[Bangia]{Sachet Bangia}
\address{Departments of Economics and Computer Science, Duke
  University, Durham NC 27708} \email{sachet.bangia@duke.edu}
\author[Graves]{Christy Vaughn Graves} \address{Christy Vaughn Graves,
  Program in Applied and Computational Mathematics, Princeton
  University, Princeton NJ} \email{cjvaughn@princeton.edu}
\author[Herschlag]{Gregory Herschlag} \address{Gregory Herschlag,
  Departments of Mathematics and Biomedical Engineering, Duke
  University, Durham NC 27708} \email{gjh@math.duke.edu}
\author[Kang]{Han Sung Kang} \address{Han Sung Kang, Departments of
  Electrical and Computer Engineering, Computer Science, and
  Mathematics, Duke University, Durham NC 27708} \email{h.k@duke.edu}
\author[Luo]{Justin Luo} \address{Justin Luo, Departments of
  Electrical and Computer Engineering and Mathematics, Duke
  University, Durham NC 27708} \email{justin.luo@duke.edu}
\author[Mattingly]{Jonathan C. Mattingly} \address{Jonathan Mattingly,
  Departments of Mathematics and Statistical Science, Duke University,
  Durham NC 27708} \email{jonm@math.duke.edu} \author[Ravier]{Robert
  Ravier} \address{Robert Ravier, Department of Mathematics, Duke
  University, Durham NC 27708} \email{robert.ravier@duke.edu}

\begin{document}
\maketitle

\begin{abstract} We develop methods to evaluate whether a political
  districting accurately represents the will of the people.
To explore and showcase
  our ideas, we concentrate on the congressional districts for the
  U.S. House of Representatives and use the state of North Carolina
  and its redistrictings since the 2010 census.  
Using a Monte Carlo algorithm, we randomly generate over 24,000 redistrictings that are non-partisan
and adhere to criteria from proposed legislation.
Applying historical voting data to these random redistrictings, 
we find that the number of democratic and republican representatives elected varies drastically depending on how districts are drawn.
  Some results are more common, and we gain a clear range of expected election outcomes. 
  Using the statistics of our generated redistrictings, we critique the
  particular congressional districtings used in the 2012 and 2016 NC
  elections as well as a districting proposed by a bipartisan
  redistricting commission. We find that the 2012 and 2016
  districtings are highly atypical and not representative of the will
  of the people.
  On the other hand, our results indicate that a plan produced by a bipartisan panel of retired judges
   is highly typical and representative. Since our analyses are based
   on an ensemble of reasonable redistrictings of North Carolina,
   they provide a baseline for a given election which incorporates the
   geometry of the state's population distribution.
  \end{abstract}

\section{The Will of the People}
\label{sec:introduction}

Democracy is typically equated with expressing the will of the people 
through government.  Perceived failures of democracy in representative 
governments are usually attributed to the voice of the people being 
muted and obstructed by the actions of special interests or the sheer 
size of government.  The underlying assumption is that the will of the 
people exists as a clear, well defined voice which only needs to be 
better heard.
Yet the will of the people is not monolithic. It is not always so
simple to obtain a consensus or even a clear majority opinion. We rely on our
elections as a proxy to express our collective opinions and our political
will, which leads to the question, how effective a given election is at capturing this
will?

In the United States, district representation schemes are used to
divide the population into distinct groups, each of which carries a
certain amount of representation. This districting acknowledges that the people's voice is
geographically diverse and that we value the expression of that
diversity in our government. We take election results to represent the
people's will,  giving the elected
officials a mandate to act in the people's name.
Hence, it is reasonable to ask if and how this representation is affected by
the choice of district boundaries. Just how sensitive are election
results, and by extension, our impression of the people's will, to our
choices for geographic divisions? 
 The method we use to reveal this will is simple. We take the
actual votes cast by North Carolinians at the 2012 and 2016 congressional elections and then
change the boundaries of the congressional districts to see how the
partisan results of the elections change. 
Our results show that the will of the people is not a single election outcome but rather a distribution of possible outcomes. The exact same vote counts can lead to drastically different outcomes depending on the choice of districts.

compMany discussions of fraudulent elections emphasize voter suppression or
voter fraud. However if the results of an election vary so widely over
different choices of redistrictings, then it is paramount that the
districts employed produce results that are an accurate proxy of the
``will of the people.''  The courts have given some guidance in this
direction by promoting the ``one person, one vote''
standard,\footnote{Wesberry v. Sanders (1964)} requiring districts
have roughly an equal number of eligible voters so that each elected representative
is a proxy for the same number of constituents. We will demonstrate that the
commonly promoted criteria for the construction of redistricting
(equal population apportionment, geographical compactness, and
preservation of historical constituencies) still leave a lot of
variability in the election results for a given set of precinct level
votes.  Given this variability, we can ask if a given redistricting leads to a common and expected outcome, or if it gives one party an unlikely advantage.

In the 2012 congressional elections, which were based on the 2010 
redistricting, four out of the thirteen 
congressional seats were filled by Democrats. 
However, in seeming contradiction, the majority of 
votes were cast for Democratic candidates on the statewide level. The election results hinged on the 
geographic positioning of congressional districts. 
While this outcome is clearly the result 
of politically drawn districts, perhaps it is not the result of 
excessive tampering. Our country has a long 
history of balancing the rights of urban areas with high population 
with those of more rural, less populated areas. Our federalist and 
electoral structures enshrined the idea that majority rule must be balanced with regionalism. It might be that in North Carolina, the 
subversion of the results of the global vote count would happen 
in any redistricting which balances the representation of the urban with the 
rural or the beach with the mountains, and each with the Piedmont.  Maybe 
the vast majority of reasonable districts which one might draw would 
have these issues due to the geography of the population's 
distribution.\footnote{In \cite{ChenRodden13}, it was shown that redistrictings may favor a party simply due to the geography of the state.} We are left asking the basic question: how much does 
the outcome depend on the choice of districts? This can be further 
refined by asking ``what are the outcomes for a typical choice of 
districts,'' or ``when should a redistricting be considered 
outside the norm?'' These last two refinements require some way 
of quantifying what the typical outcomes are for a given set of 
votes. 
We therefore set the vote counts based on historical data and ask how changing the districts these votes were counted in leads to different results.
Since we will explore these questions in the context of the American political system, we will 
assume that people vote for parties, not people.
In these polarized times this is a reasonable approximation and we find the results 
extremely illuminating.

In order to change the district boundaries, we use a Markov
Chain Monte Carlo algorithm to produce about 24,000 random but reasonable redistrictings.\footnote{The algorithm is similar to the ones presented in \cite{MattinglyVaughn2014,QuantifyingGerrymandering,fifield2015}.}  The redistrictings are constructed using non-partisan design criteria
from a proposed piece of legislation. We then re-tally the actual
historic votes from the 2012 or 2016 elections to produce about 24,000
election outcomes, one for each of our generated redistrictings.  We
observe that the number of representatives elected from a party can
vary drastically depending on the redistricting used, yet some
outcomes are more frequent than others.

Once we understand the extent to which election results can vary over a collection of possible redistrictings, we quantify how representative a
particular redistricting is by observing its place in this collection of
results. Similarly, with statistics of typical redistrictings in hand,
we devise measures of gerrymandering where the effects of packing (concentrating voters to lower their political power)
and cracking (fragmenting voting blocks to lower their political power) can be
better identified.
Since all of our analysis is based on the interaction of actual votes
with the collection of over 24,000 reasonable redistrictings, it
provides a baseline for the election which is informed by both the
geometry of the state and the distribution of the electorate though
out the state. 
It is an important feature of our analysis that our
techniques incorporate the effect of the state's geometry when
developing the baseline.

We apply our metrics to analyze and critique the
North Carolina U.S. Congressional redistrictings used in the
2012 and 2016 elections, as well as the redistrictings developed by a bipartisan
group of retired Judges as part of the ``Beyond Gerrymandering''
project spearheaded by Thomas Ross. We refer to these
redistrictings of interest as NC2012, NC2016, and Judges
respectively. See
Figures~\ref{fig:Districting2012}--\ref{fig:DistrictingJudges} in the
Appendix for visualizations of these redistrictings.
Our analysis uses the actual votes cast in the 2012 and 2016 N.C. congressional elections to illuminate the structure and features of a redistricting.

Using a related methodology, we also assess the degree to which the
three redistrictings (NC2012, NC2016, and Judges) are engineered. This
is done by seeing how close their properties are to the collection of
redistrictings that can be obtained by small changes. It seems
reasonable that the character of an election should not be overly
sensitive to small changes in the redistricting if the concept of the
``will of the people'' is to have any meaning.

The results of our analysis repeatedly show that the
NC2012 and  NC2016 redistrictings are heavily engineered and produce
results that are extremely atypical and at odds with the will of
the people. 
Finer analysis clearly shows
that the Democratic voters are clearly packed into a few districts,
decreasing their power, while Republican voters are spread more evenly, thus increasing their power. In contrast, election results from the Judges redistricting are
quite typical, producing results consistent with what is typically
seen. 
We emphasize that all of these conclusions come
from asking what the typical character and result of an election is
if we use a ``reasonable'' redistricting adhering to proposed legislation and drawn at random without any
partisan input, save the possible effect of ensuring a few districts
contain a sufficient minority population to comply with the Voting
Rights Act (VRA).

\section{Main Results: Where do you draw the line?}
\label{sec:main-results}
We emphasize from the start that in contrast to some works (see, for
example, we are
not proposing an automated method of creating redistrictings to be used in practice. Rather,
we are proposing a class of ideas for evaluating whether a redistricting is
truly representative or gerrymandered. 
We hope this helps draw the
line between fair and biased redistricting so that the will of the
people can be better heard.

Our analysis begins by generating over 24,000 ``reasonable''
redistrictings of North Carolina into thirteen U.S. House
congressional districts. For each redistricting, we tabulate the votes
from a previous election, either 2012 or 2016, to calculate the number of representatives
elected from both the Democratic and Republican Parties. We emphasize that we use
the actual votes from either the 2012 or 2016 U.S. House of
Representative elections. In using these votes, we assume that a vote
cast for a Republican or Democrat remains so even when district
boundaries are shifted.

By ``reasonable,'' we mean districts which are drawn in a nonpartisan fashion, guided only by the desire to:
\begin{itemize}
\item Divide the state population evenly between the thirteen districts.
\item Keep the districts geographically connected and compact.
\item Refrain from splitting counties as much as possible.
\item Ensure that African-American voters are sufficiently
  concentrated in two districts to give them a reasonable chance to
  affect the winner.
\end{itemize}
The precise meaning of
``reasonable'' is given in Section~\ref{sec:sampl-rand-redistr}, along
with the method we used to generate the over 24,000 ``reasonable''
redistrictings. We construct our districts by taking Voting Tabulation
Districts (VTD) from NC2012 as the fundamental atomic element used as our building
blocks. North Carolina is composed of over 2,600 VTDs.

The first criterion above enforces the ``one-person-one-vote'' doctrine, which
dictates that each representative should represent a roughly equal number of
people. The second criterion reflects the desire to have districts
represent regional interests. The third criterion embodies the idea
that districts should not fracture historical political constituencies
if possible; counties provide a convenient surrogate for these
constituencies. The last criterion, which is dictated by the Voting
Rights Act (VRA), asks that two districts have enough African-American voters
that they might be reasonably expected to choose the winner in that
district. In particular, we emphasize that no voting or registration
information is used, nor is any demographic information except for
what is dictated by the VRA. 

The exact choice of these criteria for our study comes from House Bill
92 (HB92) of the North Carolina General Assembly, which passed the House during the 2015 legislative
session. This bill proposed establishing a bipartisan commission guided solely by these principles to create redistrictings. 
Since the
companion legislation did not pass the North Carolina Senate, the
provision never became law. In fact, it is just the latest in a chain of
bills which have been introduced over the years with similar criteria
and aims.

\subsection{Beyond One-Person-One-Vote}
\label{sec:beyond-one-person} 
\begin{figure}[ht]
  \centering
\includegraphics[width=8cm]{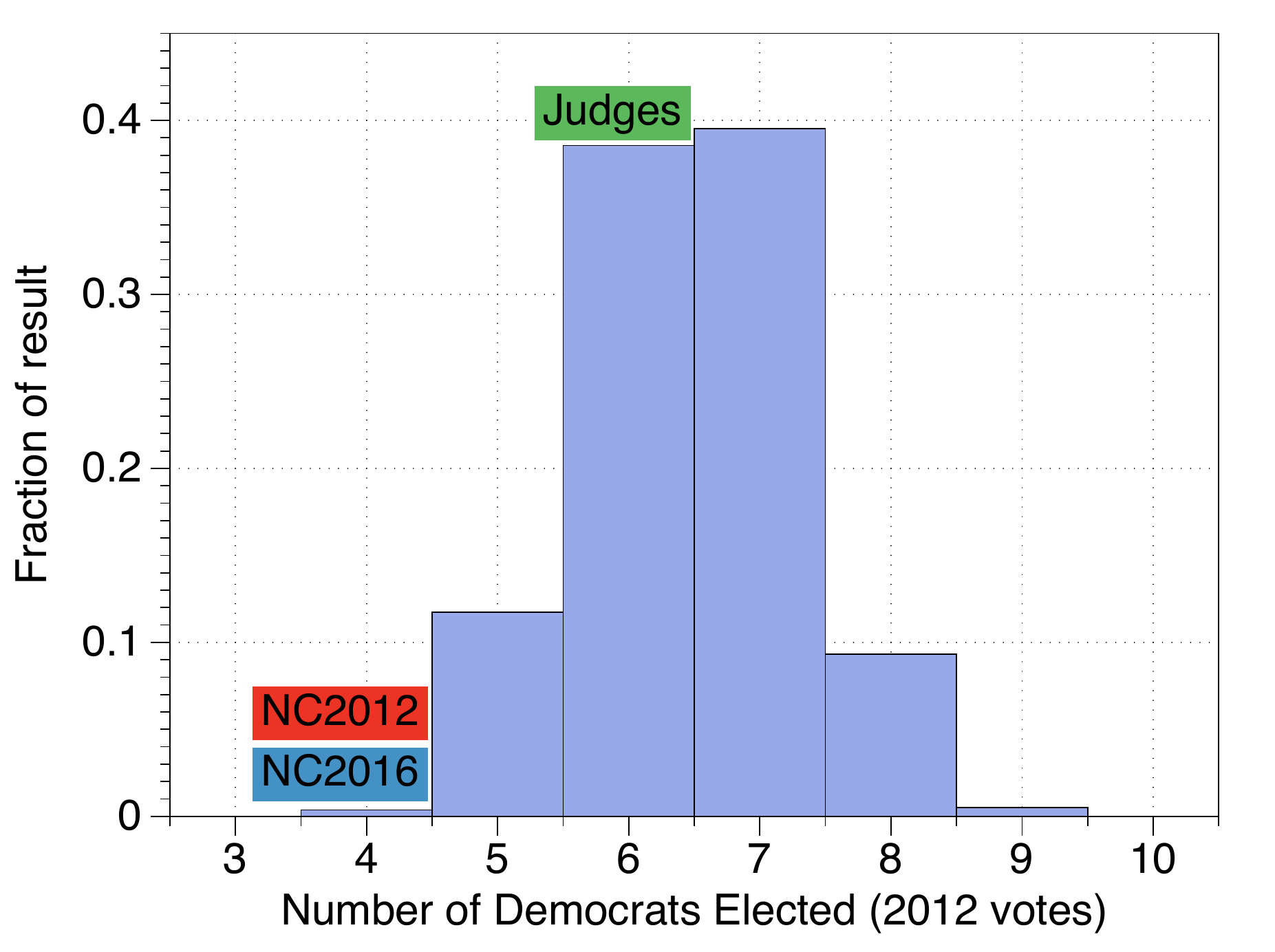}
\includegraphics[width=8cm]{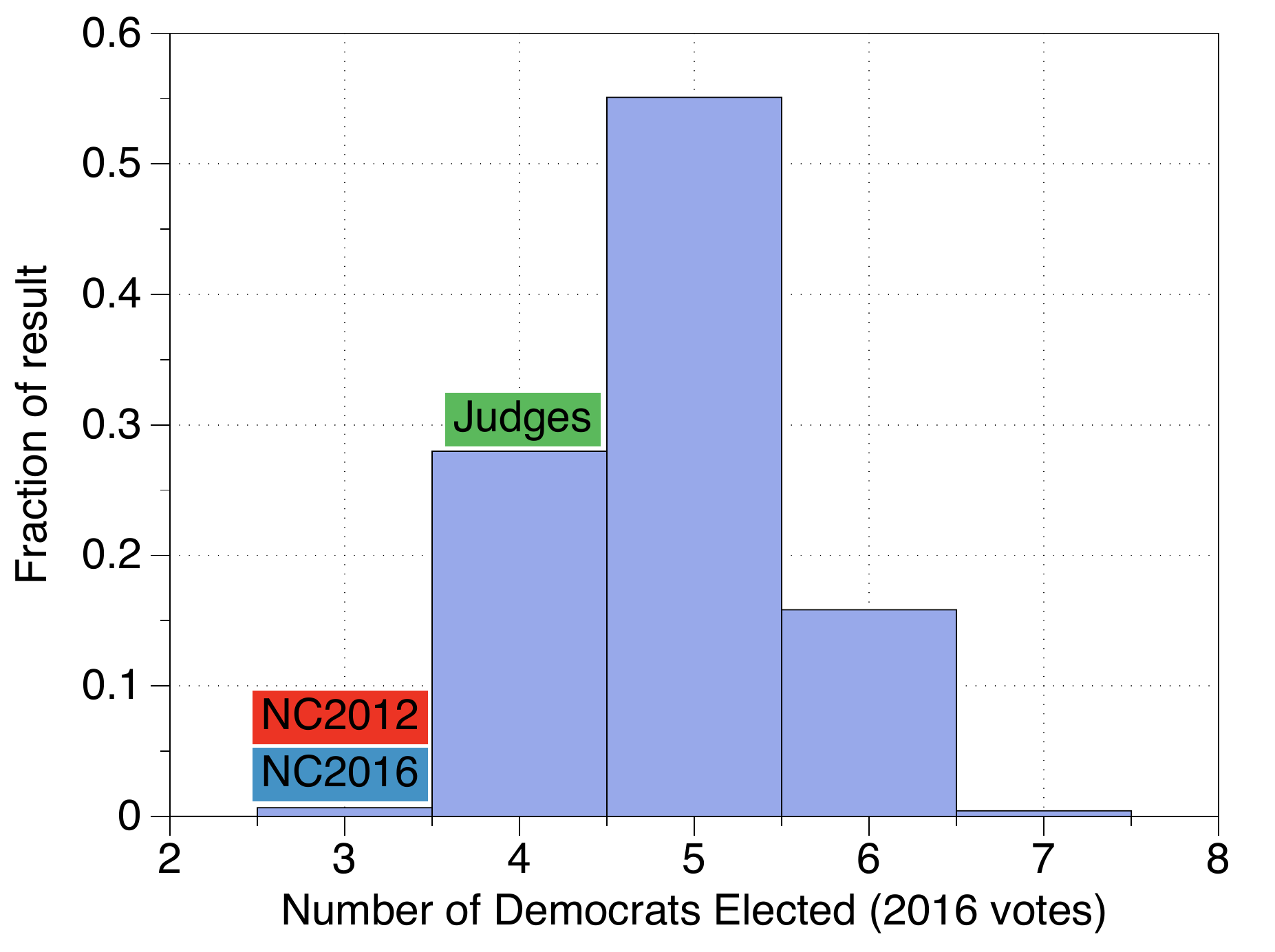}
   \caption{\capSize Probability of a given  number of Democratic wins among the
     13 congressional seats using votes from the 2012 election (left) and 2016 election (right).
}
  \label{fig:basicDemWinners}
\end{figure}

There is a large amount of variation in the outcome of an election depending on the
districts used.  The simple criteria from HB92 are not enough
to produce  a single preferred outcome of the elections. Rather, there is a
distribution of possible outcomes. Our
findings in this direction, summarized in Figure~\ref{fig:basicDemWinners}, clearly show that the results generated by the redistrictings
NC2012 and NC2016 are extremely biased towards the Republicans, while
the Judges redistricting produces acceptably representative results.
The NC2012 and NC2016  redistrictings produce results that are highly atypical of the 
non-partisan redistrictings we have randomly drawn according to HB92.

Over 24,000 random, but reasonable, redistrictings were used to
generate the probability distributions shown in
Figure~\ref{fig:basicDemWinners}.  We emphasize that the two plots use
the actual votes cast by the electorate in the 2012 and 2016
Congressional elections, respectively, to determine the outcomes for
each redistricting. For the 2012 vote counts, the NC2012 and NC2016
redistrictings both result in four Democratic seats, a result that
occurs in less than 0.3\% of our collection of over 24,000
redistrictings. The Judges redistricting results in the election of
six Democrats, which occurs in over 39\% of redistrictings. For the
2016 vote counts, the NC2012 and NC2016 redistrictings results in
three Democratic seats, a result that occurs in less than 0.7\% of
redistrictings. The Judges redistricting results in the election of
six Democrats, which occurs in 28\% of redistrictings.

\subsection{Measuring Representativeness and Gerrymandering}
\label{sec:meas-gerrrym}
While Figure~\ref{fig:basicDemWinners} is already quite compelling, it
is useful to develop quantitative measures of how representative the
results of a given election are. Gerrymandering goes beyond
just affecting the results; it also makes districts so safe that
representatives are less responsive to the will of the people, as their legislative choices will unlikely effect the result of an election.  To
measure these effects, we propose two indices. The first, which we call
the Gerrymandering Index, is based on the plots used to visualize
gerrymandering introduced in Section~\ref{sec:vis-gerrrym}. It
quantifies how packed or depleted the collection of districts is
relative to what is expected from the ensemble of ``reasonable''
redistrictings we have created. The second, which we call the Representativeness Index, is the measure of how
typical the election results produced by the redistricting are in the
context of what is seen in the ensemble of ``reasonable''
redistrictings.  It is based on the refinement of
Figure~\ref{fig:basicDemWinners} given in
Figure~\ref{fig:finehist2012} and described in
Section~\ref{sec:DetailsRep}. Later in Section~\ref{sec:EffGap}, we consider a third index,
the Efficiency Gap, which has recently been employed in the 
decision Whitford Op. and Order, Dkt. 166, Nov. 21, 2016.

In summary, in this section we consider
\begin{itemize}
\item {\bf Gerrymandering Index:}
 Measures the degree to which the
  percentage of Democratic votes in each district deviates from what is
  typically seen in our collection  of  ``reasonable'' redistrictings. 
  The squareroot of the sum of the
  square deviations is the index.
Relatively large scores are less balanced
  than the bulk of the ``reasonable'' redistrictings in our
  ensemble. These large indexed redistrictings typically have some districts with many more voters
  from one party than is normally seen or generally have a higher percentage of one
  party  in many districts than is normal, or both. 
 How the term
  ``normal'' is understood is partially explained in
  Section~\ref{sec:vis-gerrrym} and completely explained in
  Section~\ref{sec:DetailsGerry}.
\item {\bf Representativeness Index:} Measures how typical the results
  obtained by a given redistricting are in the context of the
  collection of ``reasonable'' redistrictings we have
  generated. Redistrictings with relatively large values produced an
  election outcome which is farther from the typical election outcome
  in the collection of ``reasonable'' redistrictings. Details
  are given in Section~\ref{sec:DetailsRep}.
\end{itemize}
Both of these indices are adapted to the geometry of the votes and
population density of the state as reveiled by the ensamble of
``reasonable'' redistrictings. In this sense, we expect them to be
more nuanced than other metrics which are not informed by the local
structure of the state.

As these indices are most useful when values for different
redistrictings are compared, we place each redistricting of interest on
the plot of the complementary cumulative distribution function for each of the
three above measures. This allows us to judge the relative size of
each index in the context of our collection of ``reasonable''
redistrictings. 

\begin{figure}[ht]
  \centering 
\includegraphics[width=8cm]{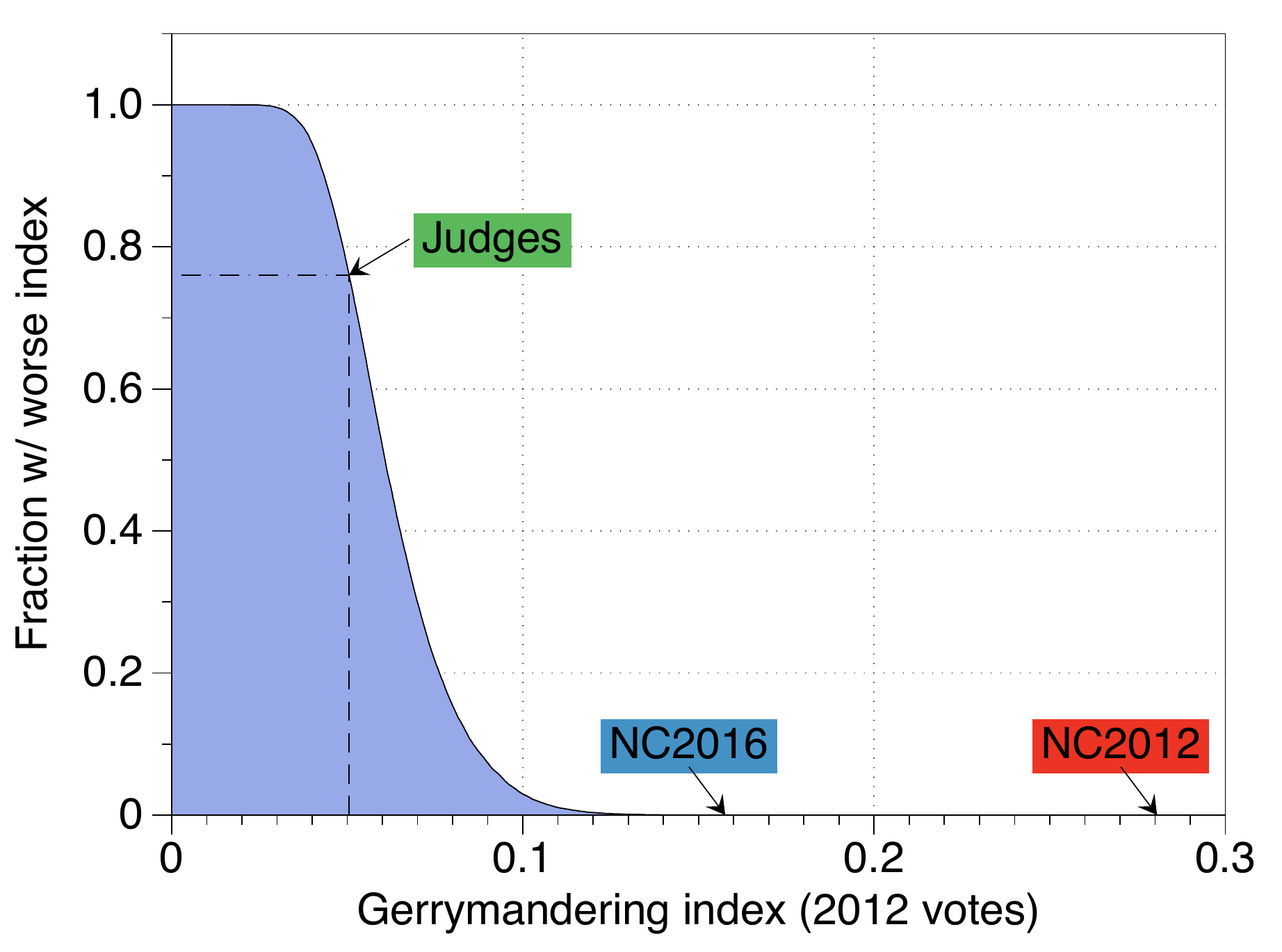}
\includegraphics[width=8cm]{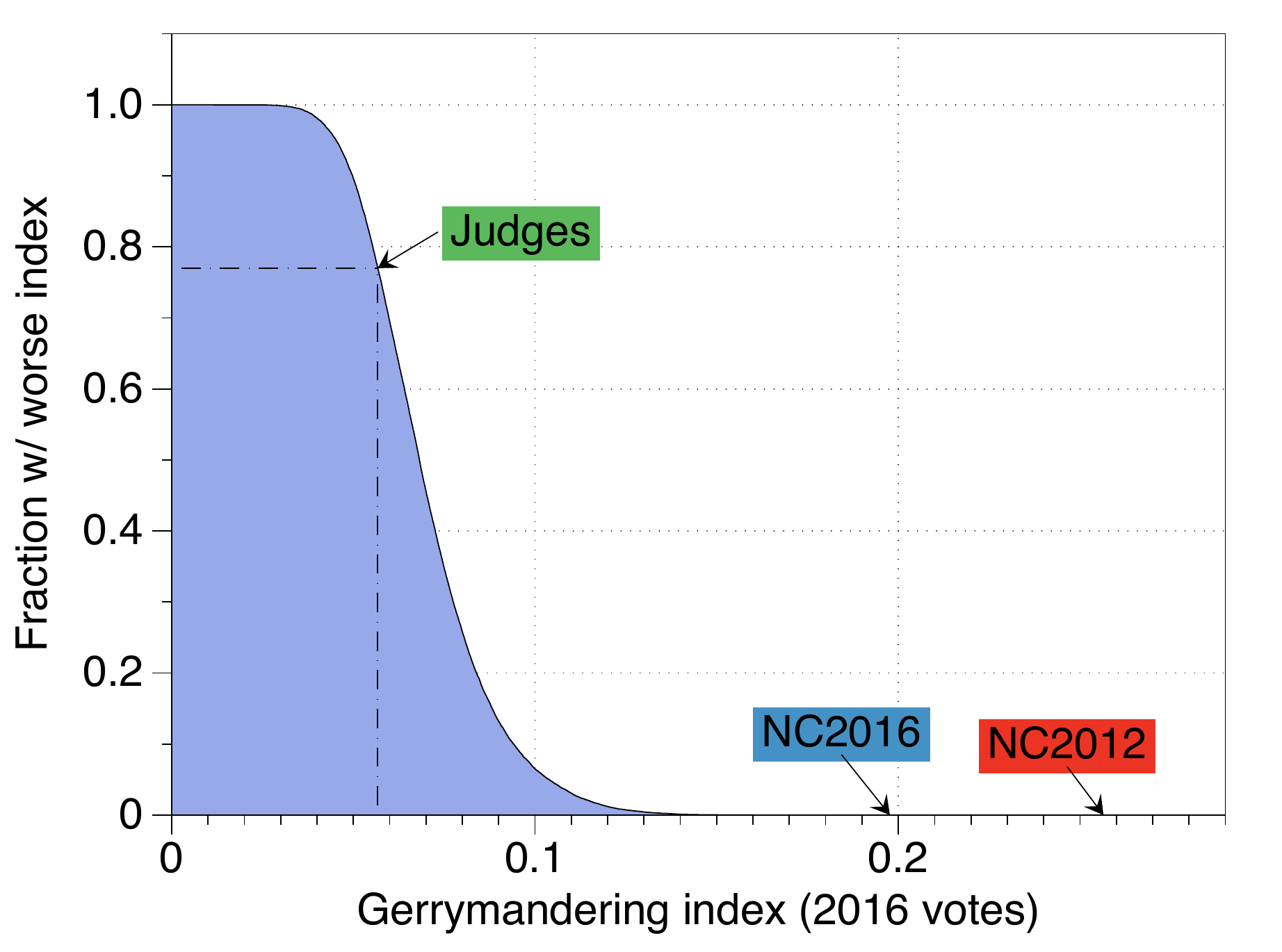}
   \caption{\capSize Gerrymandering Index for the three districts of
     interest based on the congressional voting data from 2012 (left)
     and 2016 (right). No generated redistrictings had a Gerrymandering Index higher than either the NC2012 or the NC2016 redistrictings.  The Judges redistricting plan was less gerrymandered than over 75\% of the random districts in both sets of voting data, meaning that it is an exceptionally non-gerrymandered redistricting plan.}
  \label{fig:gerrymanderingsCDF}
\end{figure}

\begin{figure}[ht]
  \centering 
\includegraphics[width=8cm]{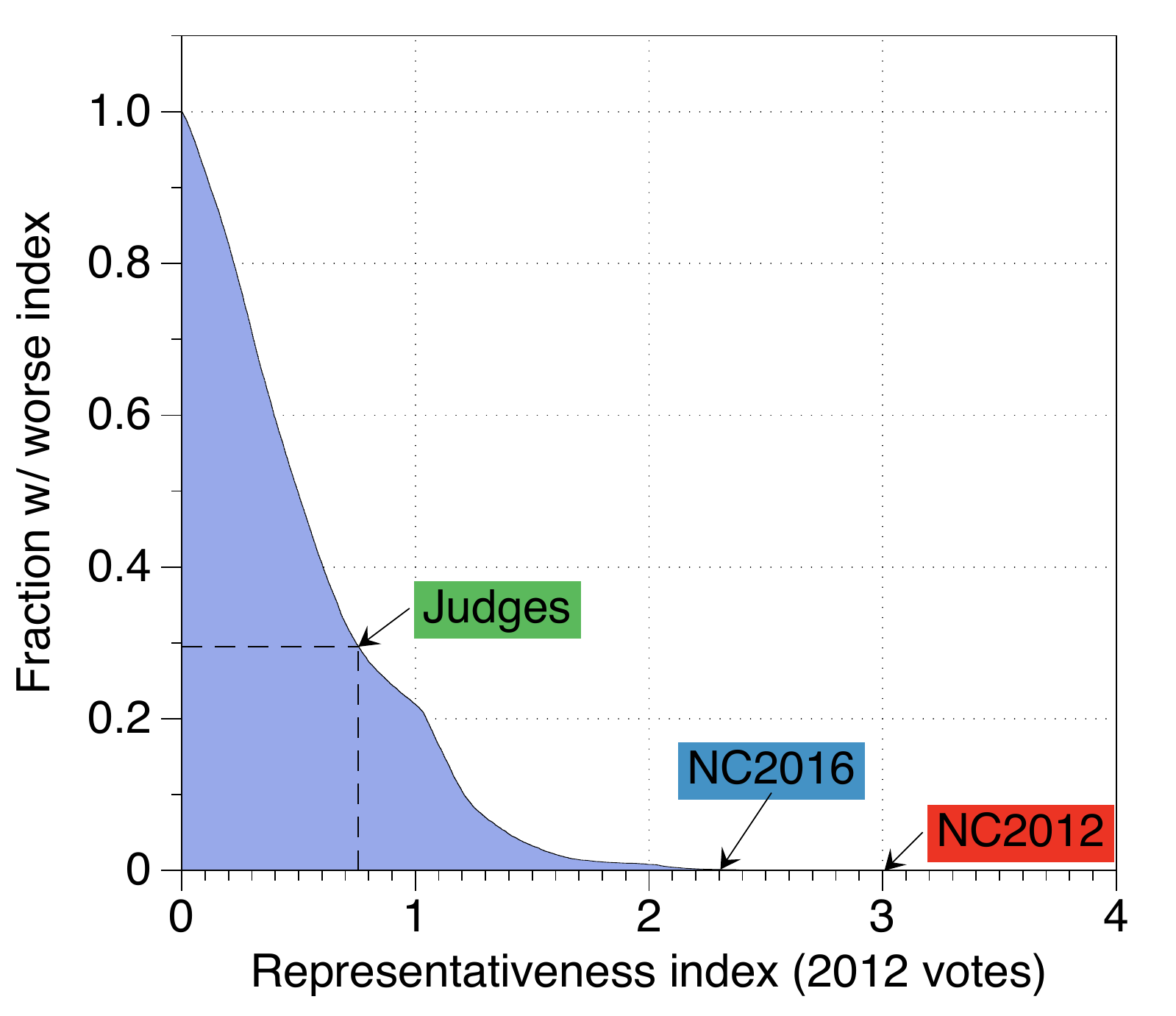}
\includegraphics[width=8cm]{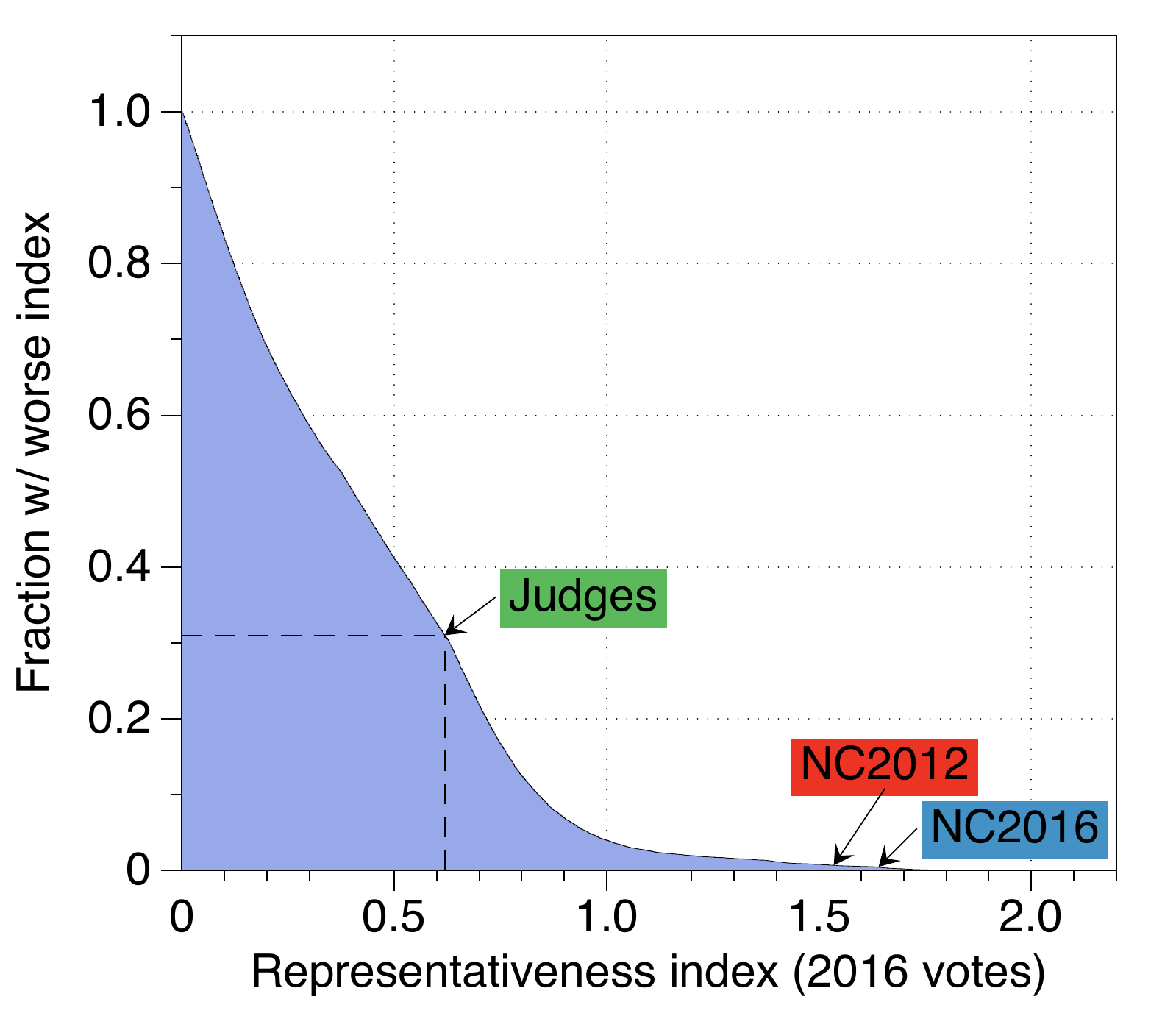}
\caption{\capSize Representativeness Index for the three districts of
  interest using congressional voting data from 2012 (left) and 2016
  (right).  No redistrictings was less representative than the NC2012
  nor NC2016 redistricting plans.  Roughly 30\% of redistricting plans
  were less representative than the Judges redistricting plan in both
  sets of voting data, meaning that the Judges plan was reasonably
  representative.}
  \label{fig:representativenessCDF}
\end{figure}

In a complementary cumulative distribution function, the vertical axis shows the fraction of random
redistrictings which have a larger index value than a redistricting with a given index
on the horizontal axis.  We plot results for the Gerrymandering Index
and 
the Representativeness Index in
Figures~\ref{fig:gerrymanderingsCDF} and  ~\ref{fig:representativenessCDF}, respectively.  We calculate the probability of
each index obtaining a value greater than a given value based on our
random redistrictings. We then situate each of our redistrictings of
interest (NC2012, NC2016, and Judges) on the plot indicating the
fraction of random redistrictings which have a larger index.

Figures~\ref{fig:gerrymanderingsCDF} and
\ref{fig:representativenessCDF}, show that the NC2012 and NC2016
redistrictings are quite atypical in both the  Gerrymandering
Index and Representative Index,  regardless if the votes from 2012 or
2016 are used in the analysis. 
None of the over 24,000 reasonable redistrictings constructed had a Gerrymandering
Index bigger than NC2012, regardless whether 2012 or 2016 votes were
used. Similarly, none of the reasonable redistrictings had a
Representativeness Index greater than NC2012 when the 2012 votes are
used and only 172 (or 0.7\%) had a greater Representativeness Index when the 2016 votes are used. Again, none
of the reasonable redistrictings had a  Gerrymandering
Index bigger than NC2016 under both the 2012 and 2016 votes. Only 34
redistrictings 
(or 0.14\%) and 105 redistrictings (or 0.43\%) had a
Representativeness Index greater than NC2016 under the 2012 and 2016
votes, respectively.

In stark contrast, 18,670 redistrictings (or 76.15\%) and 18,891
redistrictings (or 77.05\%) had larger  Gerrymandering
Index than the Judges plan under the 2012 and 2016 votes, respectively.  And 7,250
redistrictings (or 29.57\%) and 7,625  redistrictings (or 31.1\%) had
larger Representativeness Index than the Judges under the 2012 and
2016 votes, respectively. 

Our indicies indicate that the Judges plan is a very typical
plan. It has a comparatively low  level of gerrymandering 
and seems to represent the will of the people. The NC2012 and NC2016
are partially unrepresentative and have a high level of gerrymandering
in terms of both indices.

\subsection{Visualizing Gerrymandering}
\label{sec:vis-gerrrym}
While the reductive power of a single number can be quite compelling,
we have also developed a simple graphical representation to summarize the
properties of  a given redistricting relative to the collection of
referenced redistrictings. The goal was to create a graphical
representation which would make visible when a particular
redistricting packed or fractured voters from a particular party to
reduce its political power.
\begin{figure}[ht]
  \centering 
\includegraphics[width=8cm]{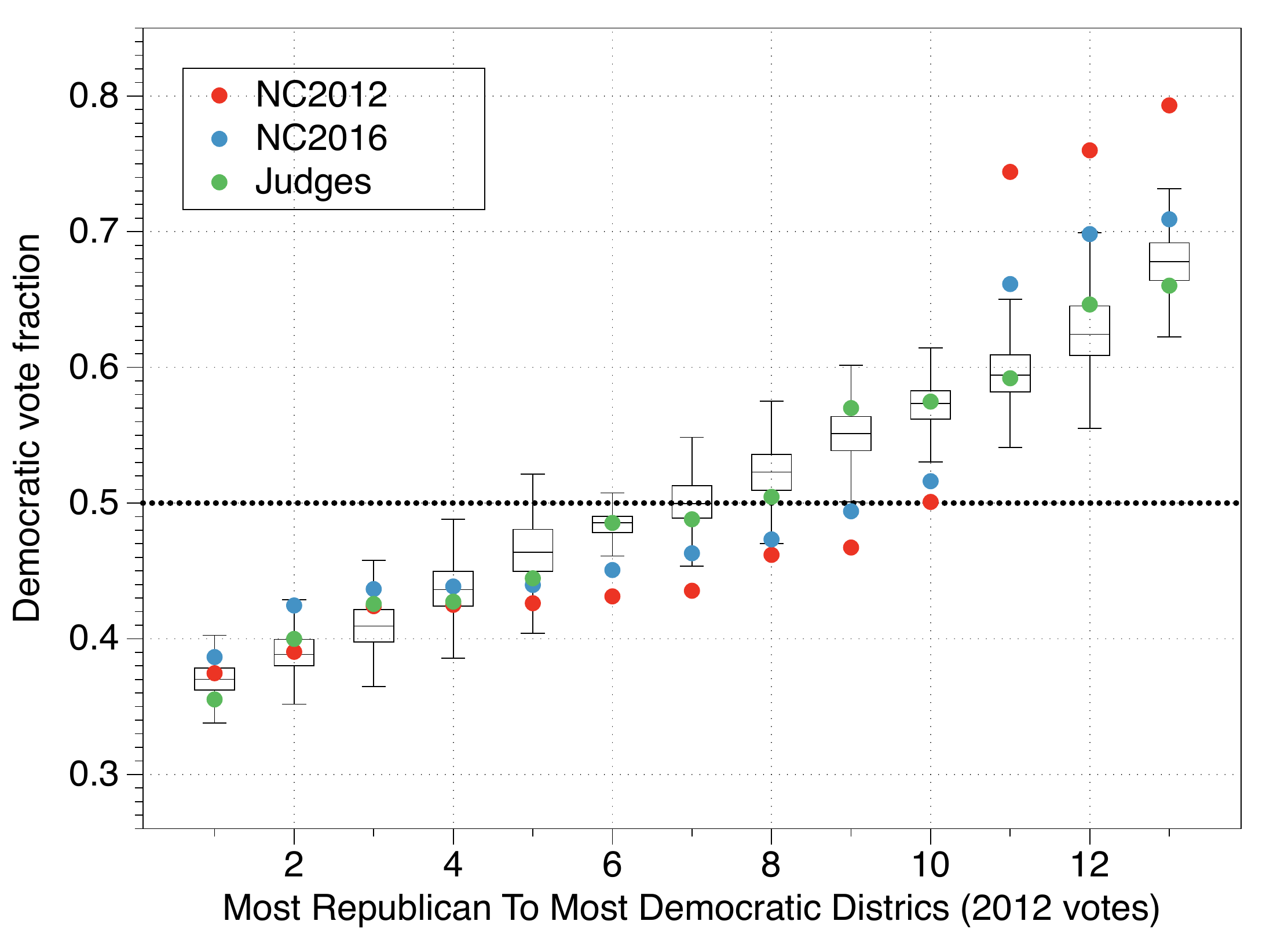}
\includegraphics[width=8cm]{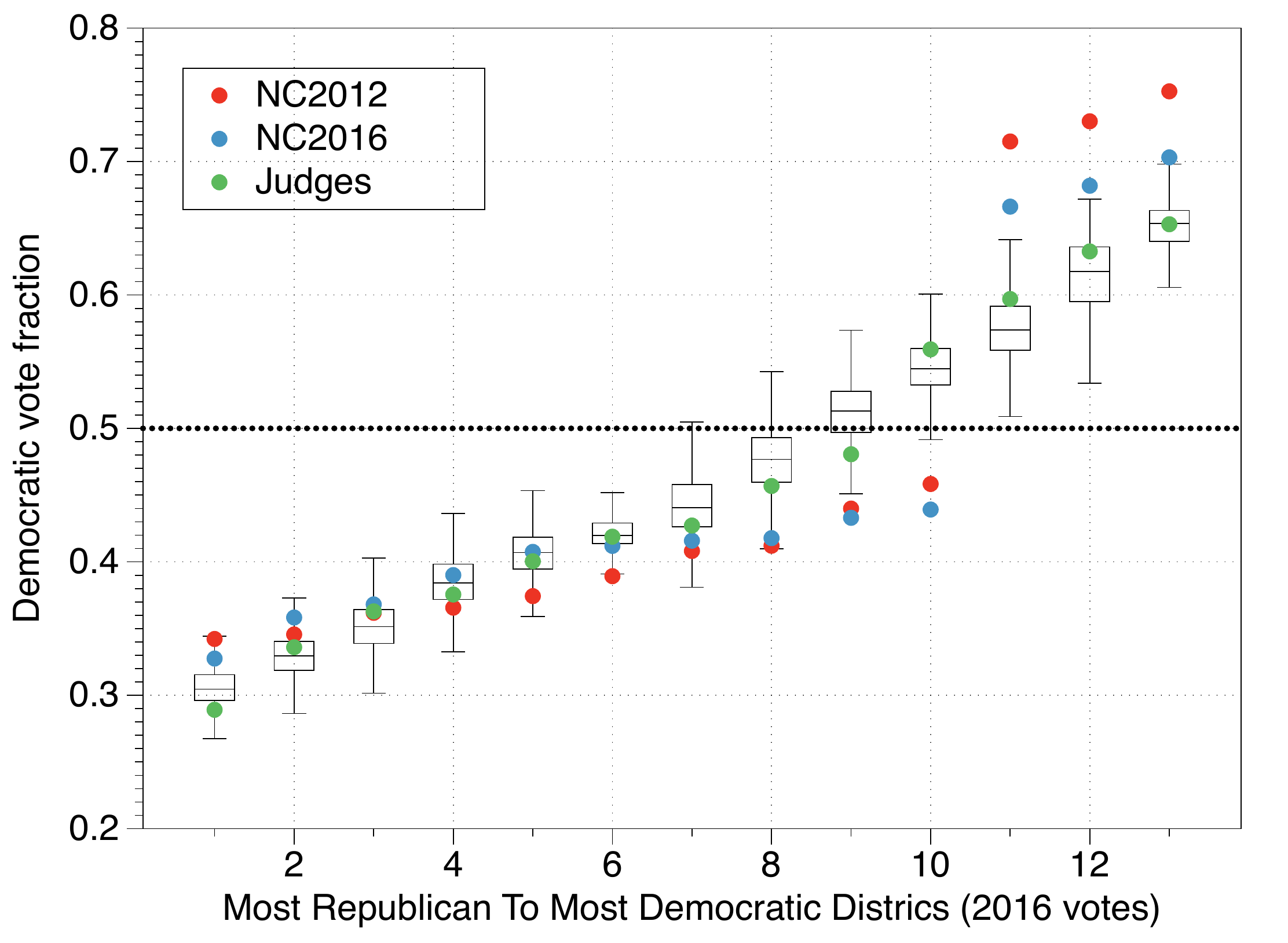}
   \caption{\capSize 
     After ordering districts from most Republican to most Democrat,
     these box-plots summarize the distribution of the percentage of
     Democratic votes for the district in each ranking position for
     the votes from 2012 (left) and 2016 (right).  We compare our
     statistical results with the three redistricting plans of
     interest. The Judges plan seems to be typical while NC2012 and
     NC2016 have more Democrats than typical in the most Democratic
     districts and fewer in districts which are often switching between
     Democrat and Republican, depending on the redistricting.
   }
  \label{fig:boxPlotsCDF}
\end{figure}

One first needs to begin by discovering the natural structure of the
geographical distribution of votes in the state when viewed through
the lens of varying over ``reasonable'' redistrictings. We begin by ordering the
thirteen congressional districts which make up a redistricting from
lowest to highest based on the percentage of Democratic votes in each
district. Since there are essentially only two parties, nothing would
change if we instead considered the percentage of Republican votes. 

We are interested in the random distribution of this
thirteen dimensional vector. Since it is difficult to visualize such a
high dimensional distribution, we summarize the distribution by
considering the marginal distribution of each position in this vector
and summarize it in a classical box-plot for each component of
the thirteen dimensional vector in
Figure~\ref{fig:boxPlotsCDF}. That is to say, we examine the
distribution of votes that make up the percentage of Democratic votes
in the most Republican district. Then we repeat the process for the
second most Republican district.  Continuing for each of the rankings,
we obtain thirteen box-plots which we arrange
horizontally on the same plot.

\begin{figure}[ht]
  \centering 
\includegraphics[width=8cm]{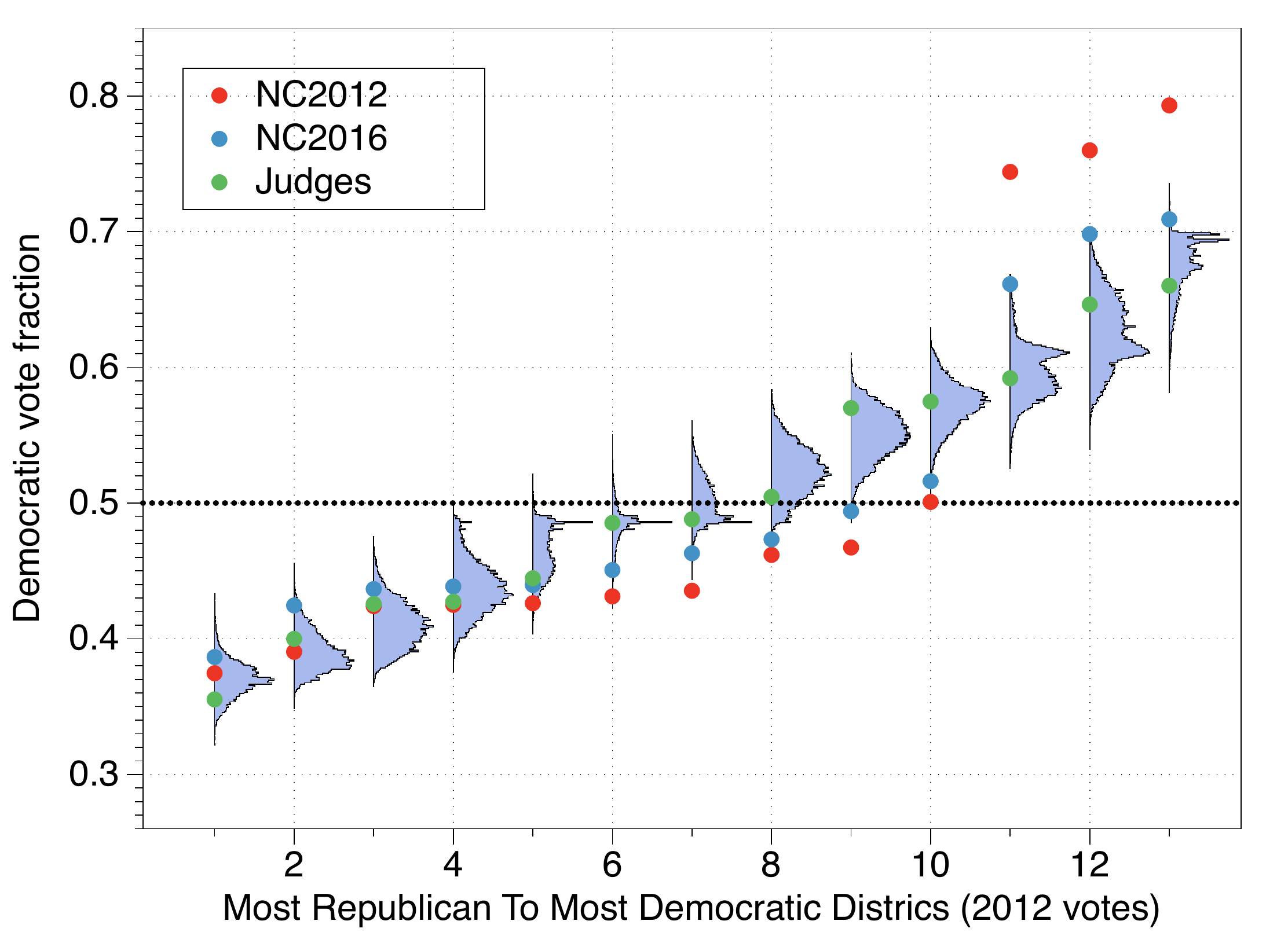}
\includegraphics[width=8cm]{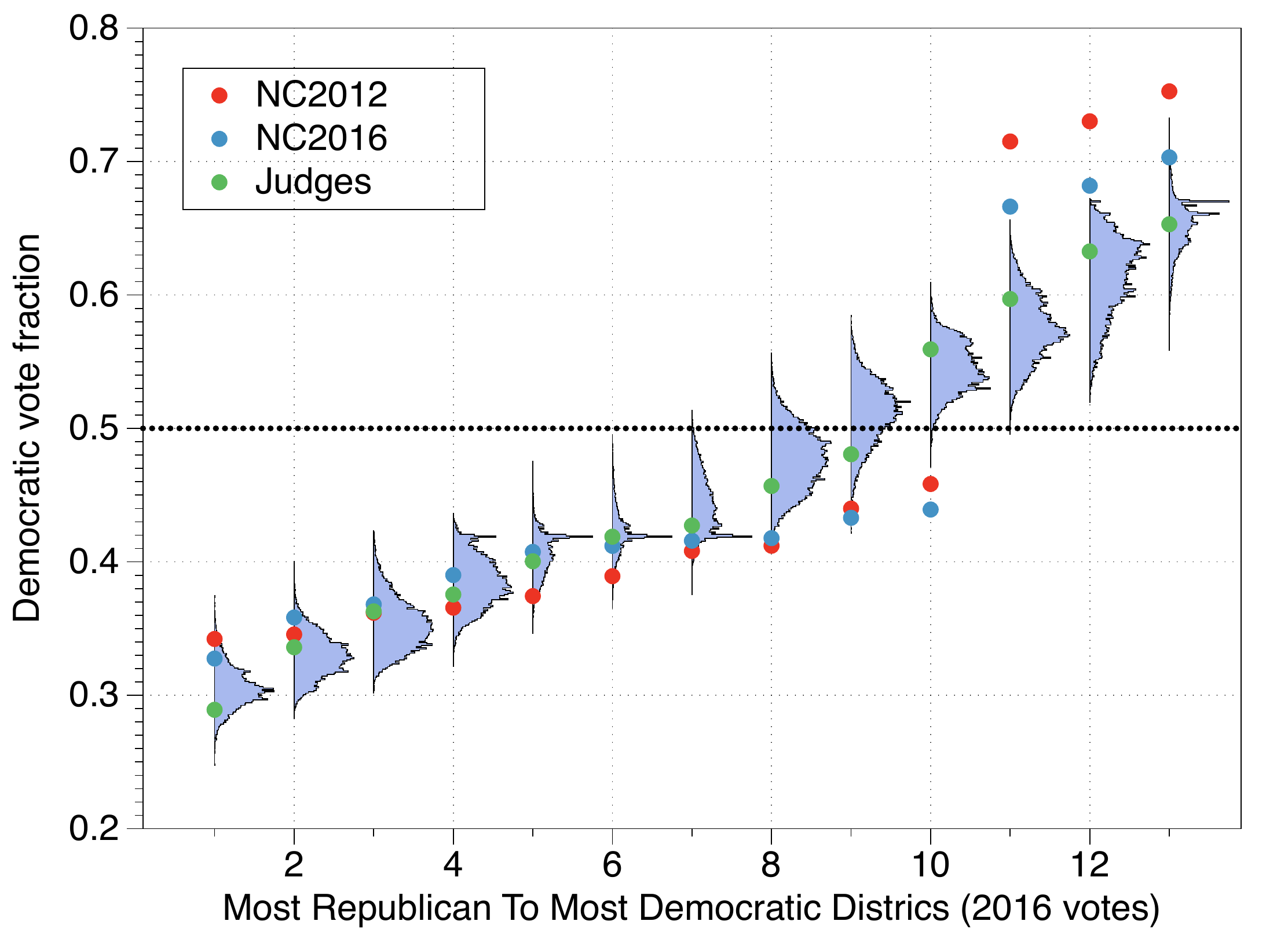}
\caption{\capSize We present the same data as in 
  Figure~\ref{fig:boxPlotsCDF}, but, with histograms replacing the box-plots. Note that the distribution of the sixth most
Republican district (district with label 6 on the plots) is quite
 peaked in both the 2012 and 2016 votes, the Judges results are centered directly
 on this peak while the NC2012 and NC2016 lie well outside the extent
 of the distribution. }
  \label{fig:DensityVersion}
\end{figure}

The box-plots are standard, meaning that within each box-plot
the central line gives the median percentage, while the ends of the box give
the location of the upper quartile and the lower quartile (25\% of the
results exist below and above these lines). The outer bracketing line
defines an interval containing either the maximum and minimum values,
or three halves the distance of the quantiles from the mean, whichever is smaller. In the interest of visual clarity, we have not plotted
any outliers. On top of these box-plots, we have overlaid the
percentages for the NC2012, NC2016, and the Judges redistricting.  In
Figure~\ref{fig:DensityVersion}, we also include plots that displays
histograms rather than box-plots.  These plots are richer in detail.
Yet, the detail makes it harder to estimate confidence intervals.

The structure of these plots is shaped by the typical structure of the
redistrictings in our ensemble and by extension, the spatial-political
structure of the voters in North Carolina.  It can then be used to
reveal the
structure of our three redistrictings of interest. Observe that for
both the 2012 and 2016 votes, the centers of the box-plots form a
relatively straight, gradually increasing line from the most Republican district (labeled 1)
to the most Democratic (labeled 13). The Judges districts mirror this
structure. Furthermore, most of the percentages from the Judges
districts fall inside the box on the box-plot which marks the central
50\% of the distribution.  The NC2012 and NC2016 have a different
structure. There is a large jump between the tenth and eleventh most
Republican district (those with labels 10 and 11, respectively). In the NC2012
redistricting, the fifth through tenth most Republican districts have
more Republicans than one would typically see in our ensemble of
``reasonable'' redistrictings. In the NC2016 redistricting, the
shifting starts with the sixth most Republican district and runs through the
tenth most Republican district (labeled 6-10). In both cases, the votes removed
from the central districts have largely been added to the three most
Democratic districts (labeled 11-13). In the 2012 votes, this moved
three to four districts that typically would have been above the 50\% line
to below the 50\% line, meaning that these districts elected the
Republican rather than the Democrat. With the 2016 votes, the changes in
structure only moved the tenth most Republican district across the
50\% line. The geographic districts associated with these rankings are given
in Table~\ref{tab:table1} from Section~\ref{sec:raw}.

Forgetting
 about the election outcomes, the structure has implications for the
 competitiveness of districts and likely political polarization. Rather than a gradual 
 increase at a constant rate from left to right as the Judges
 redistricting and the  ensemble  of box-plots, the NC2012 and NC2016
 redistrictings have significantly more Democrats in the three most
 Democratic districts and fairly safe Republican majorities in the first
 eight most Republican districts. There are established hypotheses
 that claim safe districts lead to a polarized legislative delegation
 with fewer centrist representatives on both sides of the political
 spectrum. 

 Figure~\ref{fig:boxPlotsCDF} can be used to motivate and explain the
 Gerrymandering Index for our redistrictings of interest.  For example, to calculate the
 Gerrymandering Index for NC2012, one sums the square of the distance
 from the red dots to the mean in each distribution from 1 to 13. The
 Gerrymandering Index is the square root of this
 sum. To aid with visualization, recall that the line though the
 center of each box is the median which, in these cases, is close to
 the mean.  Clearly, this index captures some of the features of
 Figure~\ref{fig:boxPlotsCDF} discussed in the previous paragraph.

It is remarkable how stable the structures in
 Figure~\ref{fig:boxPlotsCDF} are across the 2012 and 2016 votes. The
 2016 plot is largely a downward shift of the 2012 plot. This
 stability largely explains why the two plots in
 Figure~\ref{fig:gerrymanderingsCDF} of the Gerrymandering Index look
 so similar. It also speaks to the stability of the Representativeness
 Index in Figure~\ref{fig:representativenessCDF}. The efficiency gap
 (introduced in Section~\ref{sec:EffGap})
 does not seem to share this stability as it changes both  its values and probabilities in  Figure~\ref{fig:wastedCDF} across the two elections.

\subsection{Stability of Election Results}
\label{sec:stab-elect-results}
It would be unsettling to think that relatively small changes in a
redistricting would dramatically change the results of an election or
the properties of the redistricting. If this were true, it would call into question the legitimacy of the election results derived from the
redistricting. In particular it might lead one to question the extent
to which an election captures the intent of the votes.

To examine this question, we explored the degree to which the NC2012,
NC2016 and Judges redistrictings are representative of the nearby
redistrictings, where we interpret nearby to mean that roughly 10\% of
the VTDs are swapped between districts. (See the next paragraph for
more precise description.)  By switching nearby VTDs among districts
we are able to assess whether small changes impact the characteristics
of the districts or not.  We found that the districts within the
NC2012 and NC2016 redistricting plans had a Gerrymandering Index which
was significantly larger than the nearby redistrictings while the 
Judges plan had a Gerrymandering Index which was in the middle of the
range produced by nearby redistrictings. In other words, switching
nearby districts made the NC2012 and NC 2016 redistrictings less
partisan but did not change the characteristics of the Judges
redistricting. This suggests that the NC 2012 and NC2016
redistricting, in contrast to the Judges’ redistricting, were
precisely engineered and tuned to achieve a partisan goal and that the
components of the NC2012 and NC 2016 redistrictings were
not randomly chosen.
\begin{figure}[ht]
  \centering 
\includegraphics[width=5.25cm]{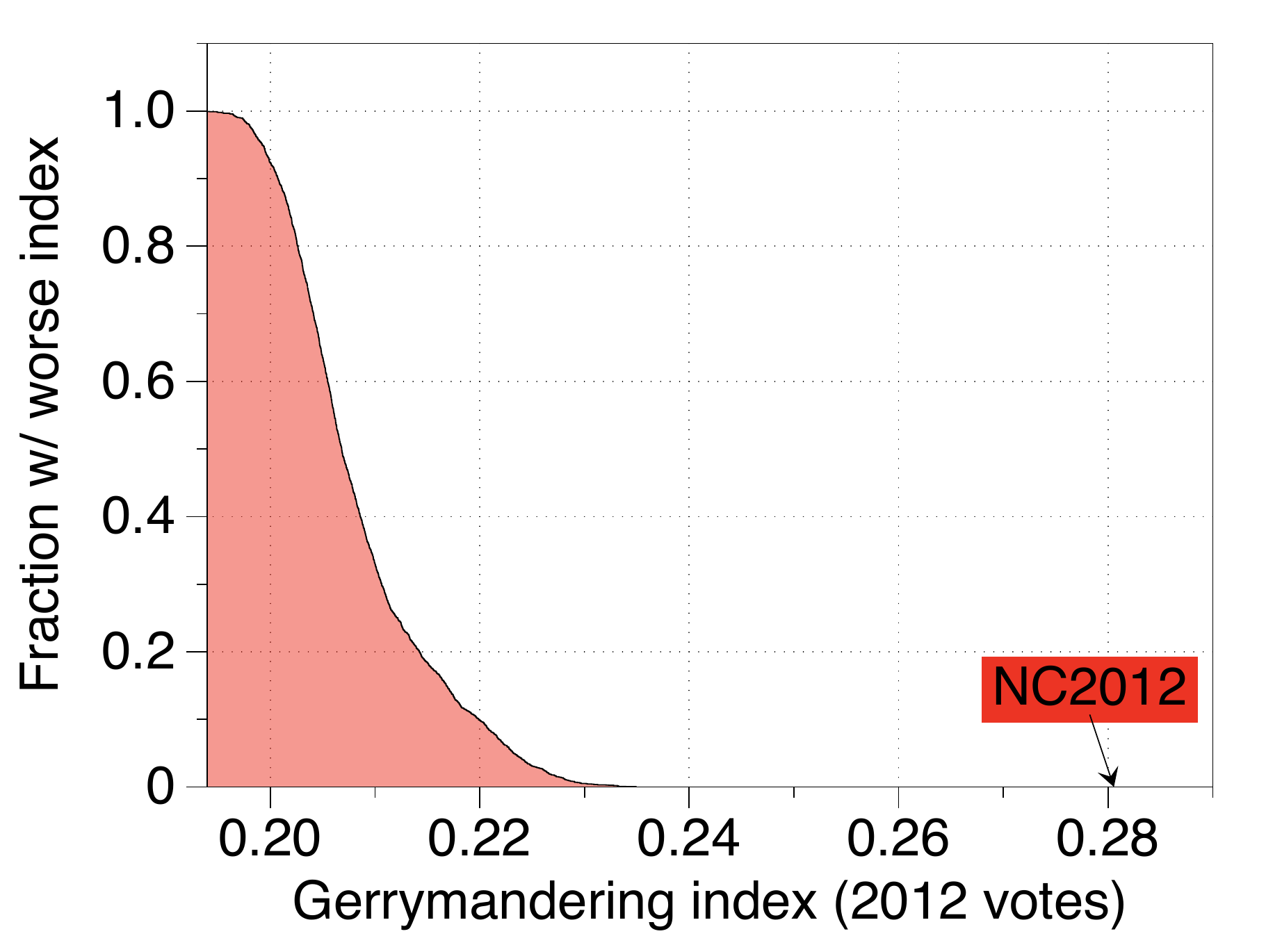}
\includegraphics[width=5.25cm]{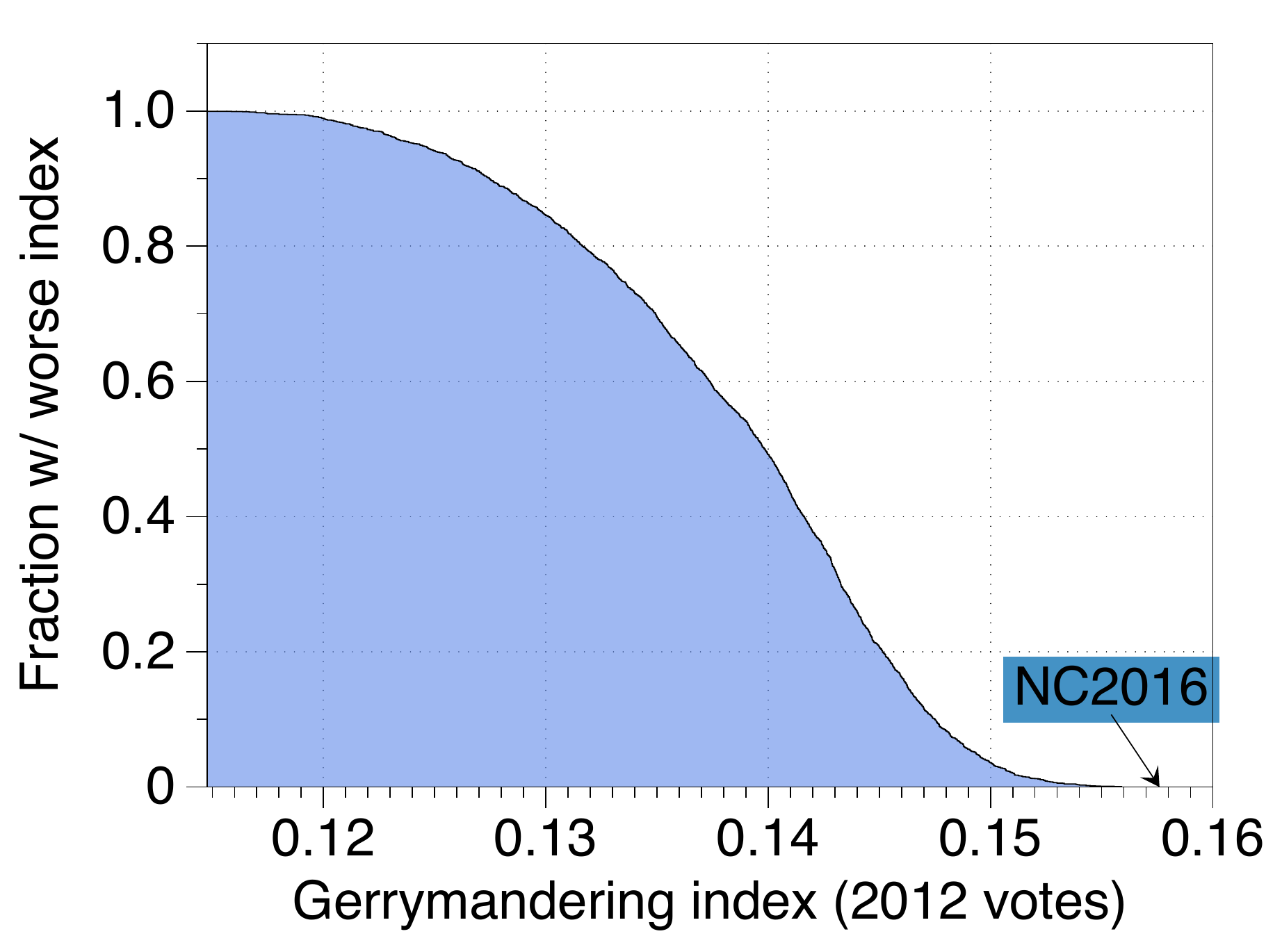}
\includegraphics[width=5.25cm]{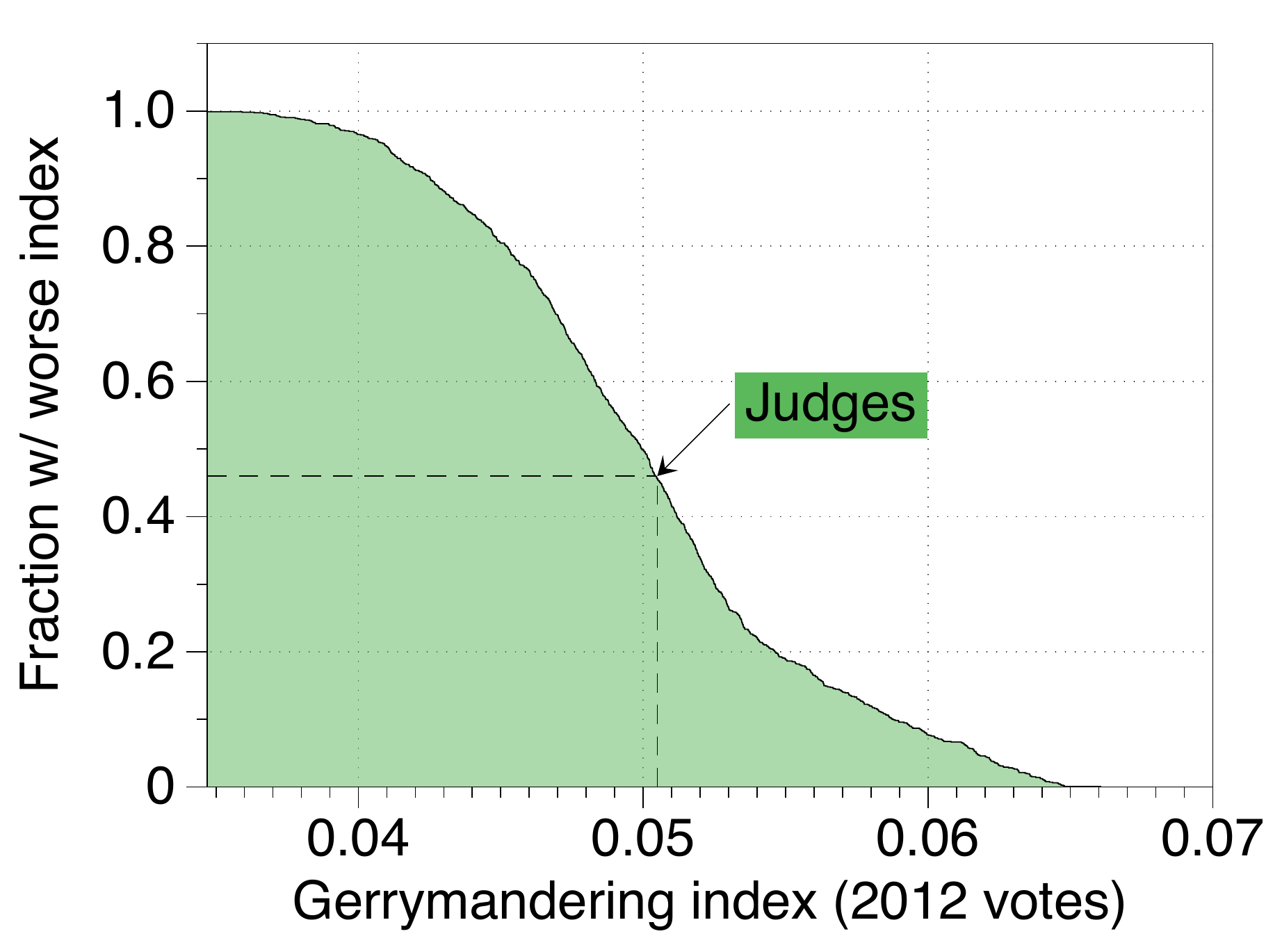}
   \caption{\capSize  Gerrymandering Index based on random samples
     drawn from nearby the three redistrictings of interest: NC2012
     (left), NC2016 (center), and Judges (right). Only for the Judges
     are the other points in the neighborhood similar to the
     redistricting of interest. All plots use the 2012 votes.}
  \label{fig:HDBoxPlots}
\end{figure}

More precisely, we randomly sample ``reasonable'' redistrictings which are
near the NC2012, NC2016, and Judges redistrictings in the sense that
no single district differs by more VTDs than a set threshold from the
redistricting understudy. To set the threshold, we observe that among the over 24,000 redistrictings we generated, the average district size
is around 210 VTDs. In a particular typical redistricting from our  ensemble, the sizes
roughly varied from 140 to 280 VTDs. 
 With these numbers in mind, we set our
threshold to be 40 VTDs. Since every VTD
switched is counted twice, once for the district
it is leaving and once for the district it is
entering, this amounts to a total of around 10\% of the VTDs
switching districts. 

Figure~\ref{fig:HDBoxPlots} shows the results of these analyses
applied to  the NC2012, NC2016, and Judges redistrictings. The
redistrictings sampled around NC2012 have markedly better
Gerrymandering Indices than NC2012 itself. The results are less dramatic for
NC2016, but telling nonetheless. This shows that  a
randomly chosen redistricting near NC2012 (or respectively NC2016)
has very different properties than NC2012 (or respectively NC2016). This
is convincing evidence that the NC2012 and NC2016 redistrictings were
deliberately constructed to have unusual properties. It would have
been unlikely to choose such a singularly unusual redistricting by chance. In contrast, the
Judges redistricting has a Gerrymandering Index which is quite typical
of its nearby redistrictings. It is  worse than
around 50\% of those nearby it and hence better than  50\% of those
nearby it. Thus, it is very representative of its nearby redistrictings.

\subsection{Efficiency  Gap and correlation between indices}
\label{sec:EffGap}

The Efficiency Gap is a third type of index that was used in the 
decision Whitford Op. and Order, Dkt. 166, Nov. 21, 2016.  It 
quantifies the difference of how many ``wasted votes'' each party 
cast; a larger number means that one party wasted more votes 
than another. 
More precisely, the Efficiency Gap is the difference of the wasted votes for the Democrats and Republicans divided by the number of the total votes in the election (for both parties). The wasted votes for 
  each party is the sum of the fraction of votes in districts the party 
  loses plus the sum over the percentage points above 50\% in the 
  districts won.
  \footnote{The original 
    used actual votes, but when the population of each district is 
    equal then the two measures are exactly equivalent. If the actual 
    votes is close to equal then they are almost the same.}

\begin{figure}[ht]
  \centering 
\includegraphics[width=8cm]{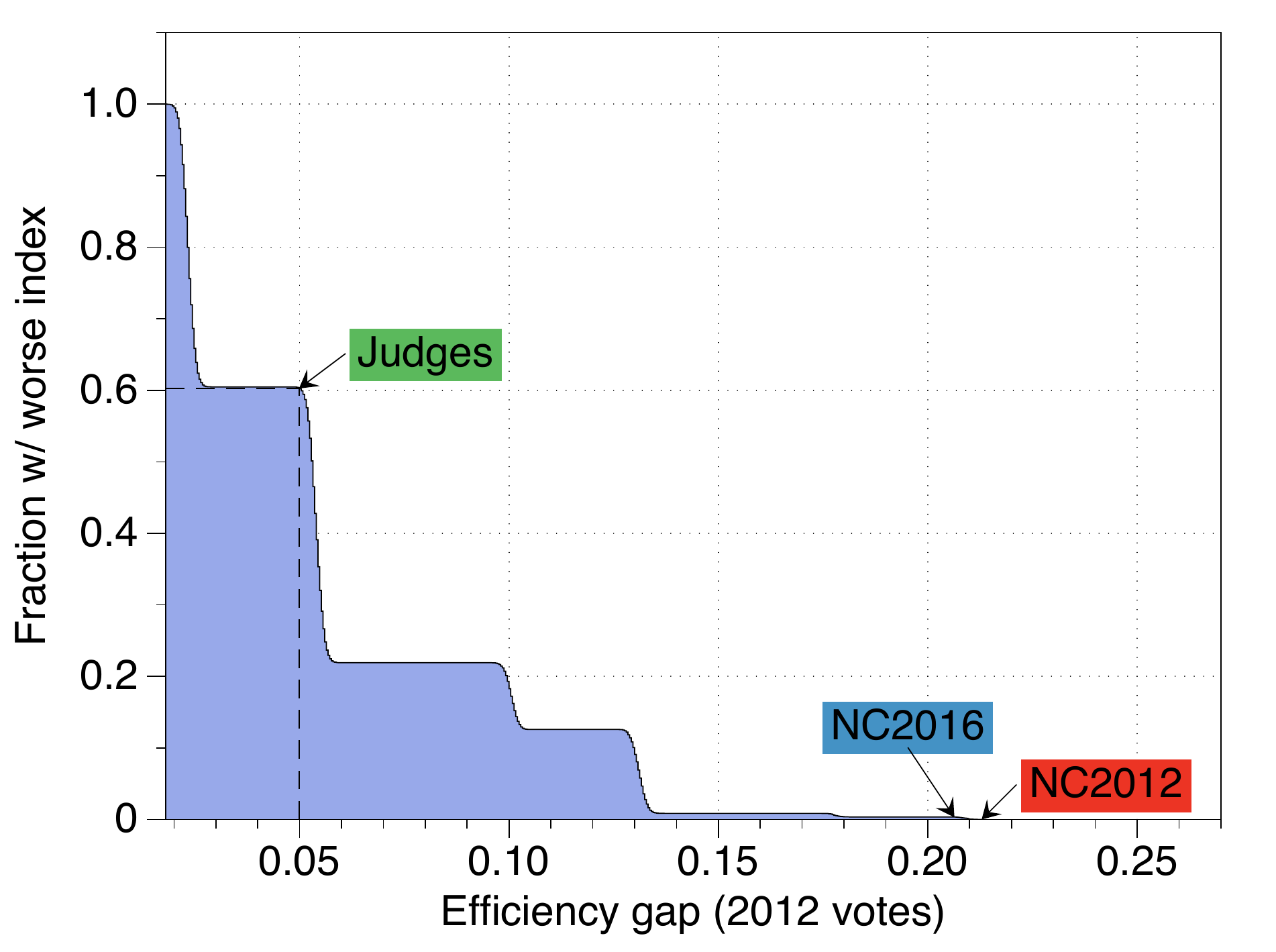}
\includegraphics[width=8cm]{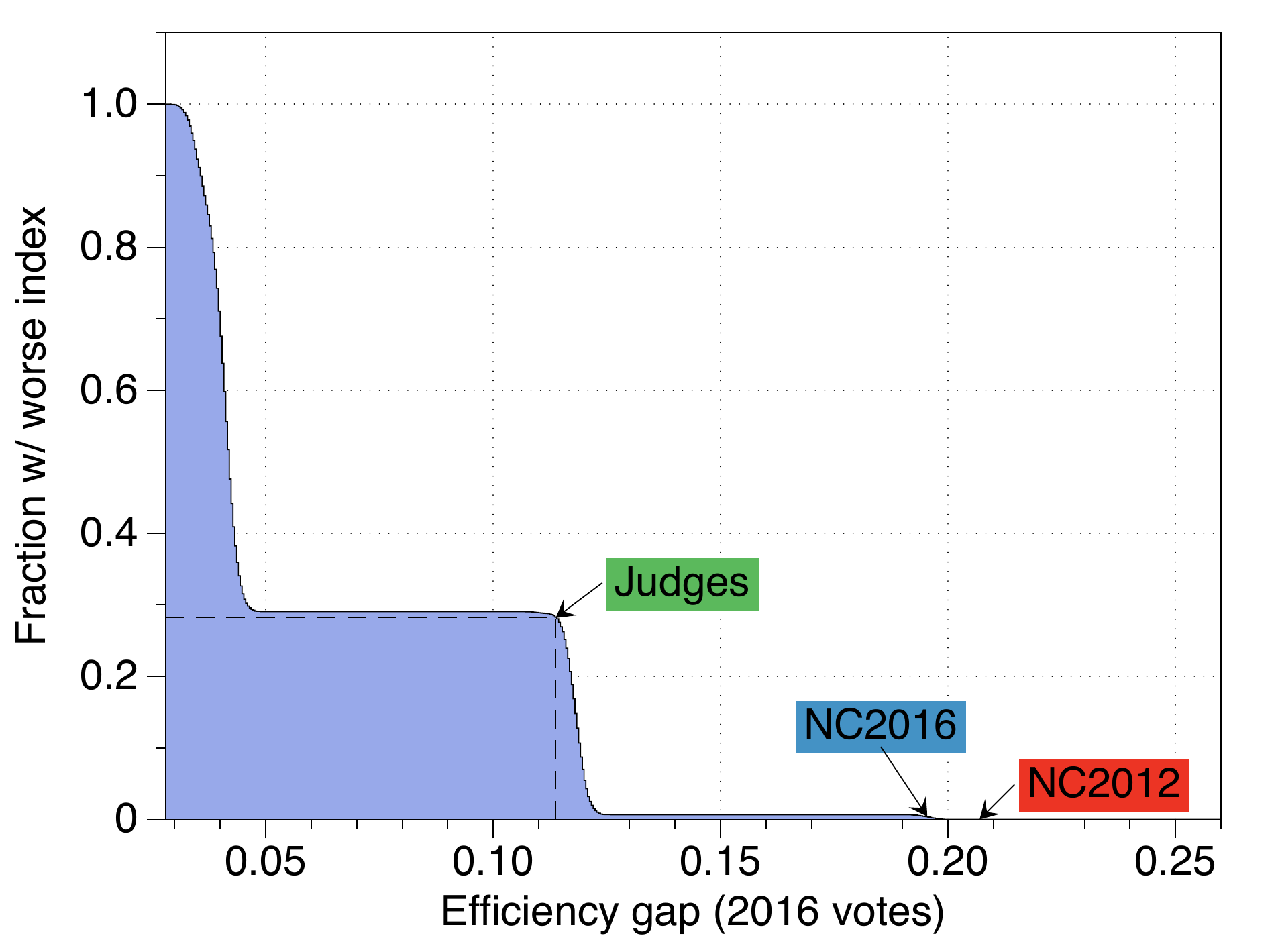}
   \caption{\capSize Efficiency gap for the three districts of 
     interest based on the congressional voting data from 2012 (left) 
     and 2016 (right). No random redistrictings had a greater 
     disparity in voter efficiency than the NC2012 redistricting plan. 
     Only about 0.3\% of the random redistrictings had a greater 
     disparity in voter efficiency than the NC2016 redistricting in both 
     sets of voting data.  Roughly 60\% and 30\% of the random 
     redistricting plans had a greater disparity in voter efficiency 
     than the Judges redistricting plan for the 2012 and 2016 election 
     votes, respectively.}
  \label{fig:wastedCDF}
\end{figure}

As with the other indices we are most interested in where a 
particular Efficiency  Gap score sits relative to Efficiency  Gap 
scores of our ensemble of 24,000 generated 
redistricts. In a graph completely analogous to the previous plots of 
complementary cumulative distribution functions, 
Figure~\ref{fig:wastedCDF} gives the fraction of the random ensemble 
with an Efficiency Gap greater than a given score. 

At first glance, the Efficiency Gap seems to have some similarity with
the Gerrymandering Index. Both try to detect the packing and cracking
of a particular political constituency. While the Efficiency Gap
concentrates on the winner of a given district by using a threshold of
50\% to determine what are excess votes, the Gerrymandering Index
tries to develop the unique signature of how a given collection of votes
interacts with the geographic distribution of the votes.

To understand how related the Gerrymandering and Representativeness
Indices are to each other and to the  Efficiency Gap, we consider the Pearson product-moment correlation coefficient. 
 Using the voting data of 2012, we find that the correlation between the Gerrymandering Index and the 
Representative Index is 0.466, meaning the two indices are moderately
correlated. This suggests that gerrymandered redistrictings tend to be
less representative and vice versa. We find that the Gerrymandering Index and the Efficiency
Gap are less positively correlated with a correlation coefficient of
0.333.  On the other hand, the 
Representative Index and the efficiency gap are more highly correlated with 
a correlation coefficient of 0.752. Hence, despite superficial
similarity, the  Gerrymandering Index and the Efficiency
Gap contain the most distinct information while the Efficiency
Gap and  Representative Index are more related. Though we only have
two sets of election data (namely 2012 and 2016), the Gerrymandering
and Representativeness Indices complementary CDF plots are much more
stable across elections than the Efficiency
Gap appears to be. One can not conclude much from two examples, but it
is suggestive that the Efficiency
Gap may be a less stable index.

\subsection{Summary of Main Results}
\label{sec:summary-main-results}
By sampling over 24,000 reasonable redistrictings, we explore the distribution
of different election outcomes by estimating the probabilities of
the numbers of  Democrats elected from North Carolina to the
U.S. House of Representatives. Our sampling of reasonable
redistrictings also allows us to estimate the distribution of winning
margins in each district as well as the value of three indices
representing gerrymandering, representativeness, and relative voter
efficiency.  In every one of our tests, we have found that the NC2012
and NC2016 redistricting plans are extraordinarily anomalous,
suggesting that (i) these districts are heavily gerrymandered, (ii)
they do not represent the will of the people and (iii) they dilute
the votes of one party.  We have also uncovered evidence that these
two redistricting plans employ packing and cracking.  On the contrary,
the  redistricting plan  produced by a bipartisan redistricting commission of retired judges from the Beyond Gerrymandering
project produced results which were highly typical among our  24,000 reasonable
redistrictings. The Judges plan was exceptionally non-gerrymandered,
was a typical representation of the people's will, and does
not seem to pack or crack either party.

We also explored the degree to which the NC2012, NC2016 and Judges
redistrictings were  representative of the nearby redistrictings,
where we interpret nearby to mean that roughly 10\% of the VTDs
are switched out of any given district. We found that the  NC2012 and
NC2016 redistrictings were significantly more gerrymandered than those
around them while the Judges redistricting was similar to those
nearby. This seems to imply that the  NC2012 and
NC2016 redistrictings were carefully engineered and tuned, and not randomly chosen among those
with a certain basic structure. 

The remainder of the paper is organized as follows. In the remainder
of this section we describe the Beyond Gerrymandering project and
situate this work in previous reports produced by this and associated
teams. In Section~\ref{sec:sampl-rand-redistr}, we describe how we
construct our distribution of ``reasonable'' redistrictings and sample
it using Markov chain Monte Carlo, { as well as provide some brief
  historical context concerning similar efforts}. In
Section~\ref{subsec:threshold}, we describe how we further sub-select
our samples based on a series of thresholds to better reflect the
proposed bipartisan redistricting legislation HB92. In
Section~\ref{ssec:determine-param}, we discuss how the parameters of
our distribution are chosen to produce ``reasonable''
redistrictings. In Section~\ref{sec:characteristics} we describe
characteristics of our generated ``reasonable'' redistrictings.  In
Section~\ref{sec:VRA}, we explore the effect of the Voting Rights Act
provision in HB92 on the outcome of the elections. In
Section~\ref{sec:DetailsIndicies}, we give the missing details in the
construction of the Representativeness and Gerrymandering Indices.  In
Section~\ref{sec:test-sens-results}, we show that our results are
insensitive to our choice of parameters. We also provide evidence that
our Markov chain Monte Carlo is running sufficiently long to produce
results from the desired distribution. In
Section~\ref{sec:more-details-results}, we provide some details about
the data used and some of the more technical choices made in the
preceding analysis. Finally in Section~\ref{sec:conclusions}, we make
some concluding remarks and discuss future directions. The Appendix of
the paper gives some sample maps drawn by our algorithm.

\subsection{The Beyond Gerrymandering Project} 
\label{sec:beyondGerrymandering}The Beyond
Gerrymandering\footnote{ For more information see
  \texttt{https://sites.duke.edu/polis/projects/beyond-gerrymandering/}}
project was a collaboration between UNC system
President Emeritus and Davidson College President Emeritus Thomas
W. Ross, Common Cause, and the POLIS center at the Sanford School at
Duke University. The project's goal is to  educate the public on how
an independent, impartial redistricting process would work. The
project  formed
an independent redistricting commission made up of ten retired jurists, five Democrat and five Republican. 
The
commission used
strong, clear criteria to create a new North Carolina congressional
map based on NC House Bill 92 from the 2015 legislative season. All
federal rules related to the Voting Rights Act were followed but no
political data, election results or incumbents’ addresses were
considered when creating new districts. The commission met twice over
the summer of 2016 to deliberate and draw maps. The maps resulting from this
simulated redistricting commission were released in August 2016. The
Judges agreed on a redistricting at the level of Voting Tabulation
Districts (VTD). This coarser redistricting was refined at the level of
census blocks to achieve districts with less than 0.1\% population
deviation. The original VTD based maps are used in our study.

\subsection{Related Works}
Ideas to 
generate redistricting plans with
computational algorithms have been being developed since the 1960's
\cite{nagel1965simplified,thoreson1967computers,gearhart1969legislative}.
There are three main classifications of redistricting algorithms:
constructive Markov Chain Monte Carlo (MCMC) algorithms
\cite{cirincione2000assessing,ChenRodden13,Chen15}, moving boundary MCMC algorithms
\cite{macmillan2001redistricting,MattinglyVaughn2014,QuantifyingGerrymandering,Wu15,fifield2015},
and optimization algorithms \cite{mehrotra1998optimization,Liu16}.  Constructive MCMC algorithms begin each new redistricting with an initial random seed and grow districts.   Moving boundary MCMC algorithms find new redistricting plans by altering district boundaries.  In
\cite{fifield2015}, the authors demonstrate that moving boundary MCMC
algorithms are better at sampling the  redistricting space than
constructive algorithms.  It is proven that the former will
theoretically sample the space with the correct probability
distribution, where as the latter may construct many similar
redistrictings of one kind while not generating as many equally likely
redistricting plans, leading to a skewed distribution.  Optimization
algorithms are primarily concerned with generating one or a collection
of `elite' districts, as opposed to sampling the space of all
districtings.  Recently an evolutionary algorithm has been proposed
which begins with a constructive method, but then use mixing to find
either elite or `good enough' redistrictings \cite{Liu16}; it used the
collection of `good enough' redistrictings to make statistical
predictions, however it is still unclear how evolutionary algorithms
compare with Monte Carlo models in terms of sampling the space
properly.  One advantage of the moving boundary MCMC approach is that we
sample form a explicitly specified and constructed probability
distribution on redistrictings.
  We remark that all of the above works have considered minimizing population deviation and compactness; a few of the works have considered minimizing county splitting; none of these works have included Voting
Rights Act requirements (see Section~\ref{sec:VRA} below). We present our algorithm for sampling the space of redistrictings in Section~\ref{sec:sampl-rand-redistr}, which is a moving boundary MCMC algorithm.  

Once a sampling of the space redistrictings is produced, there are a
variety of existing indices that have been used as a comparison metric
for a given districting (see \cite{ChoLiu16} and references therein
for a summary and history of these indices).  Such indices include
competitiveness, responsiveness, biasedness, dissimilarity, and
efficiency.  Similar to our work, the typical idea found in the
literature is to contextualize a given redistricting plan in the
context of these indices. Each of these indices provides a valid
method of determining if a given districting plan is typical in the
context of the ensemble of generated redistrictings.  None of these
previous indices, however, account for the underlying geography of a
given state.  As pointed out in \cite{ChenRodden13}, maps may
naturally be less competitive simply based on geography rather than
partisan tampering.  The gerrymandering and representative indices we
have introduced in this work are the first, to our knowledge,  that account for the geography of a given state; in particular, the Gerrymandering Index gives the first metric of how gerrymandered a given redistricting plan is, in the context of drawing plans without partisan data.  This coupling of indices may provide a more robust metric.  We have seen some preliminary evidence of this in Sections~\ref{sec:EffGap}~and~\ref{sec:meas-gerrrym}, however more study is required to test this hypothesis.

\subsection{Evolution of This Project}
This work originated as a PRUV project and subsequent senior thesis of Christy
Vaughn (now Christy Vaughn Graves), both of which were supervised by Jonathan Mattingly during the summer of 2013 and the
academic year 2013-2014. This initial phase of the project
concentrated on North Carolina and was
summarized in the technical report  ``Redistricting and the Will of
the People.'' (See \cite{MattinglyVaughn2014} from references.)

That work
grew into a summer undergraduate research project as part of the
Data+ program in the Information Initiative at Duke\footnote{see
  \texttt{http://bigdata.duke.edu/data}} during the summer of 2015. This work analyzed a number
of different states and introduced VRA and county fragmentation
considerations. The research team consisted of Duke undergraduates
Christy Vaughn Graves, Sachet Bangia, Bridget Dou, and Sophie Guo and
was again mentored by Jonathan Mattingly.
The work is summarized at the online resource
``Quantifying Gerrymandering'' (See \cite{QuantifyingGerrymandering} from references.).

During the
summer of 2016, a second Data+ team was formed with the intention of analyzing the
redistrictings produced by the judges in the Beyond
Gerrymandering project. (See
Section~\ref{sec:beyondGerrymandering}  from references.). The Summer 2016 Data+ team
consisted of Duke undergraduates Hansung Kang and Justin Luo,
graduate mentor Robert Ravier, post-doc mentor Greg Herschlag, and
faculty mentor Jonathan Mattingly.

This report is based on the new code base developed by the Summer 
2016 Data+ team. It uses a refined formulation that builds on the work
of the previous teams.  Christy Vaughn Graves and  Sachet Bangia provided
important continuity between the years. The 2016 Data+ team developed
new analytic tools, some based on the useful visualizations provided by
the box-plots developed by the summer 2015 Data+ team. Some of the text
and arguments from the 2014 report have been integrated into this new report.

\section{Random Sampling  of Reasonable Redistrictings}
\label{sec:sampl-rand-redistr}
Central to our analysis is the ability to generate a large number of
different ``reasonable'' redistrictings. This is accomplished by
sampling a probability distribution on possible
redistrictings of North Carolina. The distribution is constructed so
that it is concentrated on  ``reasonable'' redistrictings. We 
then filter the randomly drawn redistrictings, using only those which
satisfy our criteria for being ``reasonable.''

As already mentioned, we  take our definition of ``reasonable''
redistrictings from the unratified House Bill 92 (HB92) from the 2015
Session of the North Carolina General Assembly \footnote{Nonpartisan Redistricting Commission. House Bill 92. General Assembly of North Carolina Session 2015. House DRH10039-ST-12 (02/05)} which stated that a bipartisan commission should draw up
redistrictings while observing the following principles:
\begin{itemize}
\item \S120-4.52(f): Districts must be contiguous; areas that meet only at points are not considered to be contiguous.
\item \S120-4.52(c): Districts should have close to equal populations, with deviations from the ideal population division within 0.1\%.
\item  \S120-4.52(g): Districts should be reasonably compact, with (1) the maximum length and width of any given district being as close to equal as possible and (2) the total perimeter of all districts being as small as possible.
\item \S120-4.52(e): Counties will be split as infrequently as possible and into as few districts as possible.  The division of Voting Tabulation Districts (VTDs) will also be minimized.
\item \S120-4.52(d): Redistrictings should comply with pre-existing federal and North Carolina state law, such as the Voting Rights Act (VRA) of 1965.
\item \S120-4.52(h): Districts shall not be drawn with the use of (1) political affiliations of registered voters, (2) previous election results, or (3) demographic information other than population.  An exception may be made only when adhering to federal law (such as the VRA).
\end{itemize}
We  restrict our probability distribution to redistrictings which
have connected districts. The remaining principles are encoded in
a \textit{score
function} which is minimized by redistrictings that are most
successful at satisfying the remaining design principles. We introduce some mathematical formalisms in order to describe the score function.

We represent the state of North Carolina as a graph $G$ with
edges $E$ and vertices $V$. Each vertex represents a Voting
Tabulation District (VTD) and an edge between two vertices exists if
the two VTDs are adjacent on the map. This graph representing the
North Carolina voting landscape has over 2500 vertices and over
8000 edges. 

Since North Carolina has thirteen seats in the U.S. House of Representatives, we define a redistricting plan to be a function from the set of vertices to the
integers between one and thirteen. More formally, recalling that $V$ was
the set of vertices, we  represent a redistricting plan by a
function $\xi: V \rightarrow \{1,2,\dots,13\}$.  If a VTD is represented by a vertex $v \in V,$ then $\xi(v)=i$ means that the VTD in question belongs to district $i.$
Similarly for $i \in \{1,2,\dots, 13\}$ and a
plan $\xi$, the
$i$-th district, which we  denote by $D_i(\xi)$, is given by the set $\{v \in
V: \xi(v)=i\}$. 
 We wish to only consider  redistricting plans $\xi$ such
that each district $D_i(\xi)$ is a single connected component. We
will denote the collection of all redistricting plans  with connected
districts by $\dist$.

\subsection{The Score Function}
We now wish to define a function $J$ that assigns a nonnegative number $J(\xi)$ to every redistricting $\xi \in \dist.$ To do this, we  employ functions $J_{p}, J_{I}, J_{c},$ and $J_{m}$ that measure how well a given redistricting satisfies the individual principles outlined in HB92. The \textit{population score}
$J_{p}(\xi)$ measures how well the redistricting $\xi$ partitions
the population of North Carolina into 13 equal parts. The
\textit{isoperimetric score} $J_{I}(\xi)$
measures how compact the districts are by returning the sum of the
isoperimetric constants for each district, a quantity which is minimized
by a circle.  The \textit{county score} $J_{c}(\xi)$ measures the number of counties split
between multiple districts; the minimum is achieved when there are no split counties. Lastly, the \textit{minority score}  $J_{m}(\xi)$ measures
the extent to which the districts with the largest percentage of African-Americans achieve  stipulated target percentages. With these, we then define our score function $J$ to be a weighted sum of $J_{p}, J_{I}, J_{c},$ and $J_{m};$ we use a weighted combination so as to not give one of the above scores undue influence, since all of the
score functions do not necessarily change on the same scale. Specifically, we define:
\begin{align}
  J(\xi) = w_{p} J_{p}(\xi)+w_{I} J_{I}(\xi)+ w_{c} J_{c}(\xi) + w_{m} J_{m}(\xi),
  \label{eqn:score}
\end{align}
where $w_{p}$, $w_{I}$, $w_{c}$,  and $w_{m}$ are a collection of
positive weights. 

To describe the individual score functions, we attach to our
graph $G=(V,E)$ some data which gives relevant features of each VTD.
We define the positive functions $\pop(v)$, $\area(v)$, and $\mino(v)$ for a vertex
$v \in V$ as respectively the total population, geographic area, and
African-American population of the
VTD associated with the vertex $v$. We extend these functions to a
collection of vertices $B \subset V$ by
\begin{align}\label{eq:PopArea}
  \pop(B) = \sum_{v \in B} \pop(v), \quad \area(B) =
  \sum_{v \in B} \area(v), \quad\mino(B) =
  \sum_{v \in B} \mino(v)\,.
\end{align}

We  think of the boundary of a district
$D_i(\xi)$ as the subset of the edges $E$ which connect vertices
inside of $D_i(\xi)$ to vertices outside of $D_i(\xi)$. We  write
$\d D_i(\xi)$ for the boundary of the district $D_i(\xi)$. Since we
want to include the exterior boundary of each district (the section
bordering an adjacent state or the ocean), we add to $V$ the vertex $o$
which represents the ``outside'' and connect it with an edge to each
vertex representing a VTD which is on the boundary of the state. We
always assume that any redistricting $\xi$ always satisfies
$\xi(v)=0$ if and only if $v=o$. 
Since $\xi$ always satisfies $\xi(o)=0$ and hence $o \not \in
D_i(\xi)$ for $i \geq 1$, it does not matter that we have not defined
$\area(o)$ or $\pop(o)$, as $o$ is never included in the districts.

Given an edge $e \in E$
which connects the two vertices $v, \tilde v \in V$, we define
$\boundary(e)$ to be the length of common border of the VTDs
associated with the vertex $v$ and $\tilde v$.  As before, we extend
the definition to the boundary of a set of edges
$B\subset E$ by 
\begin{align}
\label{eq:Boundary}
    \boundary(B) = \sum_{e \in B} \boundary(e)\,.
\end{align}

With these preliminaries out of the way, we turn to defining the
first three score functions used to assess the goodness of a redistricting. 

\subsubsection{The population score function}
We define the population score by
\begin{align*}
  \JJ_{p}(\xi) = \sqrt{\sum_{i=1}^{13}
  \Big(\frac{\pop(D_i(\xi))}{\pop_{\text{Ideal}}} - 1\Big)^2},
  \quad \pop_{\text{Ideal}}=\frac{N_{pop}}{13}
\end{align*}
where $N_{pop}$ is the total population of North Carolina,
$\pop(D_i(\xi))$ is the population of the district $D_i(\xi)$ as
defined in \eqref{eq:PopArea}, and $\pop_{\text{Ideal}}$ is the population
that each district should have according to the `one person one vote'
standard; namely,  $\pop_{\text{Ideal}}$ is  equal to
one-thirteenth of the total state population. 
\subsubsection{The Isoperimetric score function}
The Isoperimetric  score $\JJ_{I}$, which measures the compactness of a
district, is the ratio of the 
perimeter to the total area of each district. The Isoperimetric score is
minimized for a circle, which is the most compact shape. Hence we define
\begin{align*}
  \JJ_{I}(\xi)= \sum_{i=1}^{13}\frac{\big[\boundary(\d D_i(\xi))\big]^2}{\area(D_i(\xi))}\,.
\end{align*}
where $\d D_i(\xi)$ is the set of edges which define the boundary,
$\boundary(\d D_i(\xi))$ is the length of the boundary of district
$D_i$ and $\area( D_i(\xi))$ is its area. 

This compactness measure is one of two measures often used in the legal
literature where it is referred to as \textit{the perimeter
  score} (See \cite{Pildes_Niemi_1993,Practice_Hebert_2010} from references). 
The second measure,
usually referred to as \textit{the dispersion score}, is more
sensitive to overly elongated districts, though the perimeter score
also penalizes them. The dispersion score does not penalize undulating
boundaries while the perimeter  score (our $J_I$) does.

\subsubsection{The county score function}
The county score function $J_c(\xi)$ penalizes redistrictings which contain
single counties contained in two or more districts. We refer to these
counties as split counties. The score consists of the number
of counties split over two different districts times a factor
$W_2(\xi)$ plus a large constant $M_C$ times the number
of counties split over three of more different districts
times a second factor
$W_3(\xi)$. Specifically, we define:
\begin{align*}
  J_c(\xi)=& \{\# \textrm{ counties split between 2
  districts}\}\cdot W_2(\xi) \\&+  M_C  \cdot \{\# \textrm{ counties split between
  $\geq$ 3
  districts}\}\cdot W_3(\xi) 
\end{align*}
where $M_C$ is a large constant and the weights $W_{2}(\xi)$ and $W_{3}(\xi)$ are defined by
\begin{align*}
  W_2(\xi) &= \sum_{\substack{\text{counties} \\ \text{split between}\\\text {2 
  districts}}} \Big(\parbox{18em}{Fraction of county VTDs in 2nd largest\\
  intersection of a district with the county }\Big)^{\frac12}\\
  W_3(\xi) &= \sum_{\substack{\text{counties} \\ \text{split
  between}\\\text { $\geq$ 3
  districts}}} \Big(\parbox{21em}{Fraction of county VTDs not in 1st
  or 2nd \\largest
  intersection of a district with the county }\Big)^{\frac12}\\
\end{align*}
The factors $W_2(\xi)$ and $W_3(\xi)$ make the score
function vary in a more continuous fashion, which encourages reduction
of the smaller fraction of a split county.

\subsubsection{The Voting Rights Act or minority score function}
It is less clear what it means for a redistricting to comply with the
VRA. African-American voters make up approximately 20\% of
the eligible voters in North Carolina. Since 0.2 is between
$\frac{2}{13}$ and $\frac{3}{13}$, the current judicial interpretation
of the 
VRA stipulates that at least two districts should have
enough African-American representation so that this demographic may
elect a candidate of their choice. However, the NC2012 redistricting
plan was ruled unconstitutional because two districts, each containing
over 50\% African-Americans, were ruled to have been packed too
heavily with African-Americans, diluting their influence in other
districts. The NC2016  redistricting was accepted based on
racial considerations of the VRA and contained districts that held
44.48\% African-Americans, and 36.20\% African-Americans.  The amount
of deviation constitutionally allowed from these numbers is unclear.

Based on these considerations, we chose a VRA score function which
awards lower scores to redistrictings which had one district with at
least 44.48\% African-Americans and a second district with at least 36.20\% African-Americans.
We write  
\begin{align}
J_m(\xi) = \sqrt{H(44.48\%-m_1)}+\sqrt{H(36.20\%-m_2)}, 
\end{align}
where $m_1$ and $m_2$ represent the percentage of African-Americans in
the districts with the  highest and second highest percent
of African-Americans, respectively. $H$ is the function
defined by $H(x)=0$ for $x \leq 0$ and $H(x)=x$ for $x \geq 0$. We
chose this function to make the transition smoother, and we utilize
the square root function to encourage districts that are just
above the threshold to be less probable than when no square root is
included. Notice that whenever $m_1\geq 44.84\%$ and $m_2 \geq
36.20\%$ we have that $J_m=0$.


\subsection{The Probability Distributions on Redistrictings}
We now use the score function $J(\xi)$ to assign a probability
to each redistricting $\xi \in  \dist$ that makes redistrictings with
lower scores more likely. Fixing a $\beta >0$, we define the
probability of $\xi$, denoted by $\mathcal{P}_\beta(\xi)$, by 
\begin{align}
  \label{eq:Prob}
  \Pr(\xi) = \frac{e^{- \beta \JJ(\xi)}}{\mathcal{Z}_\beta}
\end{align} 
 where $\mathcal{Z}_{\beta}$ is the normalization constant defined so that $\mathcal{P}_\beta(\dist)=1$. Specifically,
\begin{align*}
    \mathcal{Z}_{\beta} = \sum_{\xi \in \dist} e^{- \beta J(\xi)}\,.
\end{align*}
The positive constant $\beta$ is often called the ``inverse temperature'' in analogy with statistical mechanics and gas dynamics. 
When $\beta$ is very small (the high temperature regime), different elements of $\dist$ have close to equal probability. As $\beta$ increases (``the temperature decreases''), the measure concentrates the probability around the redistrictings $\xi \in \dist$ which minimize $J(\xi)$. 

\subsection{Generating Random Redistrictings}
\label{Sampling}
If we neglect the fact that the individual districts in a
redistricting need to be connected, then there are more than $13^{2500}\approx 7.2 \times 10^{2784}$ different
redistrictings, larger than both the current estimate for the number of atoms in
the universe (between $10^{78}$ and $10^{82}$) and the estimated number of
seconds since the Big Bang ($4.3\times 10^{17}$). While there are
significantly fewer redistrictings in $\mathcal{R}$ (the set of simply
connected redistrictings), it is not practical to enumerate
all redistrictings to find those with the lowest values of
$\J$ (i.e. the most probable ones).

The standard, very effective way to escape this curse of
dimensionality is to use a Markov chain Monte Carlo (MCMC) algorithm
to sample from the probability distribution $\Pr$. The basic idea is
to define a random walk on $\mathcal{R}$ which has $\Pr$ as its unique,
attracting stationary measure. We do this using the standard Metropolis-Hastings
algorithm.

The Metropolis-Hastings algorithm is designed to use one Markov transition
kernel $Q$ (the proposal chain) to sample from another Markov transition
kernel that has a unique stationary distribution $\mu$ (the target distribution).
$Q(\xi,\xi')$ gives the  probability of moving from the
redistricting $\xi$ to the redistricting $\xi'$ in the proposal Markov
chain and is readily computable. We use $Q$ to
draw a sample distributed according to $\mu$. 
The algorithm proceeds as follows:
\begin{enumerate}
\item Choose some initial state $\xi \in \dist$.
\item Propose a new state $\xi '$ with transition probabilities given
  by $Q(\xi,\xi')$.
\item Accept the proposed state with probability $p=\min
  \big(1,\frac{\mu(\xi ') q(\xi ', \xi)}{\mu(\xi) Q(\xi, \xi ')}\big)$.\label{acceptStep}
\item Repeat steps 2 and 3.
\end{enumerate}
The stationary distribution of this
Markov chain matches the stationary measure $\mu$. Thus, the states can be treated as samples from the desired distribution.
The stationary measure we would like to sample is $\Pr$. We sample
from three possible initial states:  the NC2012, the NC2016, and the
Judges redistricting.  Since
this algorithm is designed to converge to a unique stationary measure
$\Pr$, any results should be independent of the initial starting
point. However, this assumes the parameters have been chosen so that the time
to equilibrate is short enough to happen during our runs.\footnote{Though technical, one can rigorously prove that the Markov Chain given by this
  algorithm converges to the desired distribution if run long
  enough. One only needs to establish that the Markov Chain transition
  matrix is irreducible and aperiodic. Since one can evolve from any
  connection redistricting to another through steps of the chain, it is
  irreducible. Aperiodicity follows as there exist redistrictings which are
  connect to itself through a loop consisting of two steps and a loop
  consisting of three steps. Since
2 and 3 are prime and hence have greatest common divisor 1, the chain is
aperiodic. See the Perron–-Frobenius Theorem for more details.}
 We show that the results are
 independent of the initial condition in
 Section~\ref{sec:IndependenceOfIC}, which lends credence to the
 assertion that the algorithm is equilibrating.\footnote{As this work was being
 completed an the work in \cite{Chikina_Frieze_Pegden_2017} appeared
 which provides an interesting set of ideas to assess if samples being
 drawn are typical or outliers
 exactly in our context. We hope to explore these ideas in the near future.}

We define the proposal chain $Q$ used for
proposing new redistrictings in the following way:
\begin{enumerate}
\item Uniformly pick a conflicted edge at random. An edge, $e=(u,v)$
  is a conflicted edge if $\xi (u) \neq \xi(v)$, $\xi(u) \neq 0$,
  $\xi(v) \neq 0$.
\item For the chosen edge $e=(u,v)$, with probability $\frac12$, either:
\begin{equation*}
   \xi '(w) = 
     \begin{cases}
       \xi(w) & w \neq u\\
       \xi(v)& u
     \end{cases}
\qquad\text{or}\qquad
\xi '(w) =
\begin{cases}
   \xi(w) & w \neq v\\
       \xi(u)& v 
\end{cases}
\end{equation*} 
\end{enumerate}
Let $\con(\xi )$ be the number of conflicted edges for redistricting
$\xi$.  Then we have $Q(\xi,\xi ')=\frac{1}{2}\frac1{\con(\xi)}$. The
acceptance probability is given by:

$$p=\min \Big(1, \frac{\con(\xi)}{\con(\xi ')} e^{-\beta (\J(\xi ')-\J(\xi))} \Big)$$
If a redistricting $\xi '$ is not connected, then we refuse the step, which is equivalent to setting
$\J(\xi ')=\infty$.

Given a fixed set of weights $(w_p,w_i,w_c,w_m)$, one still needs to
determine an appropriate $\beta$ so that typical samples from the
distribution are ``reasonable'' redistrictings.  If $\beta$ is chosen
to be too large, the algorithm will seek out a local minimum and leave
this minimum with very low probability, meaning that it may require a
large amount of steps to switch between high quality redistrictings.
If $\beta$ is chosen to be too low, then the algorithm will never find
the locally good districts as it will choose redistrictings
indiscriminately.

There are several well established ideas in the literature to overcome
these challenges, including simulated annealing (e.g. \cite{van87}),
parallel tempering (e.g. \cite{hansmann1997}) and simulated tempering
(e.g. \cite{fifield2015}). In the present work, we examine simulated
annealing, in which $\beta$ is set to be small at first until a
certain number
of steps are accepted (in the sense of step (\ref{acceptStep}) from the
algorithm in Section~\ref{Sampling}). This allows the
system to explore the space of redistrictings more freely.
Next, $\beta$ is increased linearly to a maximum value over the course of a defined
number of steps.  This slowly ``cools'' the systems, hopefully relaxing
it into a redistricting  $\xi$ which has a relatively low score
$J(\xi)$. Finally, $\beta$ is kept at this fixed maximum value for a
defined number of steps so that the algorithm locally samples the
measure $\Pr$ sufficiently long enough to produce a good
redistricting. During the summer of 2016 Data+ Project, we explored
extensively the use of parallel tempering to generate Monte Carlo
sampled. We found that simulated annealing more reliably explored the
state space. Parallell tempering  does have the advantage of always
sampling from the same target distribution while simulated annealing
changes the target distribution to improve mixing.  In theory
parallell tempering should also behave well, tough we found tuning it
properly more difficult. 

The
principal results quoted in Section~\ref{sec:main-results} use the low
$\beta$ to be zero over $40,000$ steps, linearly increase $\beta$ to
one over $60,000$ steps, and fix $\beta$ to be one for $20,000$ steps
before taking a sample.  This process is repeated for each sample
redistricting.  One potential critique with using simulated annealing
is that the results may depend on the number of steps chosen above.
We make a standard test to confirm that we have taken an appropriate
number of steps by doubling each number of steps and repeating our
analysis. The results of this test, which are found in
Section~\ref{sec:IndependenceOfIC}, show that doubling the number of steps has
little effect on the results.

\subsection{Thresholding the sampled redistrictings}
\label{subsec:threshold}

It is possible for the simulated annealing algorithm to draw a redistricting with a bad score when using the MCMC algorithm from
Section~\ref{Sampling} combined with the probability distribution given in
\eqref{eq:Prob}. 
 Additionally, the use of simulated annealing also increases the chance 
that we become stuck in a local minimum with a less than desirable
score function, as such local minimum may take longer than the time
we spend at high $\beta$ to escape. These local trapping events can
often lead to  samples with less than perfect score functions. Lastly, our
score functions do not perfectly encapsulate our redistricting design
aesthetic. For example, since the isoperimetric score function is the
sum of the individual isoperimetric scores of each district, it is
still possible to have one bad district if the rest have exceptionally small isoperimetric scores.

Since
we want to maximize the degree of compliance with HB92, we  only
use samples which pass an additional set of thresholds, one for each of
the selection criteria. This additional layer of rejection sampling was
also used in reference \cite{fifield2015}, though the authors of
reference \cite{fifield2015} chose to reweigh the
samples to produce the uniform distribution over the set redistrictings
that satisfy the thresholds. We prefer to continue to bias our
sampling according to the score function so better redistrictings
are given higher weights; we note that the idea of preferring some redistrictings to others is consistent with the provisions HB92. We now detail our thresholding requirements.

It is our experience from the Beyond Gerrymandering project that
redistrictings which use VTDs as their building blocks and  have less that
1\% population deviation can readily be driven to 0.1\% population
deviation by breaking the VTDs into census tracts and performing 
minimal alterations to the overall redistricting plan.  We thus only
accept redistrictings that have no districts above 1\% population
deviation. Many of our samples have deviations considerably below this
value. It is important to emphasize that we require this of every district
in the redistricting. In Section~\ref{subsec:vary-thresholds}, we show
that the results are quantitatively extremely similar, and
qualitatively identical,  when the population threshold is decreased from
1\% to 0.75
\% and then to 0.5\%.

We have found that districts with isoperimetric scores under 60 are
almost always reasonably compact. Thus, we choose to accept a redistricting only if each district in the plan has an isoperimetric ratio less than 60. The Judges redistricting plan would be accepted under this threshold as its least compact district has an isoperimetric score of 53.5. Neither NC2012 nor NC2016 would be accepted with this thresholding as the least compact districts of each plan have isoperimetric scores of 434.65 and 80.1, respectively. We also note that only two of the thirteen districts for the NC2012 plan meet our isoperimetric score threshold, whereas eight of the thirteen districts of NC2016 fall below the threshold. Although we examine our principle results over a space of highly compact redistricting plans, we also demonstrate that our results are insensitive to lifting this restriction in Section~\ref{subsec:vary-thresholds}.

Though redistrictings which split a single county in three are
infrequent, they do occur among our samples. Since these are undesirable, we only accept
redistrictings for which no counties are split across three or more
districts. Note that, in order to satisfy population requirements, we
must allow counties to be split into two districts because of the
large populations of Wake and Mecklenburg Counties which each contain a
population larger than a single Congressional district's ideal population. We do not explicitly
threshold based on number of split counties, though redistrictings
with more split counties have a higher scores, and hence are less favored.  We remark that none of our generated redistrictings had more county splits than the NC2012 redistricting plan, and that the NC2012 plan was never critiqued or challenged based on the number of county splits.

To build a threshold based on minority requirements of the VRA, we
note that the NC2016 redistricting was deemed by the courts to
satisfy the VRA. The districts in this plan with the two highest proportion of African-Americans to total population are composed of 44.5\%  and 36.2\% African-Americans. With this in mind, we only accept redistrictings if the districts with the two highest percentages of African-American population have at least 40\% and 33.5\%, respectively.

The effect of all of these thresholds was to select around 16\% of the
samples initially produced by our MCMC runs.  Though this leads to unused samples, it ensures that all of the redistrictings
used meet certain minimal standards. This in turn allowed us to better adhere to the spirit of
HB92. The reported 24,000  
samples used in our study refer to those left after thresholding. The  
full data set of samples was in excess of 150,000.  That being said, we show in Section~\ref{subsec:vary-thresholds}
that results without thresholding were quantitatively very close and
qualitatively identical. As already mentioned, we also show
that decreasing the
population threshold from
1\% to 0.75\% and then to 0.5\% also  has little effect on the quantitative results
and no effect on the qualitative conclusions.

\subsection{Determining the weight parameters}
\label{ssec:determine-param}
As we have mentioned above, we have four independent weights
$(w_p,w_I,w_c,w_m)$ used in balancing the effect of the different
scores in the total score $J(\xi)$. In addition to these parameters, we
also have the low and high temperatures corresponding respectively to
the max and min $\beta$ used in the simulated
annealing. We set the minimum value of $\beta$ to be zero which corresponds to infinite temperature.  In this regime, no district is favored over any other, which allows the redistricting plan freedom to explore the space of possible redistrictings. The only parameter left is to set the high value of $\beta$.  Since $\beta$ multiplies the weights, one of these degrees
of freedom is redundant and can be set arbitrarily.  We chose to fix
the low temperature (high value of $\beta$) to be one. 

To select appropriate parameters, we employ the following tuning method:
\begin{enumerate}
\item{Set all weights to zero.}
\item{Find the smallest $w_p$ such that a fraction of the results are within a desired threshold (for the current work we ensured that at least 25\% of the redistrictings were below 0.5\% population deviation, however we typically did much better than this).}
\item{Using the $w_p$ from the previous step,  find the smallest $w_I$ such that a fraction of the redistrictings have all districts below a given isoperimetric ratio (we ensured that at least 10\% of the results were below this threshold; we chose a threshold of 60 (see Section \ref{subsec:threshold})).}
\item{If above criteria for population is no longer met, repeat steps 2 through 4 until both conditions are satisfied}
\item{Using the $w_p$ and $w_I$ from the previous steps, find the smallest $w_m$ such that at least 50\% of all redistrictings have at least one district with more than 40\% African-Americans and a second district has at least 33.5\% African-Americans.}
\item{If the thresholds for population were overwhelmed by increasing $w_m$, repeat steps 2 through 6.  If the thresholds for compactness were overwhelmed, repeat steps 3 through 6.}
\item{Using the $w_p$, $w_I$, and $w_m$ from the previous steps, find the smallest $w_c$ such that we nearly always only have two county splits, and the number of two county splits are, on average, below 25 two county splits.}
\item{If the thresholds for population are no longer satisfied, repeat steps 2 through 8.  If the criteria for the compactness is no longer met, repeat steps 3 through 8.  If the criteria for the minority populations is not satisfied, repeat steps 5 through 8.  Otherwise, finish with a good set of parameters. }
\end{enumerate}
With this process, we settle on parameters $w_{p}=3000$, $w_{I}=2.5$,
$w_{c}=0.4$,  and $w_{m}=800$ and have used these parameters for all
of the results presented in the main results above (Section
\ref{sec:main-results}). In Section~\ref{sec:VaryingWeights}, we show that
variations of these choices have little qualitative effect on the results.

\section{Characteristics of ``reasonable'' redistrictings ensamble}
\label{sec:characteristics}
\begin{figure}[ht]
  \centering 
\includegraphics[width=8cm]{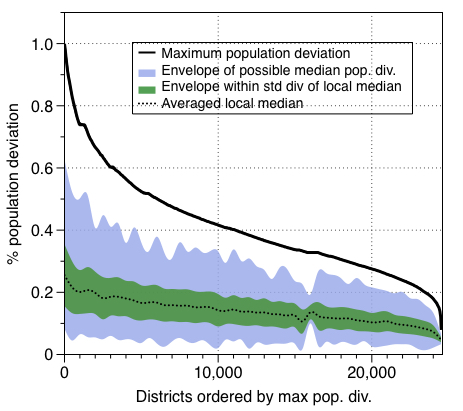}
\includegraphics[width=8cm]{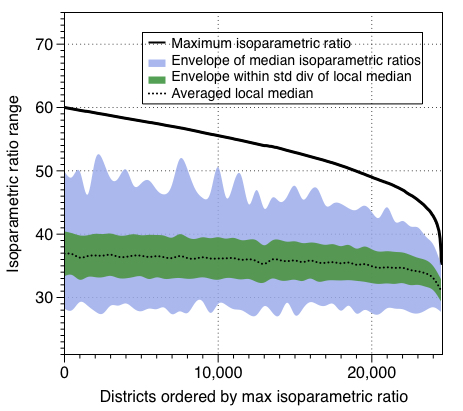}
   \caption{\capSize The redistrictings ordered by the worst case
     district in terms of either population deviation (left) or
     isoperimetric ratio (right). The solid dark line give the worst
     case districts value while the dotted line gives the average
     across redistricting with a given max value of 
     median districts value. The outer shading gives the max and min
     value of this median while the inner-shading covers one standard
     deviation above and below the mean of the  medians.}
  \label{fig:popIsoStats}
\end{figure}

We now explore the properties of the over 24,000 random restrictings we have
generated using the algorithm described in the preceding sections. All
of the random redistrictings passed the threshold test described
in Section~\ref{subsec:threshold}. As such, they all have no district
with population deviation above 1\%. However, most have a deviation
much less than 1\%: the mean population deviation taken over the
more-than $13\times 24000= 312,000$ districts is 0.16\% with a standard deviation of 0.14\%.  
Figure~\ref{fig:popIsoStats} gives a finer view for the distribution of the
population deviation.  We order each redistricting by the maximum population deviation over all districts.  To simultaneously give a sense of the median population deviation of the districts with a given maximum population deviation, we examine the local statistics of the ordered districts to find the maximum and minimum values of the local median (plotted as the blue envelope) along with the standard deviation of the median (green envelop) and expected value of the median (dotted line).  With this plot we notice that over 50\% of redistrictings have
a worst case population deviation under 0.4\% and many of these redistrictings have a median population deviation well below 0.2\%.

To compare the population deviation of our generated districts with the districtings of NC2012, NC2016 and the Judges, we note that all of these districtings all had to split VTDs in order to achieve a population deviation below 0.1\%.  Before splitting VTDs, NC2012, NC2016 and Judges had a district with maximum population deviation of 0.847\%, 0.683\%, and 0.313\% respectively, and had median population deviations of 0.234\%, 0.048\%, and 0.078\% respectively, meaning that the districts sampled by our algorithm are very similar to the three districts we have compared our results with in terms of population deviation.

Turing to the isoperimetric ratios, recall that all of the districts have an
isoperimetric constant under 60 as that was our threshold value.  The
mean isoperimetric ratio of the more-than $13\times 24000=312,000$ districts is 36.9 with a standard deviation of 9.  Examining the second part of Figure~\ref{fig:popIsoStats} gives an analogously finer view for the distribution of the isoperimetric ratios of all districts.  The figure shows that most redistrictings have a median isoperimetric ratio in the mid-thirties and that roughly 50\% of our redistrictings have a district with isoperimetric ratio no worse than 55 for an isoperimetric ratio.

 When comparing our generated districts, we note that the NC2012, NC2016 and Judges redistrictings have districts with maximum isoperimetric ratio of 434.6, 80.1, and 54.1 respectively, and have median isoperimetric ratios of 114.4, 54.5, and 38.2 respectively.  The NC2012 and NC2016 districts would be rejected under our thresholding criteria.  Below, we demonstrate that sampling from redistrictings that include less compact districts does not change our results (see section \ref{sec:test-sens-results}).

\begin{figure}[ht]
  \centering 
\includegraphics[height=5.75cm]{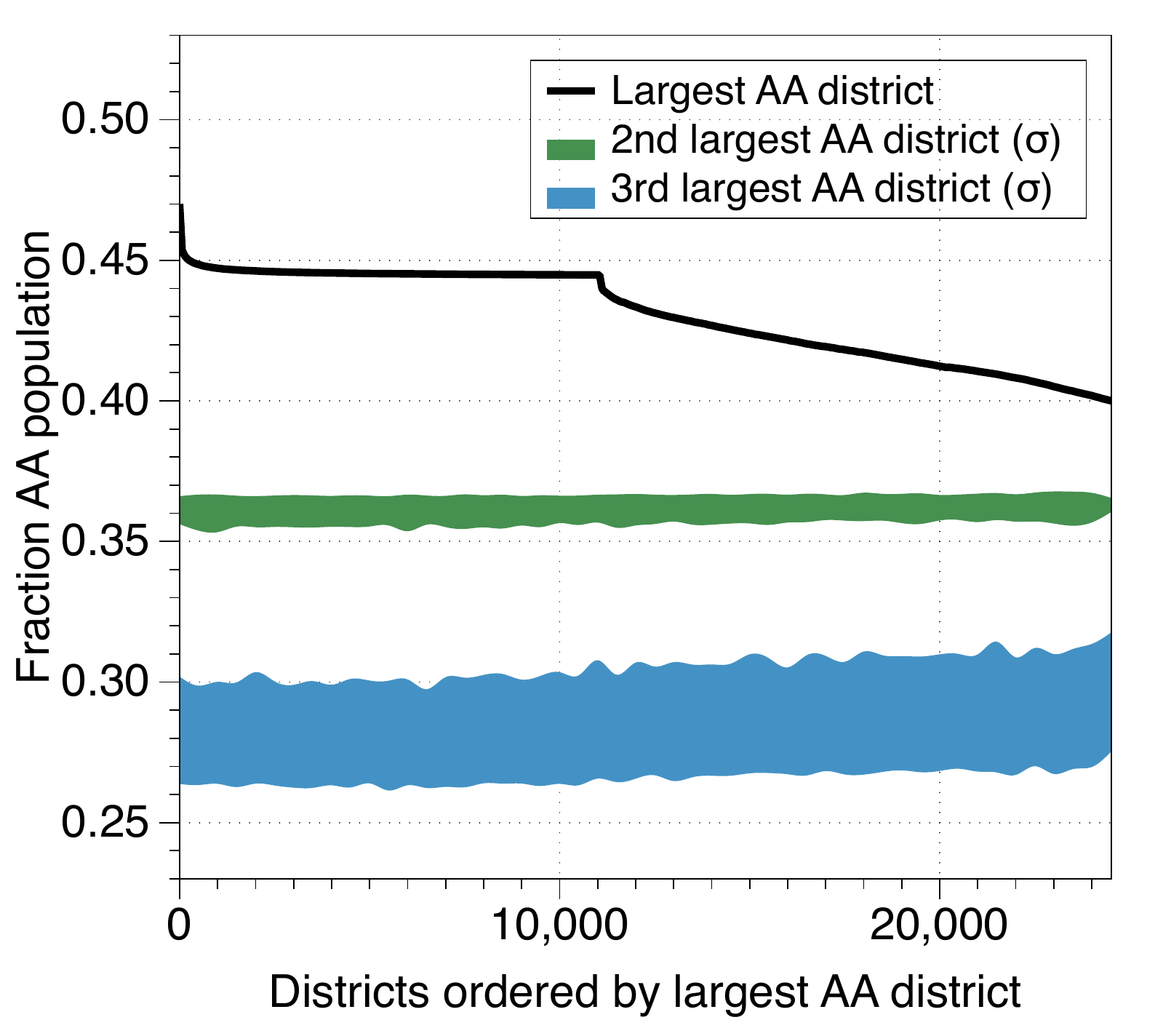}
\includegraphics[height=5.75cm]{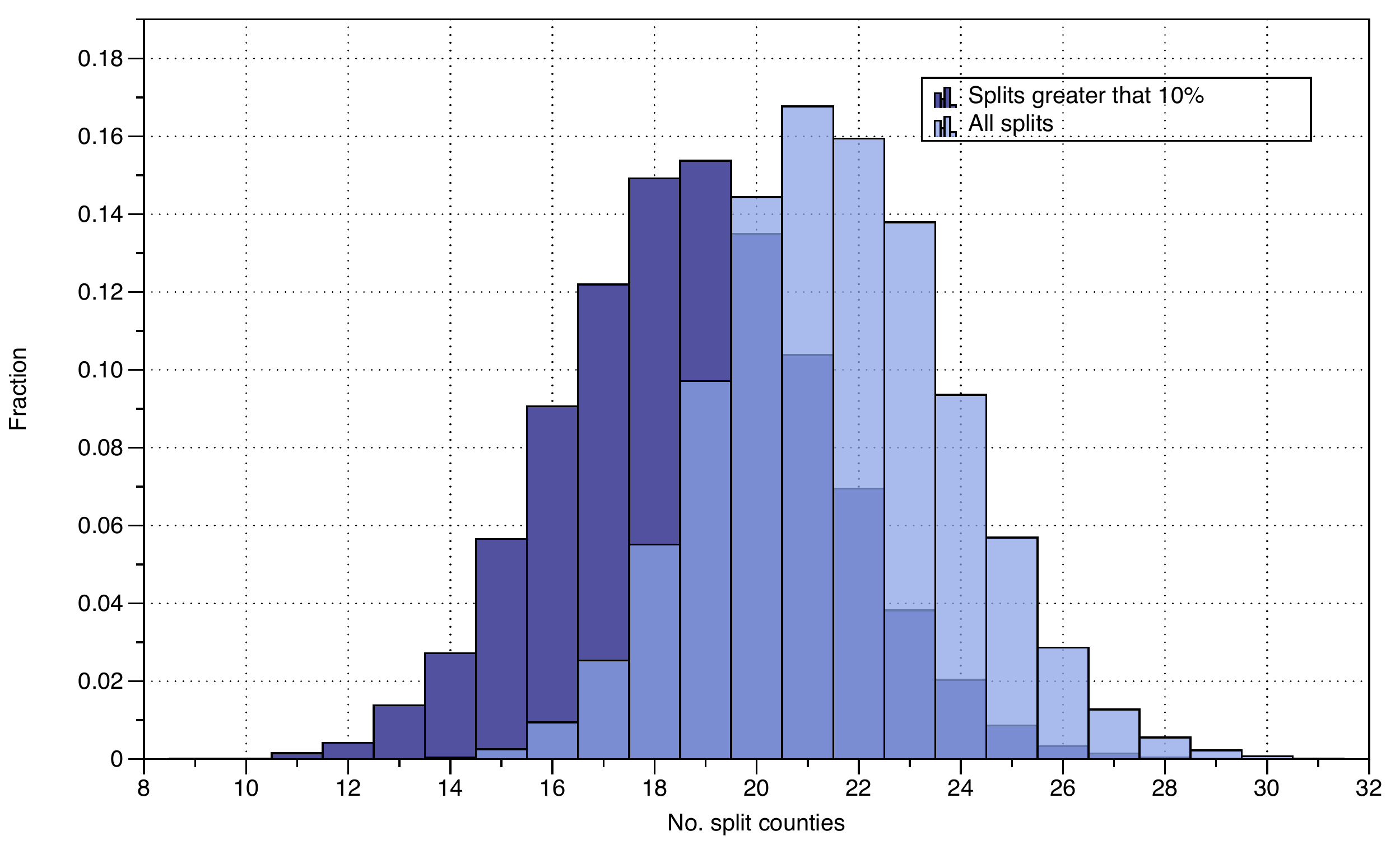}
\caption{\capSize The redistrictings are ordered by the district with
  the largest African-American percentage (left).  Subsequent ranges
  show standard deviations for districts with the second  and  third
  largest  African-American representation.  We plot the histogram of
  the number of county splits in each districting (right). The lighter
  histogram gives the number total split counties while the darker
  histogram gives only the number which splits the county into two
  parts each containing  more than 10\% of the total VTDs.
}
  \label{fig:minoCountyStats}
  \end{figure}

Next we examine the four districts with the highest minority representation in each districting.  In Figure~\ref{fig:minoCountyStats}, we order the redistrictings in decreasing order on the over 24000 accepted redistrictings.  The kink in this line at 44.46\% occurs due to the minority energy function which does not favor any population above this limit (recall this number was based on NC2016).   Roughly half of the redistrictings have a district with greater than 44.46\% of the population as African-Americans, whereas the other half has between 40\% and 44.46\% in the district with the largest number of African-Americans.  For the district with the second highest African-American representation, we remark that over 80\% of all redistrictings have more than 35\% African-American representation in the second largest district; there is not a single redistricting that has the second largest African-American district with more than 40\% of the population African-American.  

Finally we display the histogram of the number of split counties over our generated redistrictings.  We find a median of 21 split counties with a mean of 21.6, and a range from 14 to 31.  We remark that NC2012, NC2016, and Judges districtings had 40, 13, and 12 split counties respectively.

\section{The effect of the Voting
Rights Act} 
\label{sec:VRA}
\begin{figure}[ht]
  \centering 
\includegraphics[width=8cm]{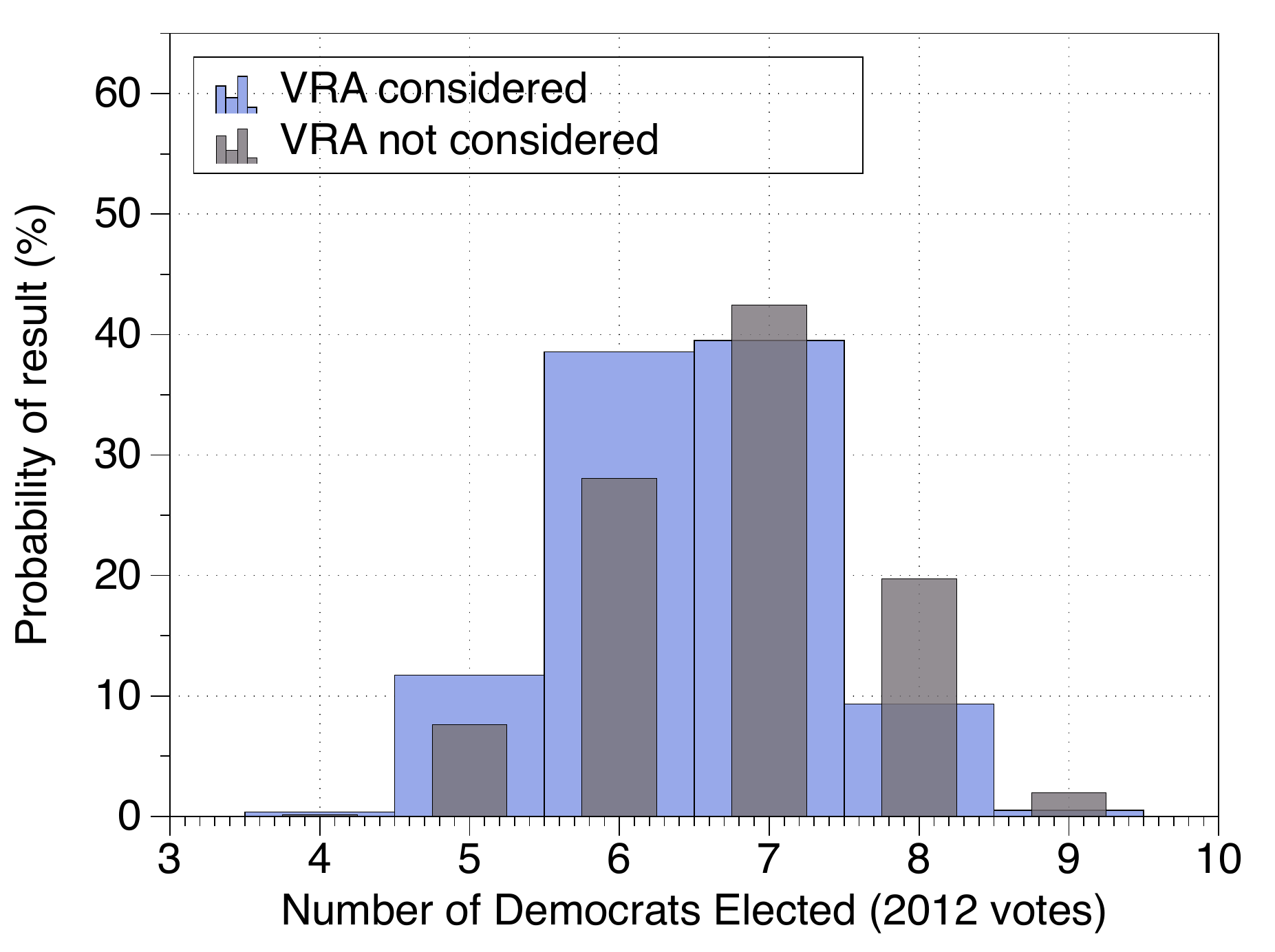}
\includegraphics[width=8cm]{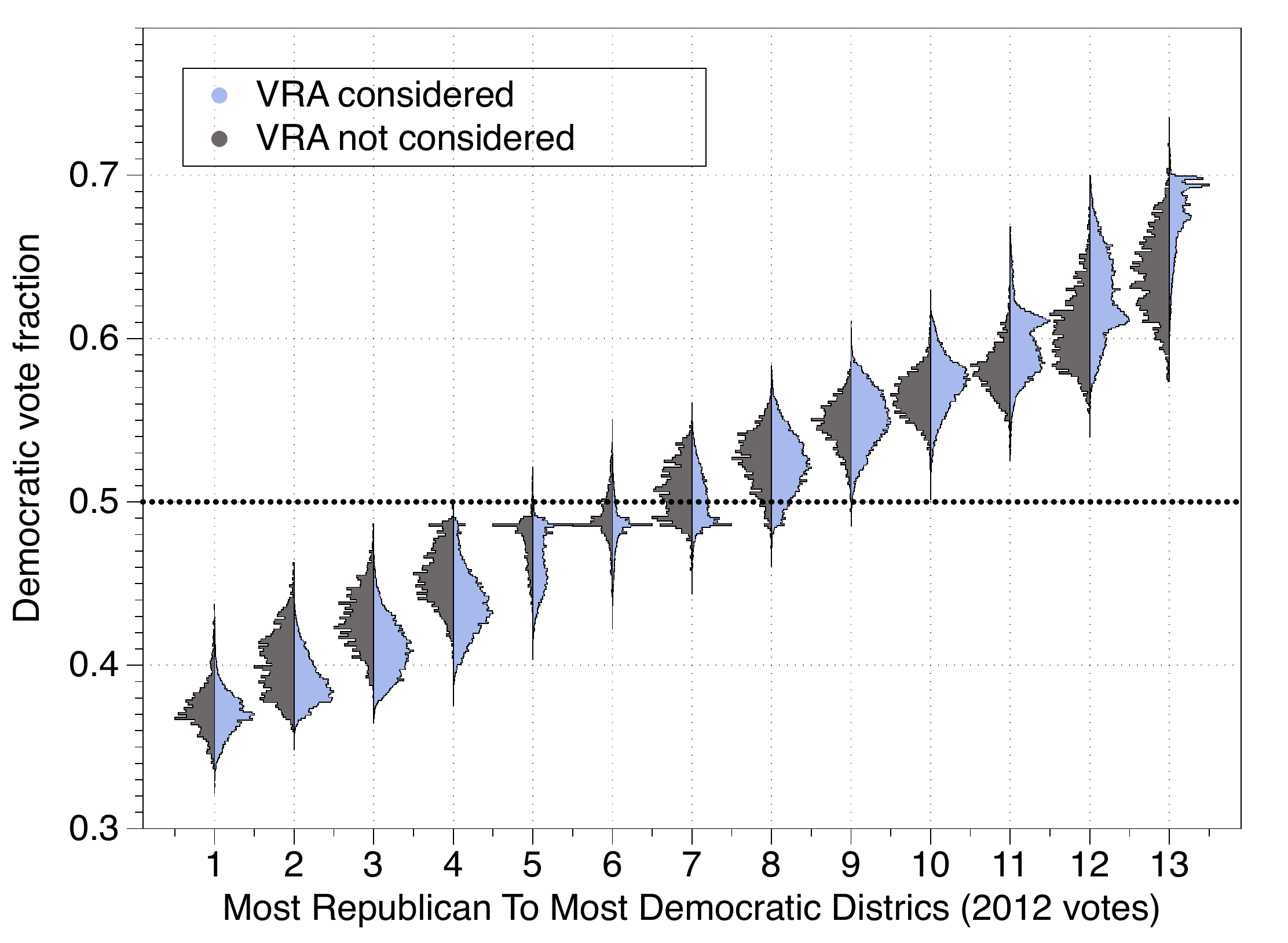}
   \caption{\capSize We display changes of the distribution of
     election results when the VRA is not taken into consideration
     (left).  The histogram formed from the distribution of our main
     results overlays this image with the gray shaded histogram.  We display changes to the histogram of the box-plot when comparing the results when VRA is considered or not (right).}
  \label{fig:NoVRA}
\end{figure}

The NC2012 districts were labeled unconstitutional for over packing
African-Americans and diluting their voice in other districts.  We
investigate the effect of the Voting
Rights Act (VRA) on election outcomes by considering
samples taken from simulations that do not take the VRA into account,
which is to say that we set $w_m=0$.  We examine the distribution of
elected Democrats along with the histogram box-plots in
Figure~\ref{fig:NoVRA}.  We find that the VRA, even with the more modest thresholds of 40\% and 33\% required African-Americans, significantly favors the Republican party.  Without the VRA, there is roughly a 65\% chance that 7 or more Democrats will be elected, with a 20\% chance that 8 Democrats will be elected; in contrast, with the VRA considered, there is a 50\% chance that 7 or more Democrats are elected, with a 10\% chance that 8 Democrats are elected.  Without commenting specifically on the VRA, we  point out that these results highlight the importance of \S120-4.52(h) in HB92, as they demonstrate that districts drawn with demographic information in mind can significantly alter the likely outcome of an election. 

\section{Details of the Indices}
\label{sec:DetailsIndicies}
 We begin by expounding
and clarifying how we compute the Gerrymandering Index and the
Representativeness Index.  We have thoroughly explained the efficiency
gap above so omit further discussion in the current section. 
\subsection{Details of  Gerrymandering Index}\label{sec:DetailsGerry} 
To compute the Gerrymandering Index, we examine the mean percentage of
Democratic votes in each
of the thirteen districts when the districts are ordered from most to least Republican (see
Figure~\ref{fig:boxPlotsCDF}).  To calculate the Gerrymandering Index
for any given redistricting plan, we take the Democratic votes
for each district when the  districts are again ordered from most to
least Republican. The differences between the mean and the observed
democratic percentage are taken for each district using a given set of
votes. These differences are then each squared and
summed over the 13 districts. The square root of this sum of squares is our
Gerrymandering Index. 

The Gerrymandering Index is 
smallest when all of the ordered Democratic vote percentages are
precisely the mean values. However, this is likely not possible as the
percentages in the different districts are highly correlated. To
understand the range of possible values, we plot the complementary cumulative
distribution function of the Gerrymandering Index of our ensemble of
randomly generated reasonable redistrictings (see Figure~\ref{fig:gerrymanderingsCDF}). This gives a context in
which to interpret any one score.

The mean percentages for the collection of redistricting we
generated is 
\begin{align*}
  (0.37,0.39,0.41,0.44,0.46,0.48,0.50,0.52,0.55,0.57,0.60,0.63,0.67)\,.
\end{align*}
If a given redistricting is associated with the sorted winning
Democratic 
percentages 
\begin{align*}
  (0.36,0.38,0.39,0.40,0.41,0.42,0.43,0.44,0.49,0.52,0.64,0.66,0.7)\,.
\end{align*}
then the Gerrymandering Index for the redistricting is the square root
of 
\begin{align*}
  (0.37&-0.36)^2 + (0.39-0.38)^2+ (0.41-0.39)^2 \\&+ (0.44-0.40)^2
                                                                   +(0.46-0.41)^2+(0.48-0.42)^2+(0.50-0.43)^2\\&+(0.52-0.44)^2
+(0.55-0.49)^2+(0.57-0.52)^2+(0.60-0.64)^2\\&+(0.63-0.66)^2+(0.67-0.7)^2=0.0291
\end{align*}
In summary, in this example the Gerrymandering Index is $\sqrt{0.0291}=0.17$.

\subsection{Details of Representativeness Index}\label{sec:DetailsRep} 
To calculate the Representativeness Index, we first construct a modified histogram of election results that captures how close an election was to swapping results.  To do this for a given redistricting plan, we examine the least Republican district in which a Republican won, and the least Democratic district in which a Democrat won.  We then linearly interpolate between these districts and find where the interpolated line intersects with the 50\% line.  For example, in the 2012 election, the 9th most Republican district elected a Republican with 53.3\% of the vote, and the fourth most Democratic district won their district with 50.1\% of the vote.  We would then calculate where these two vote counts cross the 50\% line, which will be 
\begin{align}
\frac{50-(100-50.1)}{53.3-(100-50.1)}\approx0.03,
\end{align}
and add this to the number of Democratic seats won to arrive at the
continuous value of 4.03.  This index allows us to construct a
continuous variable that contains information on the number of
Democrats elected, and also demonstrates how much safety there is in the
victory. 

\begin{figure}[ht]
  \centering 
\includegraphics[width=8cm]{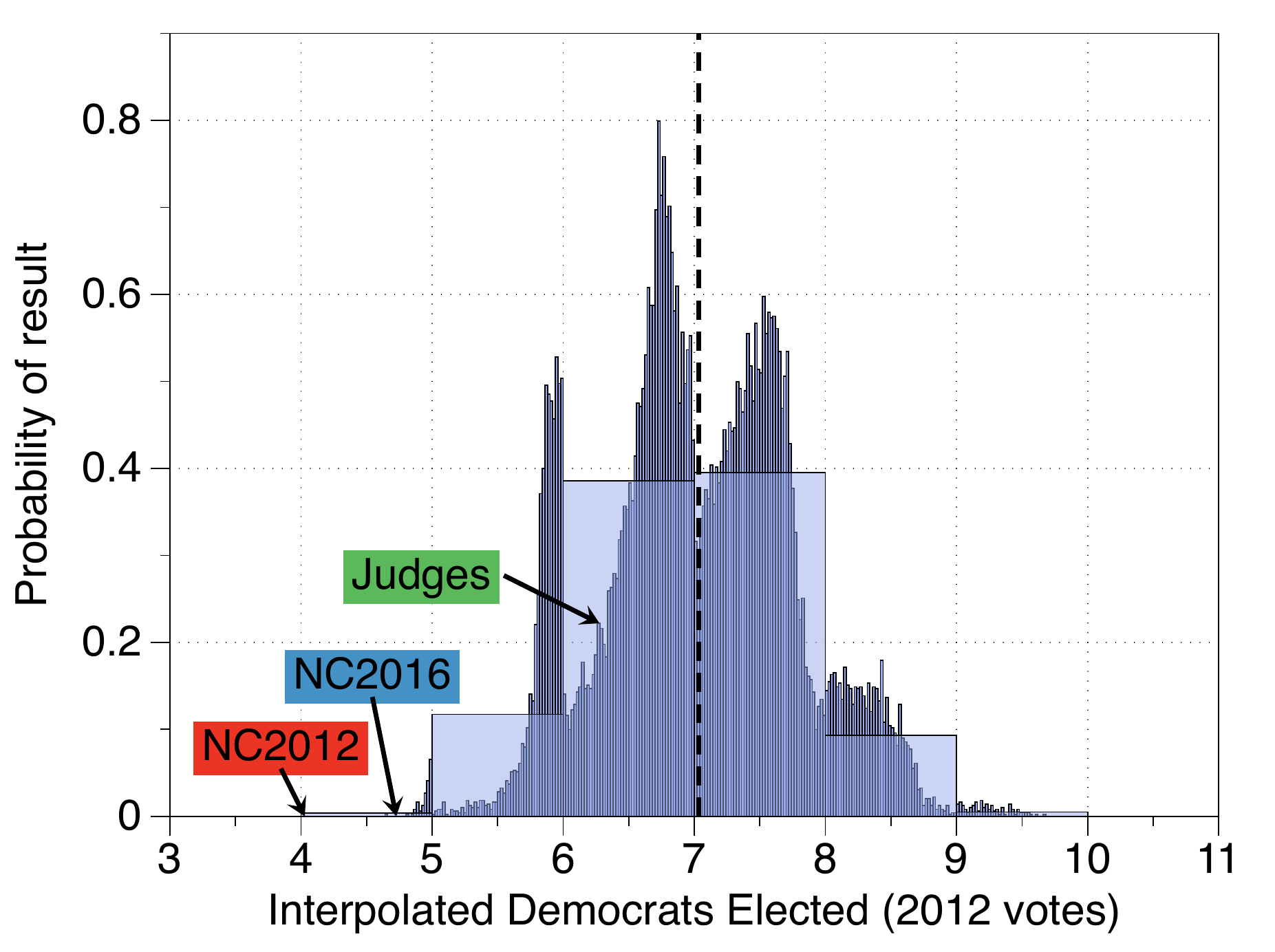}
\includegraphics[width=8cm]{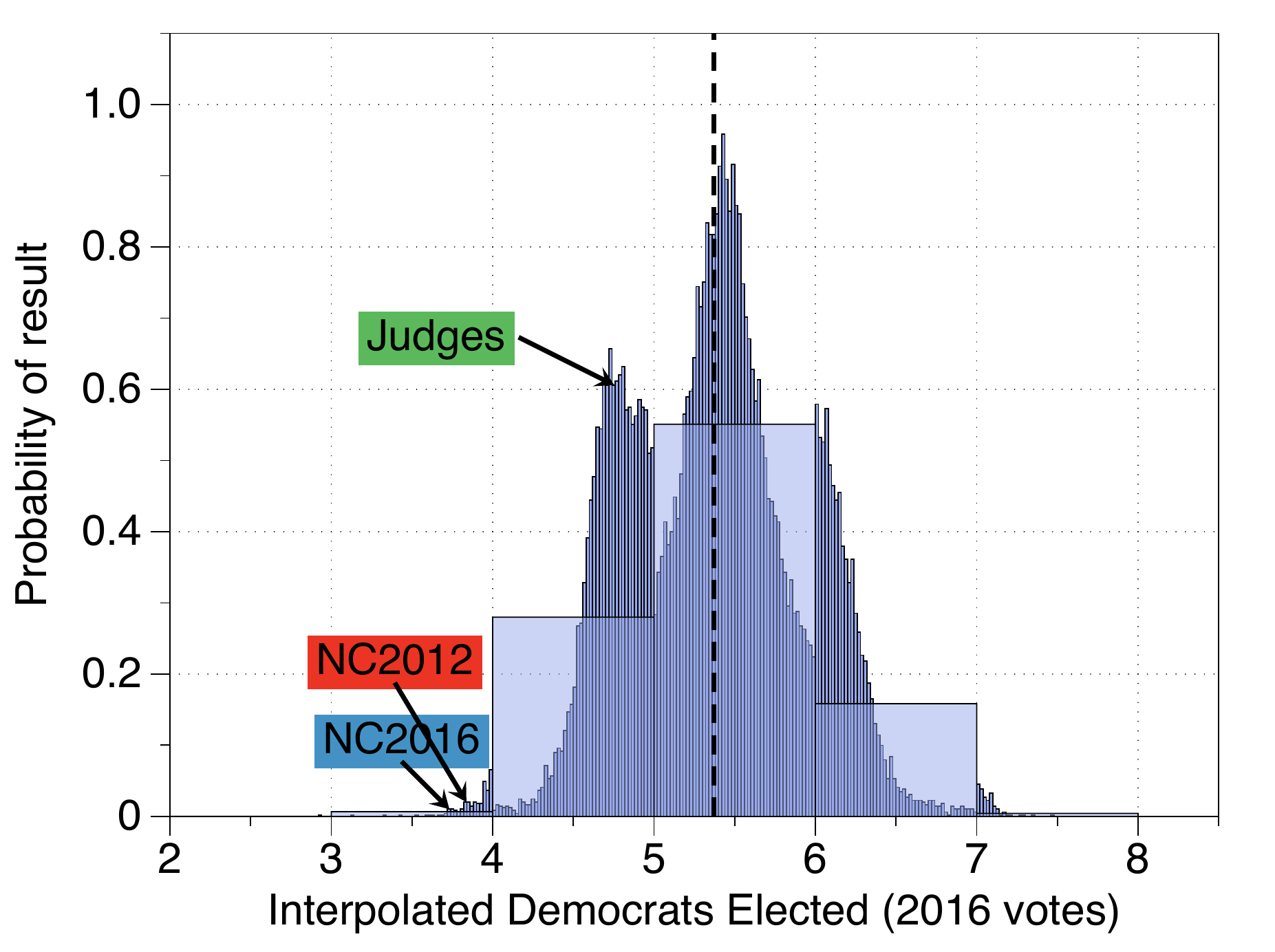}
\caption{\capSize For the 2012 votes (left) and the 2016 votes
  (right), we plot the interpolated winning margins, which give the number of seats won by the Democrats in finer detail.  We determine the mean of this new histogram and display it with the dashed line. The Representativeness Index is defined to be the distance from this mean value.  The histogram presented in Figure~\ref{fig:basicDemWinners} is overlaid on this plot for reference. }
  \label{fig:finehist2012}
\end{figure}

Fractional parts close to zero suggest that the most competitive
Democratic race is less likely to go Democratic than the most
competitive Republican race is to go Republican.  On the other hand, fractional
parts close to one suggest that the most competitive Republican race
is less likely to go Republican than the most competitive Democratic
race is to go Democratic.  Instead of simply creating a histogram of
the number of seats won by the Democrats, in
Figure~\ref{fig:finehist2012} we construct a histogram of our new
interpolated value.  We define the representativeness as the distance
from the interpolated value to the mean value of this histogram (shown
in the dashed line).  These are the values we report in
Figure~\ref{fig:representativenessCDF}.  For the 2012 vote data, we
find that the mean interpolated Democratic seats won is 7.01, and the
Judges plan yields a value of 6.28, giving a Representative Index of
$|7.01-6.28|=0.73$.  The NC2012 and NC2016 plans both have
representative indices greater than two.

\section{Testing the Sensitivity of Results}
\label{sec:test-sens-results}

We wish to ensure that our algorithm has sampled the space of redistrictings in a robust way.  We use this section to carefully study the effect of changing the number of samples used, changing the set of threshold values, changing the weights in our distribution, changing the type of energy function used for compactness, changing simulated annealing parameters on election results, and determining the possible effect of splitting VTDs to achieve zero population deviation. We also verify that the choice of the initial district does not influence our results and that this information is lost as the algorithm updates the redistrictings.  

\subsection{Varying thresholds}
\label{subsec:vary-thresholds}
\begin{figure}[ht]
  \centering
\includegraphics[width=8cm]{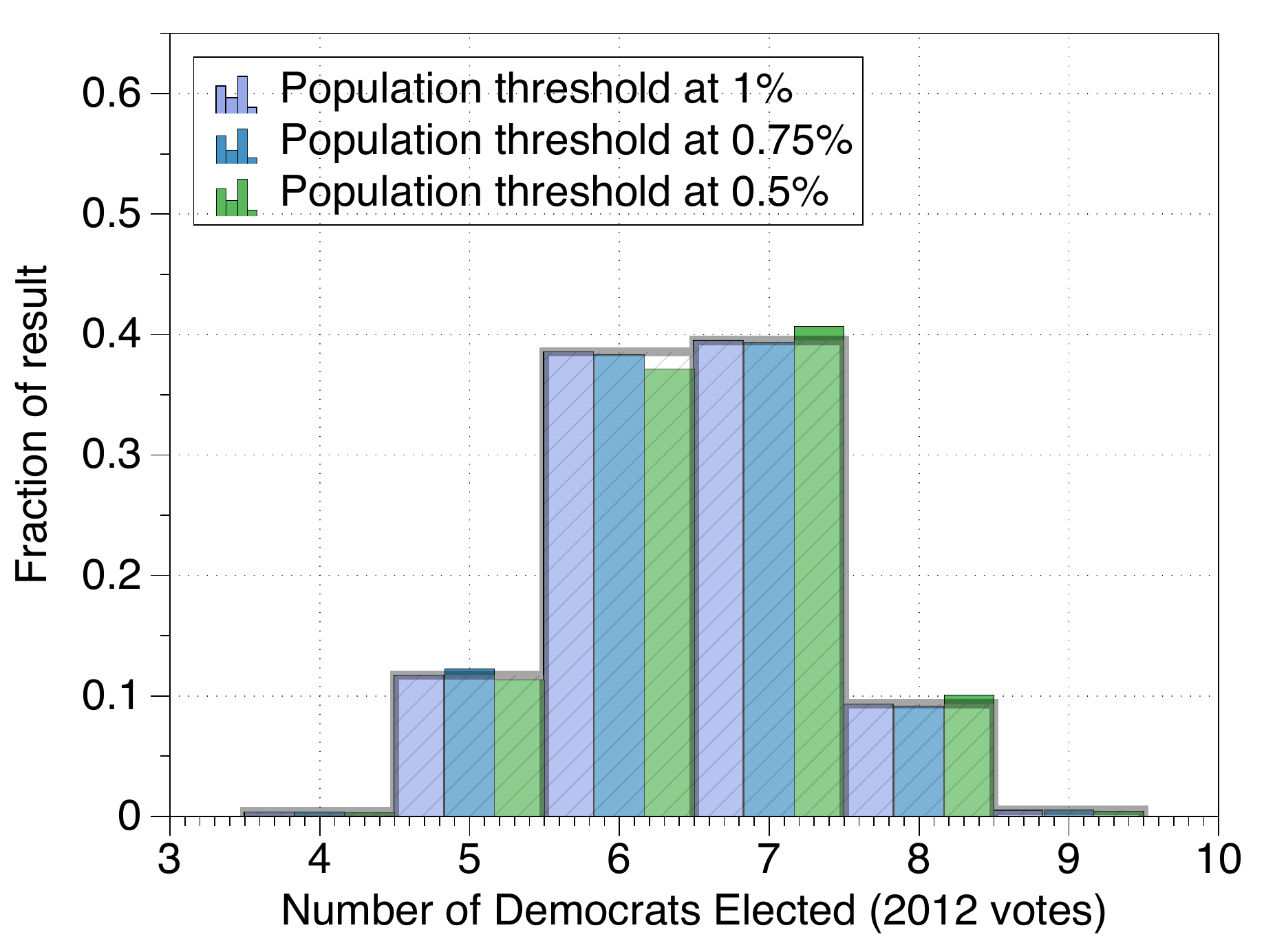}
\includegraphics[width=8cm]{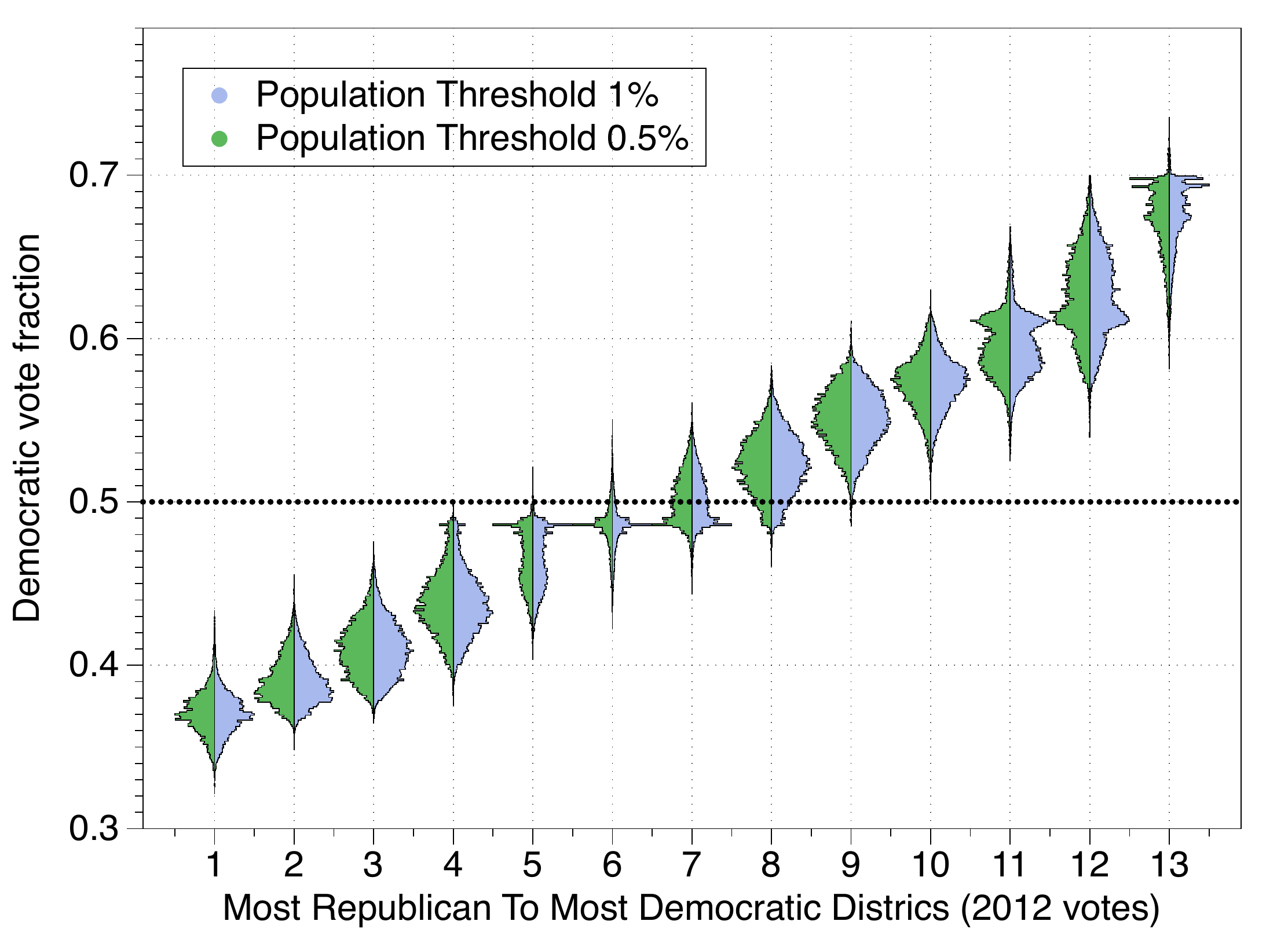}
   \caption{\capSize We display changes of the distribution of election results with changes to the population threshold (left).  The histogram formed with 1\% population deviation overlays this image with the gray shaded histogram.  We display changes to the histogram of the box-plot when comparing 1\% population deviation threshold with 0.5\% (right).}
  \label{fig:deltaPopThresh}
\end{figure}

Achieving a 0.1\% population deviation is the only statute of HB92 that we violate.  Although we have noted above that the Judges original redistrictings in the `Beyond Gerrymandering' project were all slightly over 1\% population deviation, and splitting VTDs to fall below this threshold had little impact on the election results.  We test this for our own redistrictings by changing the population threshold to 0.75\% and 0.5\%.  The results are shown in Figure~\ref{fig:deltaPopThresh}, for which we have used the 2012 vote data.  We find that tightening the population threshold has negligible impact on the number of Democrats elected, and that the variation in the histogram box-plots is barely perceptible.  In the 0.5\% population deviation threshold plots, we have discarded over half of our results and we still do not see any significant changes. These results support our claim that splitting VTDs to achieve a less than 0.1\% deviation will have a negligible effect on our conclusions.  

\begin{figure}[ht]
  \centering
\includegraphics[width=8cm]{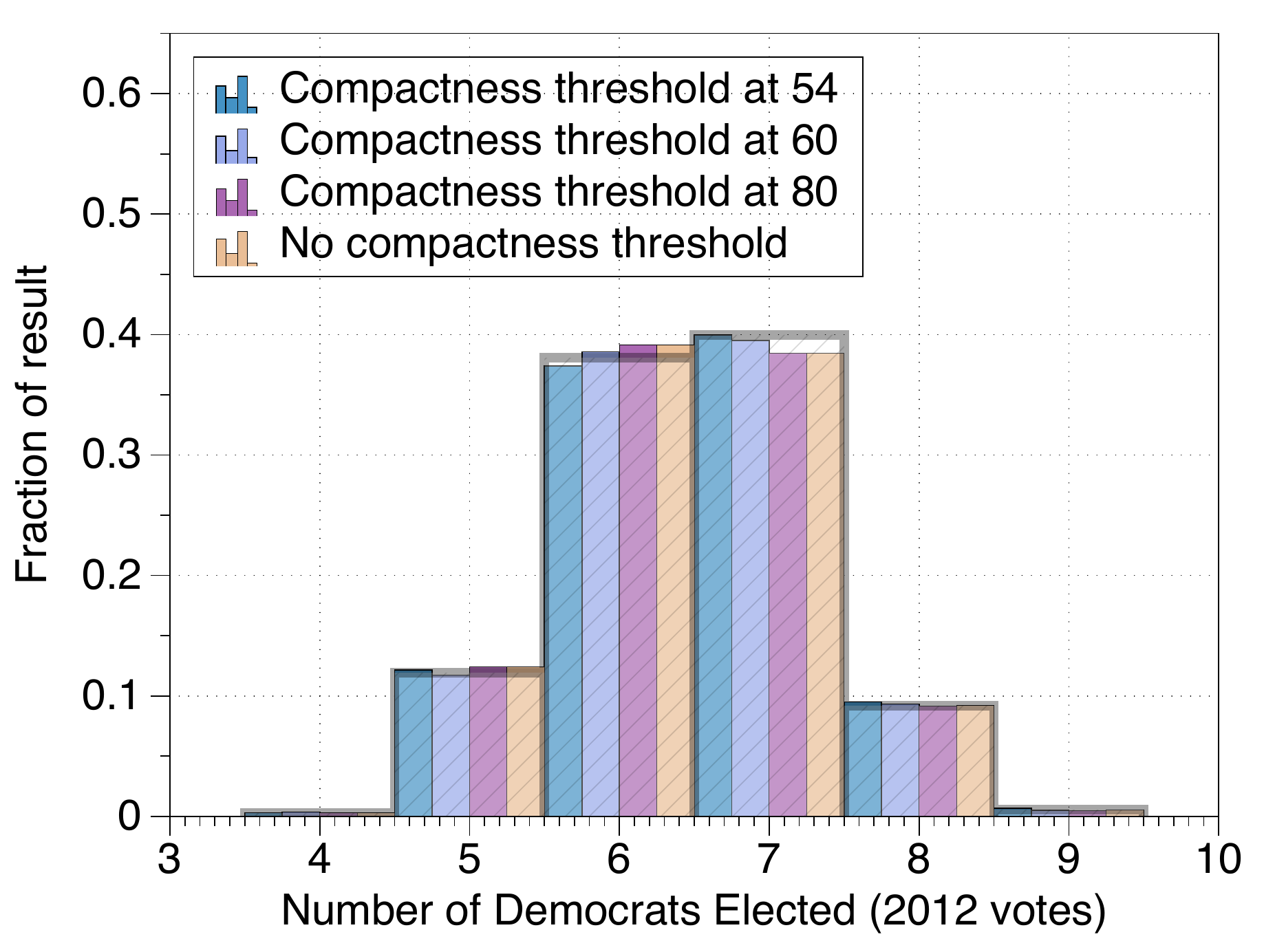}
\includegraphics[width=8cm]{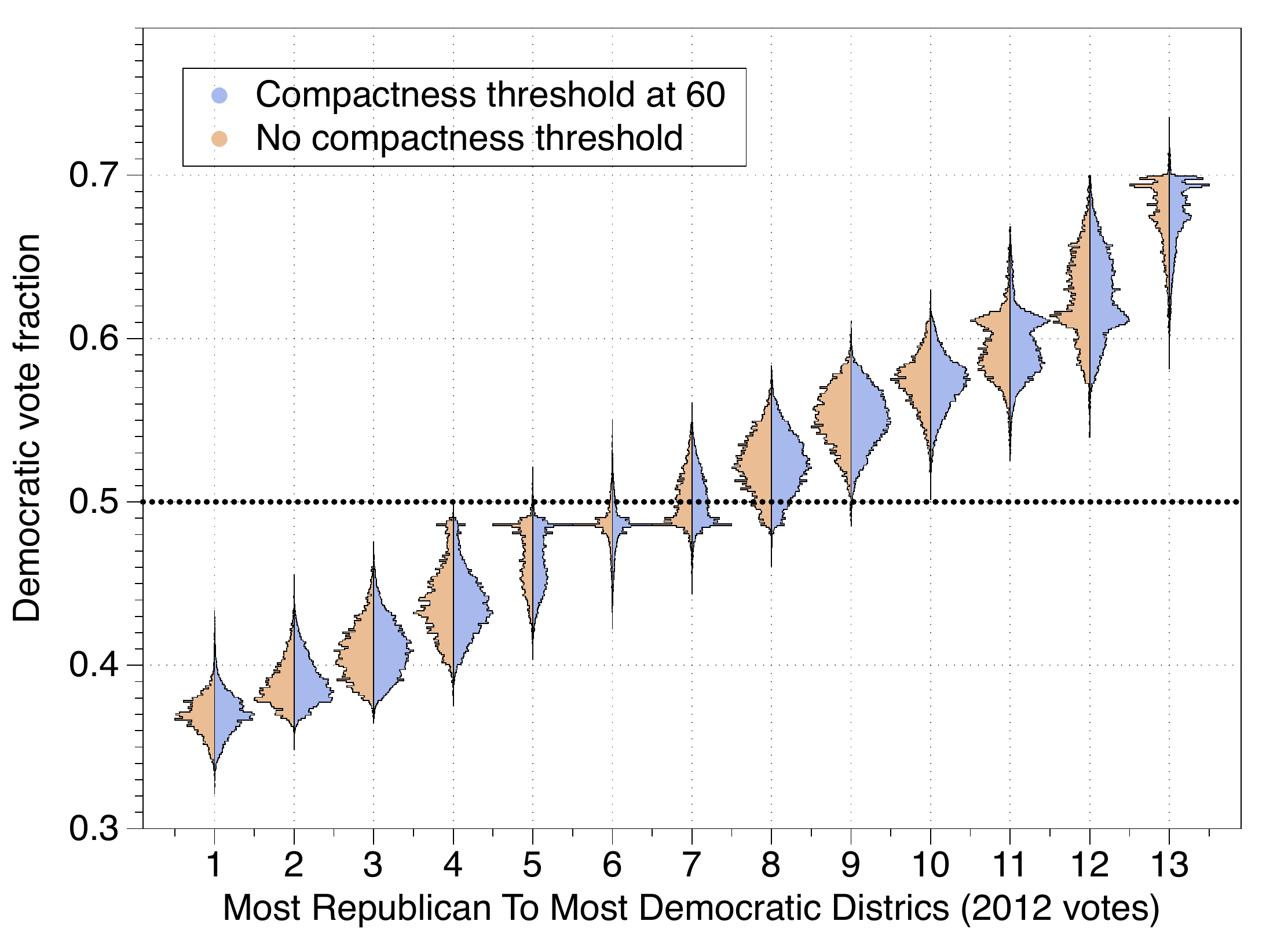}
   \caption{\capSize We display changes of the distribution of election results with changes to the compactness threshold (left).  The histogram formed with a maximum of 60 for the isoperimetric ratio overlays this image with the gray shaded histogram.  We display changes to the histogram of the box-plot when comparing a maximum of 60 in the isoperimetric ratio without any thresholding on compactness (right).}
  \label{fig:deltaCompThresh}
\end{figure}

Next, we note that there is no corresponding law to dicate a choice
of compactness threshold.     The NC2016 districts have a maximum isoperimetric ratio of around 80, and the NC2012 districts have a maximum of over 400.  The Judges redistricting has a district with maximum isoperimetric ratio of around 54.  To test the effect of setting different compactness thresholds, we repeat our analysis by choosing 54, 80 and no threshold for the maximum isoperimetric ratio of all districts within a redistricting.  We find that relaxing the compactness threshold minimally changes the election results as demonstrated in Figure~\ref{fig:deltaCompThresh}.  We note that having no threshold does not mean that we have arbitrarily large compactness values. This is because of the cooling process in the simulated annealing algorithm and the fact that we continue to penalize large compactness scores.  We find that we have an average maximum isoperimetric ratio of around 75 and that we rarely see redistrictings with maximal ratio larger than 120.
 
\subsection{Sample Maps, Details of Districts and Raw Data}
\label{sec:raw}
Table~\ref{tab:table1} gives the percentage of Democratic votes
in NC2012, NC 2016, and the Judges redistrictings. The numbers in
parentheses give the numerical label of the individual districts as
identified in the maps in the Appendix. The last set
of columns contains the mean values for each positon in the  ranked ordered vector containing the
percentage of Democratic votes for each redistricting. As already
described, the
Gerrymandering Index is Euclidean
distance from this vector of  marginal
means. One can easily identify the
particular districts which have likely been packed or cracked by comparing
the values for a given district to this vector of means.
\begin{table}[h!]
  \centering
  \begin{tabular}{r||l|l||l|l||l|l||l|l||}
    \multicolumn{1}{c||}{ }&\multicolumn{2}{|c||}{NC2012}
    &\multicolumn{2}{|c||}{NC2016}&\multicolumn{2}{|c||}{Judges}&\multicolumn{2}{|c||}{Mean}\\ 
    \cline{2-9}
    \multicolumn{1}{c||}{Rank}& 2012 & 2016  & 2012 & 2016  & 2012 & 2016  & 2012 & 2016  \\
    \hline
    1 & 37.5\pp (3)  & 34.2\pp (3)  & 38.7\pp (3) & 32.8\pp (3)  &
                                                                   35.5\pp
                                                                   (10)
                                     & 28.9\pp (10)
                                        &  37.0\pp   & 30.6\pp\\
    2  & 39.0\pp (6) & 34.6\pp (11)  & 42.5\pp (10) & 35.8\pp (11)  &40.0\pp (2)
                                     & 33.6\pp (2) & 39.1\pp &33.0\pp\\
    3 & 42.4\pp (5) & 36.2\pp (7)  & 43.7\pp (6) & 36.8\pp (10)
                              &42.6\pp (12)  & 36.3\pp (7)
                                             &41.0\pp &35.3\pp  \\
    4 & 42.5\pp (11)  & 36.6\pp (8)  & 43.9\pp (11) &  39.0\pp (7)
                              &42.7\pp (7)  & 37.6\pp (12)  &43.7\pp &38.5\pp \\
    5 & 42.6\pp (2)  & 37.4\pp (10) & 44.0\pp (2)  &  40.7\pp (6)
                              &44.5\pp (9)  &40.0\pp (9) &46.4\pp &40.6\pp  \\
    6 & 43.1\pp (10) & 38.9\pp (5) & 45.1\pp (5)   &  41.2\pp (8) &
                                                                    48.5\pp (8) &41.9\pp (3) &48.4\pp &42.2\pp   \\
    7 & 43.5\pp (13) & 40.8\pp (6)  &46.3\pp (13)  &  41.6\pp(5) &
                                                                   48.8\pp
                                                                   (11)
                                     & 42.7\pp (11) &50.2\pp & 44.3\pp\\
    8 & 46.2\pp (8) & 41.2\pp (2)  & 47.3\pp (8) &41.8\pp (9) &
                                                                50.5\pp
                                                                (4) &
                                                                      45.7\pp (4) & 52.3\pp&47.7\pp\\
    9 & 46.7\pp (9) & 44.0\pp (9)  & 49.4\pp (9) & 43.3\pp (2)  &
                                                                  57.0\pp
                                                                  (3)
                                     &48.1\pp (8) & 55.1\pp&51.2\pp  \\
    10 & 50.1\pp (7)  & 45.8\pp (13)   &  51.6\pp (7) &  43.9\pp (13)
                              & 57.5\pp (5) & 55.9\pp (1) & 57.2\pp&54.6\pp  \\
    11 & 74.4\pp (4)   & 71.5\pp (1)  & 66.1\pp (4) & 66.6\pp (12)  &
                                                                      59.2\pp
                                                                      (1)
                                     & 59.7\pp (5) &59.5\pp & 57.5\pp \\
    12 & 76.0\pp (1) & 73.0\pp (4)  & 69.8\pp (12) & 68.2\pp (4)  &
                                                                    64.6\pp
                                                                    (6)
                                     &63.3\pp (13) & 62.6\pp&61.4\pp  \\
    13 &79.3\pp (12)  &75.3\pp (12)   & 70.9\pp (1)  & 70.3\pp (1)  &
                                                                      66.0\pp
                                                                      (13)
                                     &65.3\pp (6) &67.5\pp &65.1\pp
    \\
  \end{tabular}
\vspace{1em}
  \caption{Percentage of Democratic votes in each district when
    districts are ranked from most Republican to most
    Democratic. Number in parentheses give label of actual district
    using the numbering convention from maps in the Appendix. This data is
    plotted in Figure~\ref{fig:boxPlotsCDF}
    and~\ref{fig:DensityVersion} on top of the summary box-plots. }
  \label{tab:table1}
\end{table}
In particular, the three most Democratic districts labeled 1, 4 and 12
in both the NC2012 and NC2014 plan have significantly more Democratic
votes.  Districts 9 and 13 both show evidence of having less Democrats
than one would expect from their rankings. These conclusions are consistant across the 2012 and 2016 votes. 

The raw data used to produce Figure~\ref{fig:basicDemWinners} is given
in Table~\ref{table:winnerCount}. It underscores how atypical the
results produced by the NC2012 and NC2016 redistrictings are. If one
is ready to accept four seats in the  2012  vote then one should
equally accept nine. Similarly in the 2016 votes, if one accepts three
as a legitimate outcome then one should also be willing to accept seven
seats. None of these results seem particularly representative of the
votes cast.
\begin{table}[h!]
  \centering
  \begin{tabular}{r|r|r|r|r|r|r|r|r|r|r|}
 &  \multicolumn{10}{|c|}{\# of Democratic Winners}  \\
\cline{2-11}
 & 1 & 2 &  3& 4 & 5 & 6 & 7& 8 & 9 & 10\\ 
\hline
2012 Votes& 0 & 0 & 0& 89& 2875 & 9455 &9690& 2288 & 121 & 0\\
2016 Votes&  0 &1 & 162 & 6861& 13510 & 3881 & 103 & 0  & 0 & 0\\
  \end{tabular}
\vspace{1em}
  \caption{Among the 24,518 random redistrictings generated, the
    number which produced the indicated number of Democratic seats in
    the Congressional Delegation. }
\label{table:winnerCount}
\end{table}

\subsection{Independence of initial conditions and simulated annealing
  parameters}
\label{sec:IndependenceOfIC}
\begin{figure}[ht]
  \centering
\includegraphics[width=8cm]{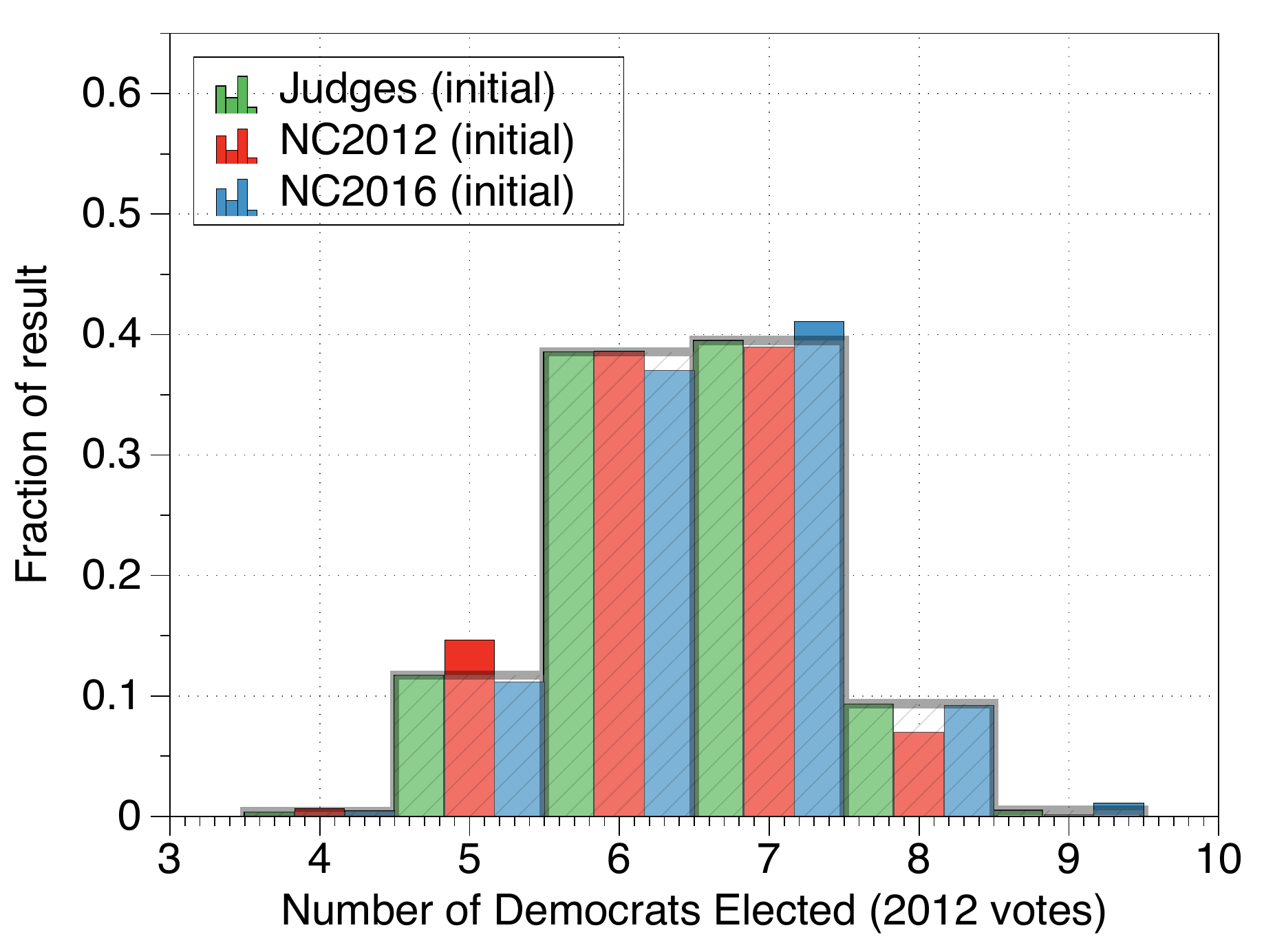}
\includegraphics[width=8cm]{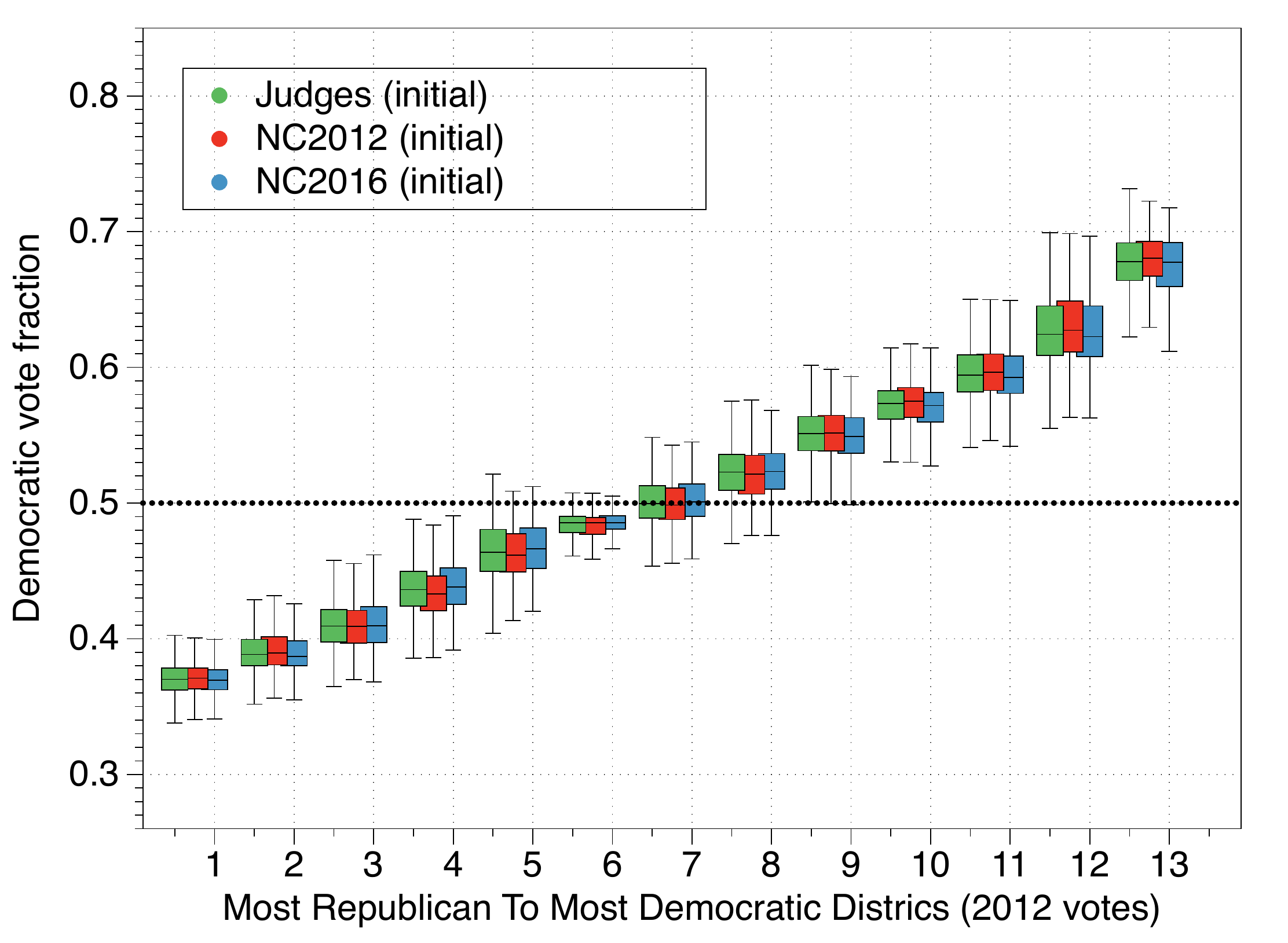}\\
\includegraphics[width=8cm]{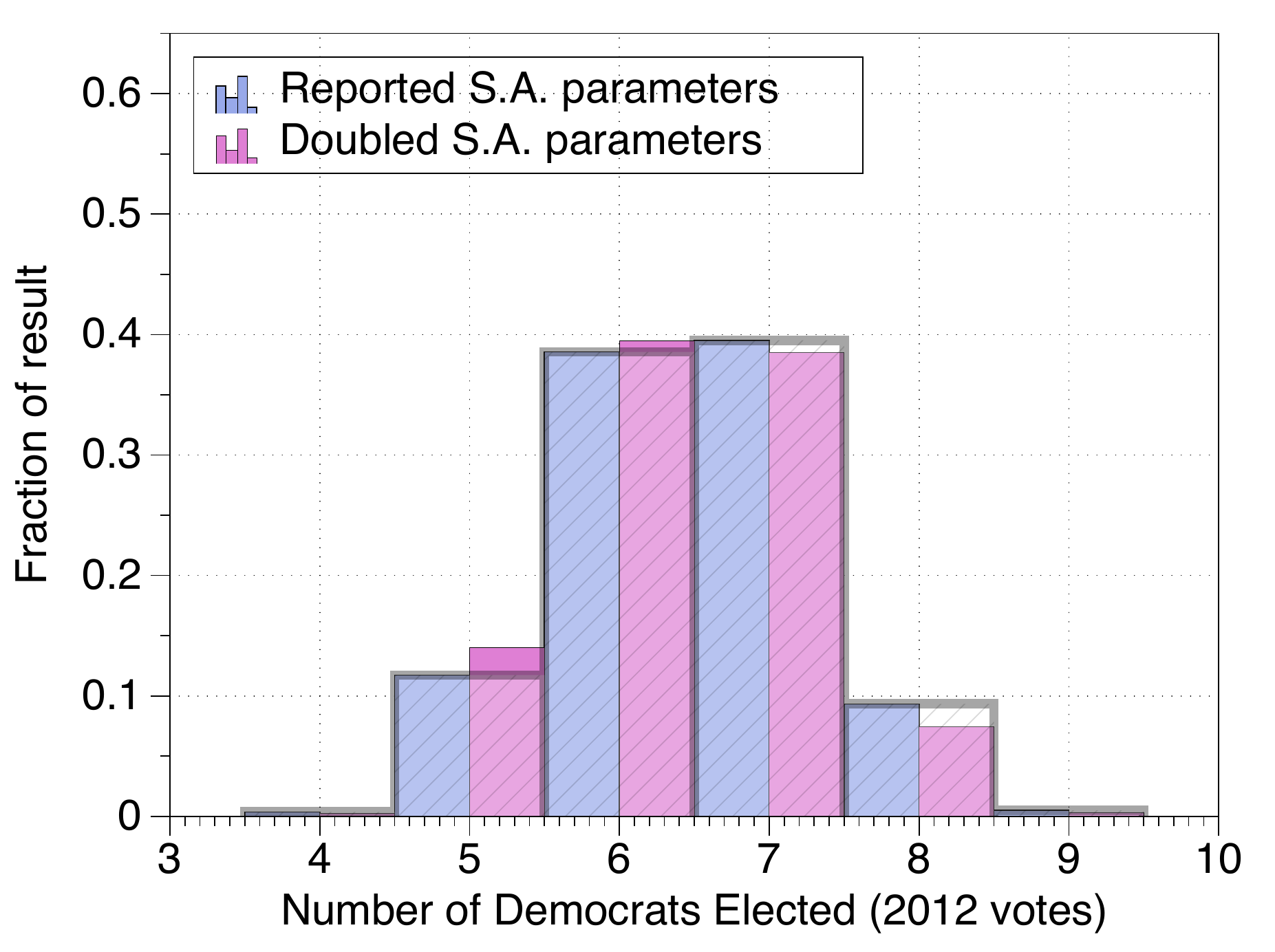}
\includegraphics[width=8cm]{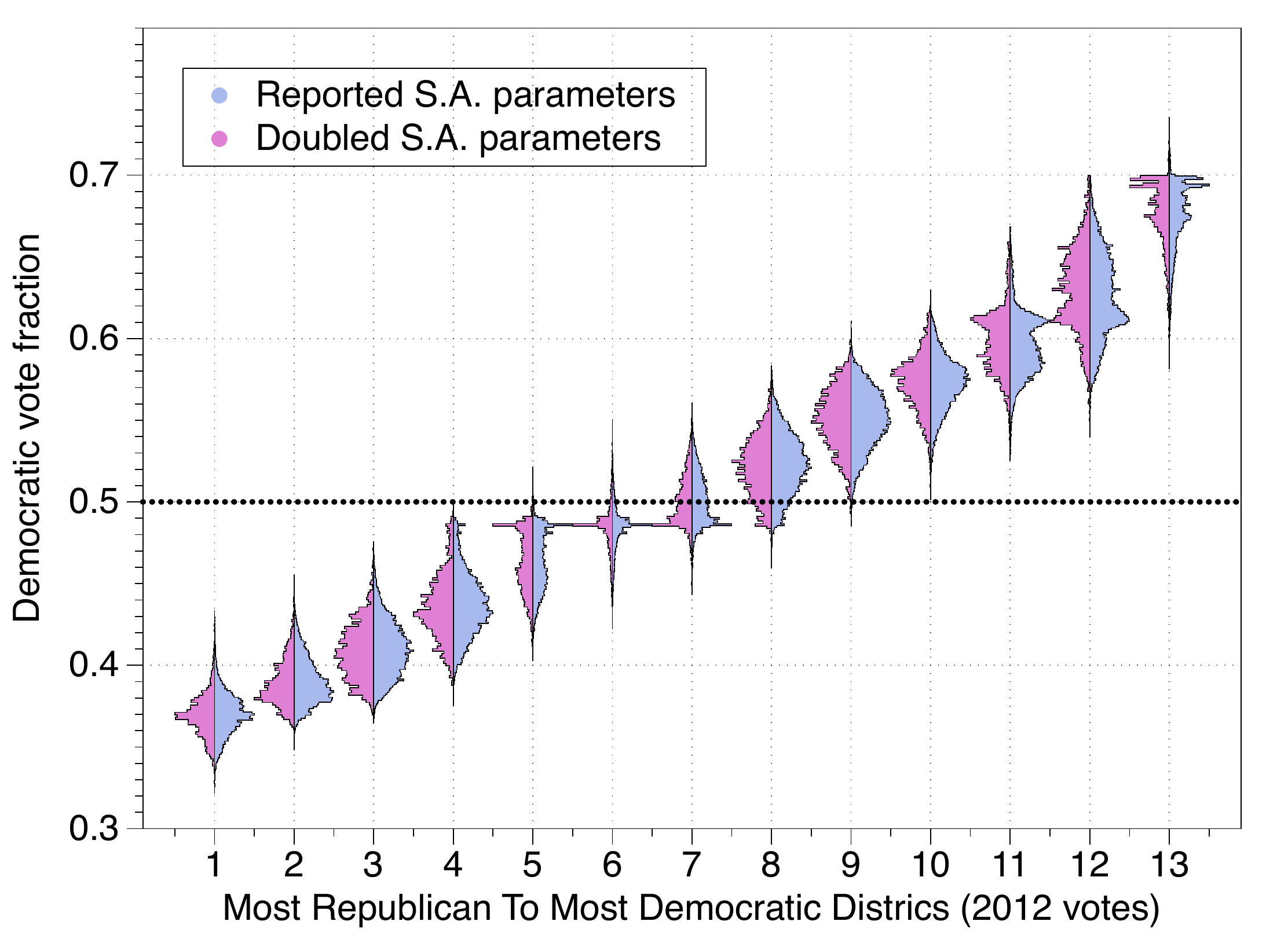}\\
   \caption{\capSize We display the probability distribution of
     elected Democrats with respect to initial conditions (top left)
     and the original versus doubled simulated annealing parameters
     (bottom left).  The histogram formed with the Judges as an
     initial condition and the previously reported simulated annealing
     parameters overlays this image with the gray shaded histogram.
     We display our standard box-plots for the three initial conditions as we need to compare three results rather than two (top right) along with the histogram box-plots to compare the effect of changing the simulated annealing parameters (bottom right).}
  \label{fig:deltaSAConditions}
\end{figure}

There is a possible pitfall of using simulated annealing: we may
become trapped in local regions, leaving us unable to explore the
entire space of redistrictings.  This may be because we have cooled the
system down too quickly, keeping it trapped in a local region, or it
may be because the likelihood of finding a path out of one local
region of redistrictings and into another is small.  We note that we
have animated our algorithm and have found that districts may travel
from one end of the state to another; such motion suggests that many
types of redistrictings are sampled, and it is reasonable to hypothesize
that as districts exchange locations, they lose information on past
configurations.  To  more fully vet this idea, we examine the effect
of (i) choosing a different initial redistricting in our algorithm, and
(ii) doubling the simulated annealing parameters, thus cooling the
system down twice as slowly.  To clarify the point (ii), instead of remaining hot ($\beta=0$) for 40,000 steps, cooling linearly for 60,000 steps, and remaining cold ($\beta=1$) for 20,000 steps, we instead remain hot for 80,000 steps, cool linearly for 120,000 steps, and remain cold for 40,000 steps. We then check to see if the election results are altered by changing these conditions and display our results in Figure~\ref{fig:deltaSAConditions}.

We find that the changes with respect to both initial conditions and
the slowdown of the annealing process have little effect on the election
results.   There are slight effects; for example, the initial
condition for the NC2012 redistricting has a 15\% chance of electing five
Democrats rather than the 12\% chance we have seen before.  We note
that these are exploratory runs, so we have less than 1000 accepted
districtings for the NC2012 and NC2016 initial conditions (each has
close to 1000) and less than 2500 runs for the increased cooling
times.  These sample sizes are robust enough to provide a general
trend but are subject to statistical variations. Hence the small sample sizes are a possible and likely culprit of these variations.

\subsection{Evidence of proper sampling}
\label{subsec:enough-samples}
\begin{figure}[ht]
  \centering
\includegraphics[width=8cm]{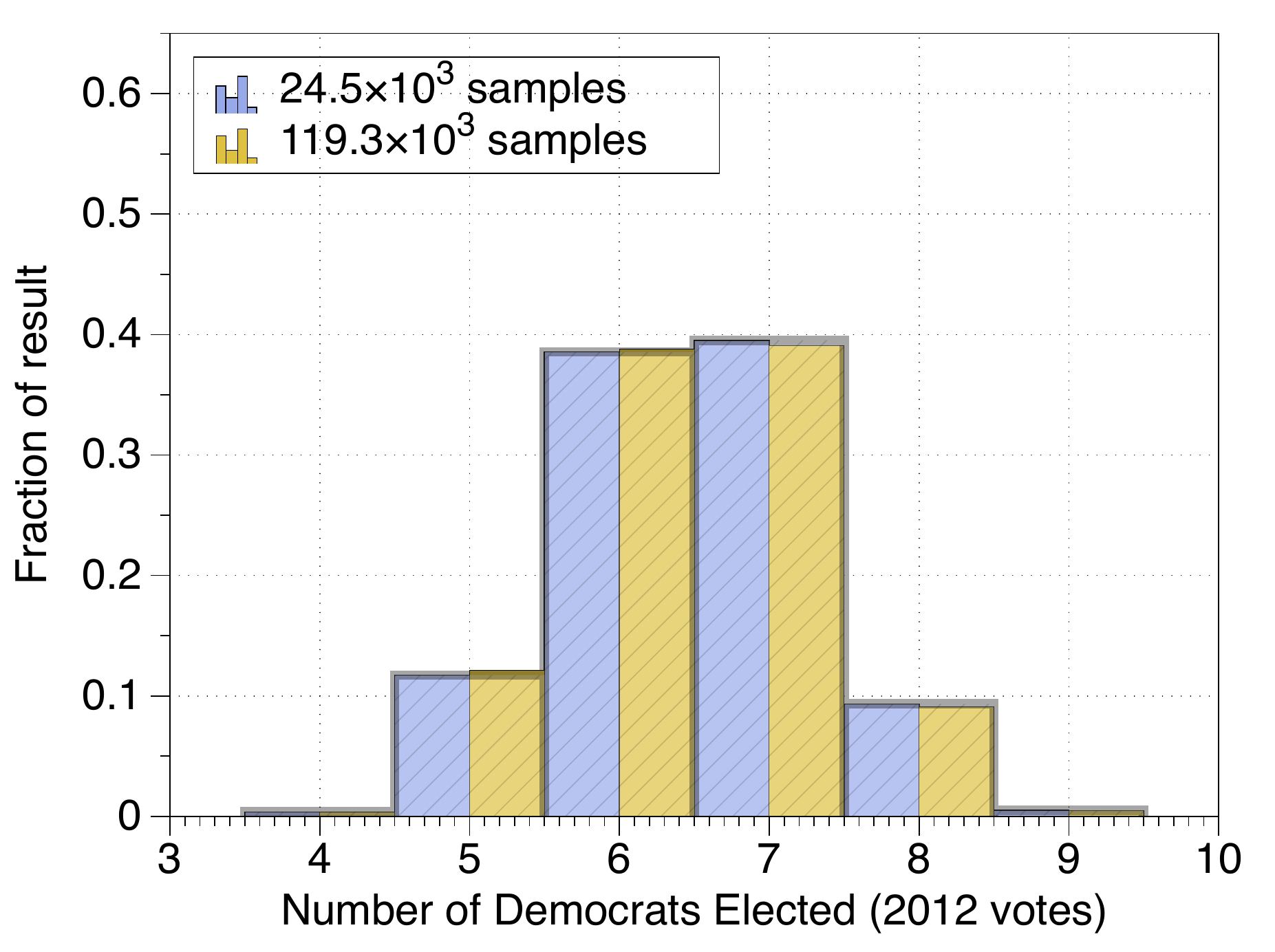}
\includegraphics[width=8cm]{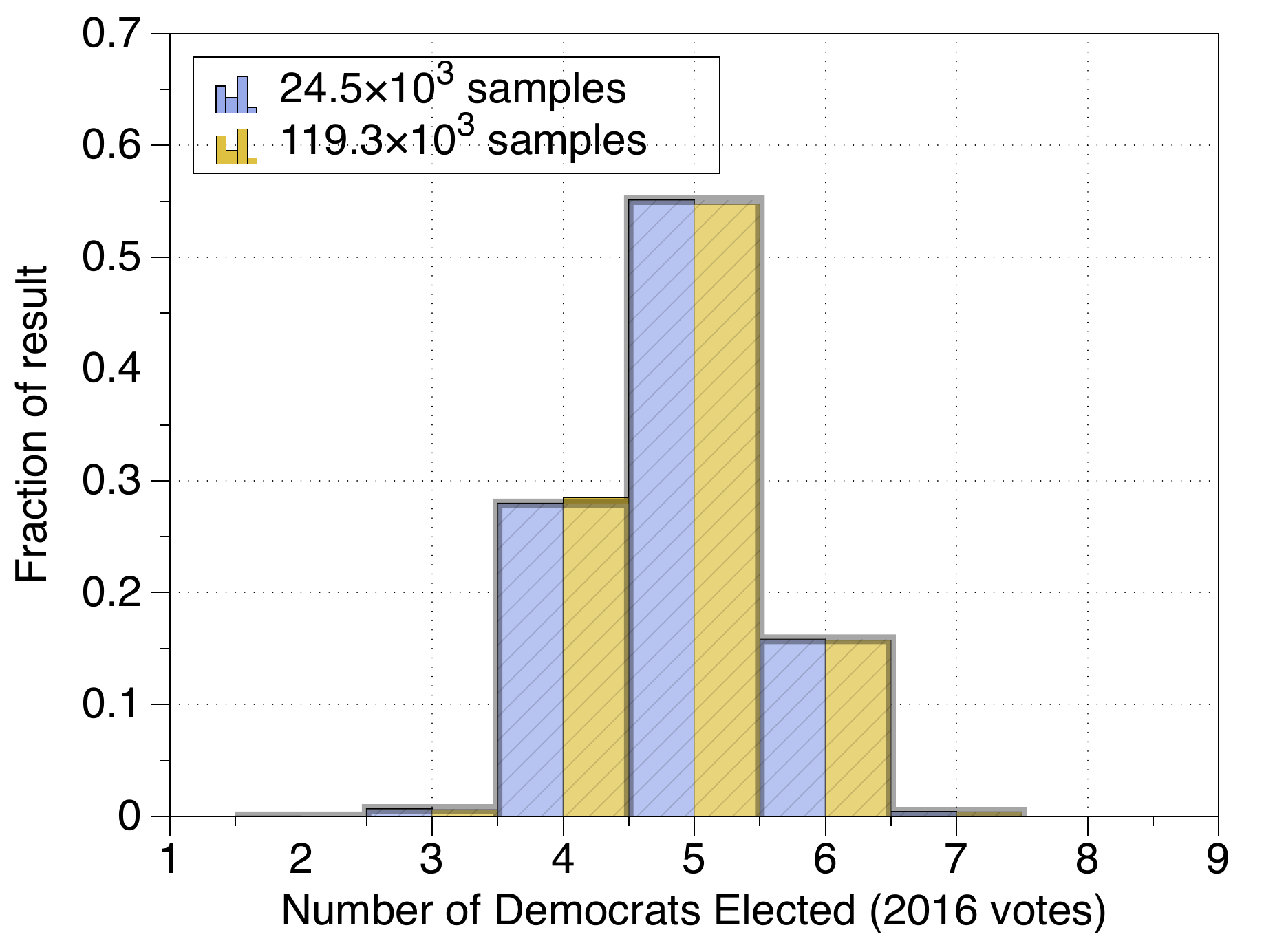}\\
\includegraphics[width=8cm]{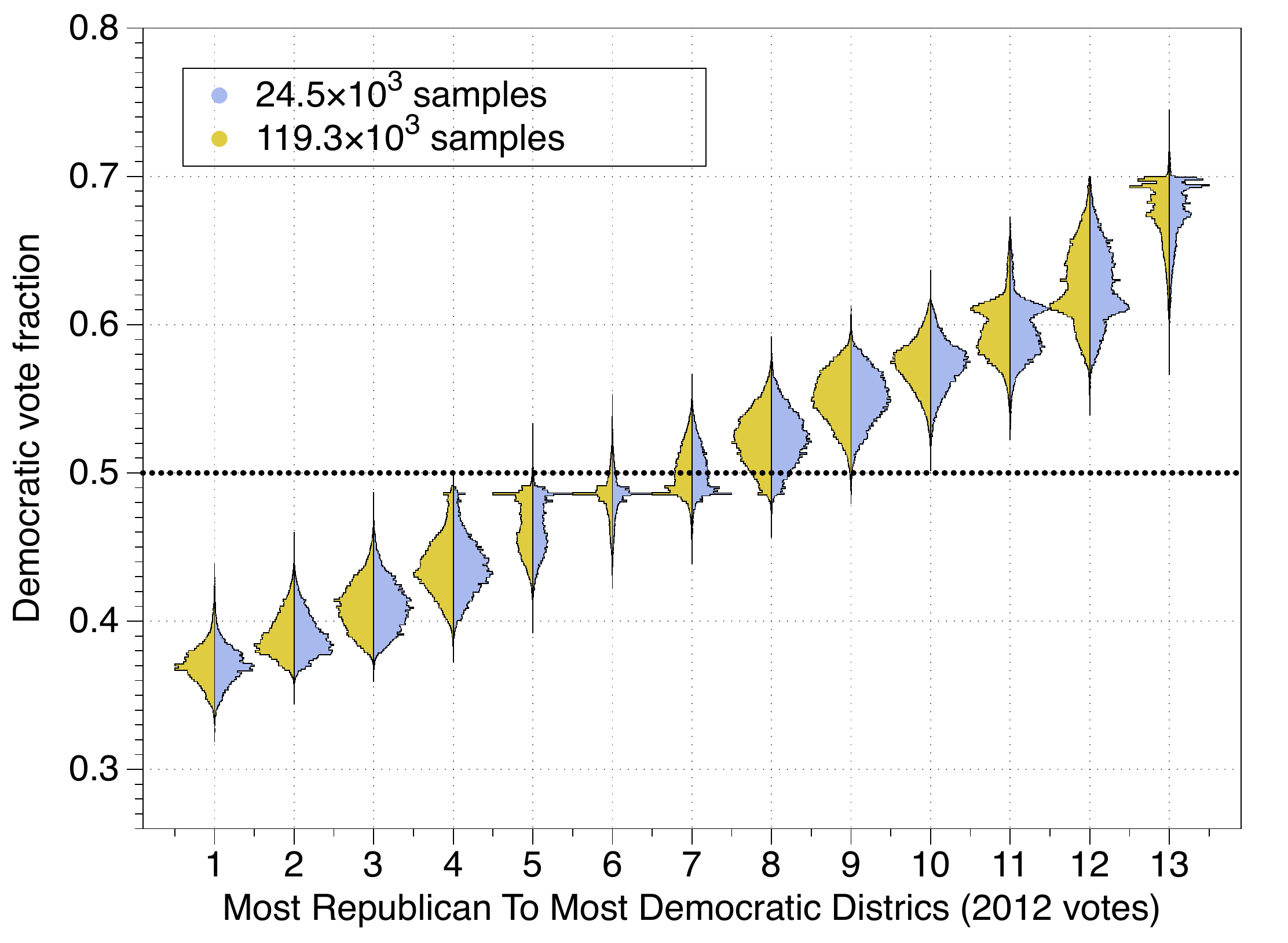}
\includegraphics[width=8cm]{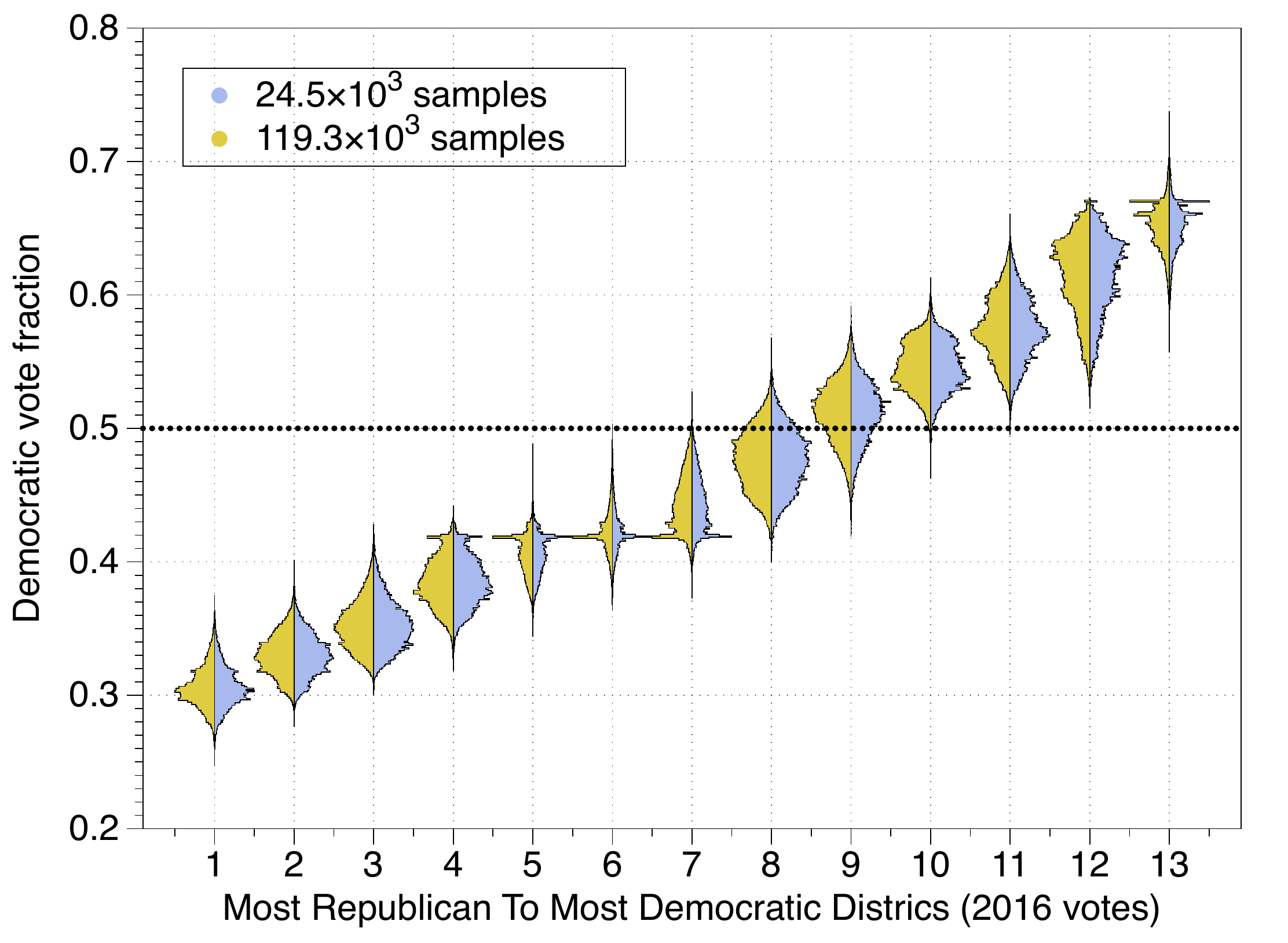}
   \caption{\capSize We extend the samples from the main text by allowing the sampling algorithm to continue until we have sampled roughly 120 thousand districts that fall below the threshold.  We find almost no difference between the distributions in the original and extended samples.}
  \label{fig:moreSamples}
\end{figure}

The above test gives strong evidence that we have properly sampled the
probability distribution of redistrictings.  To strengthen this claim,
we also continue to allow the algorithm to sample the space until we
have sampled roughly 120 thousand acceptable redistrictings as defined
by the original thresholding criteria.  We then compare the results of
the elections along with the box plots and histogram plots.  We find
that there is negligible change in the distribution of outcome both
for the overall number of elected representatives and for each ordered
district from most to least republican.  We display our results in
Figure~\ref{fig:moreSamples}.  The stability of these results together
with those presented in Section~\ref{sec:IndependenceOfIC} provides
robust evidence that we have correctly recovered the underlying probability distribution of redistrictings.

\subsection{Varying weights}
\label{sec:VaryingWeights}
We have proposed a methodology for determining the weights in the
score function that is primarily concerned with obtaining a high
percent of redistrictings below our chosen threshold values (see
Section \ref{ssec:determine-param}).  We note that other parameters
may be chosen, and here we test whether making a different choice will
affect the statistics on the election outcomes.  We are in a four
dimensional space, meaning that the parameter space is very large.
Exploring this space exhaustively  would come at an large
computational cost.  We instead perform a simple sensitivity test on
our current location in the parameter space by exploring the four
dimensional space in four linearly independent directions.  We explore
over three directions by significantly increasing and decreasing
$w_p$, $w_I$, and $w_m$.  For the fourth direction, we note that we
could simply increase or decrease $w_c$; however, we thought it might
be interesting to increase and decrease $\beta$ instead. Because
changing $\beta$ is equivalent to changing all parameters, this forms
a fourth linearly independent search direction, and provides us with
information similar to changing $w_c$. This leads us to examine eight different parameter sets, which still requires a large number of runs.  To cut down on the computational cost, we take advantage of the result presented in section \ref{subsec:vary-thresholds} above, where we conclude that ignoring the compactness threshold has a minimal effect on our results. The compactness threshold is by far the most restrictive, so omitting it will allow us to sample more redistrictings with fewer runs.

\begin{figure}[ht]
  \centering
\includegraphics[width=15cm]{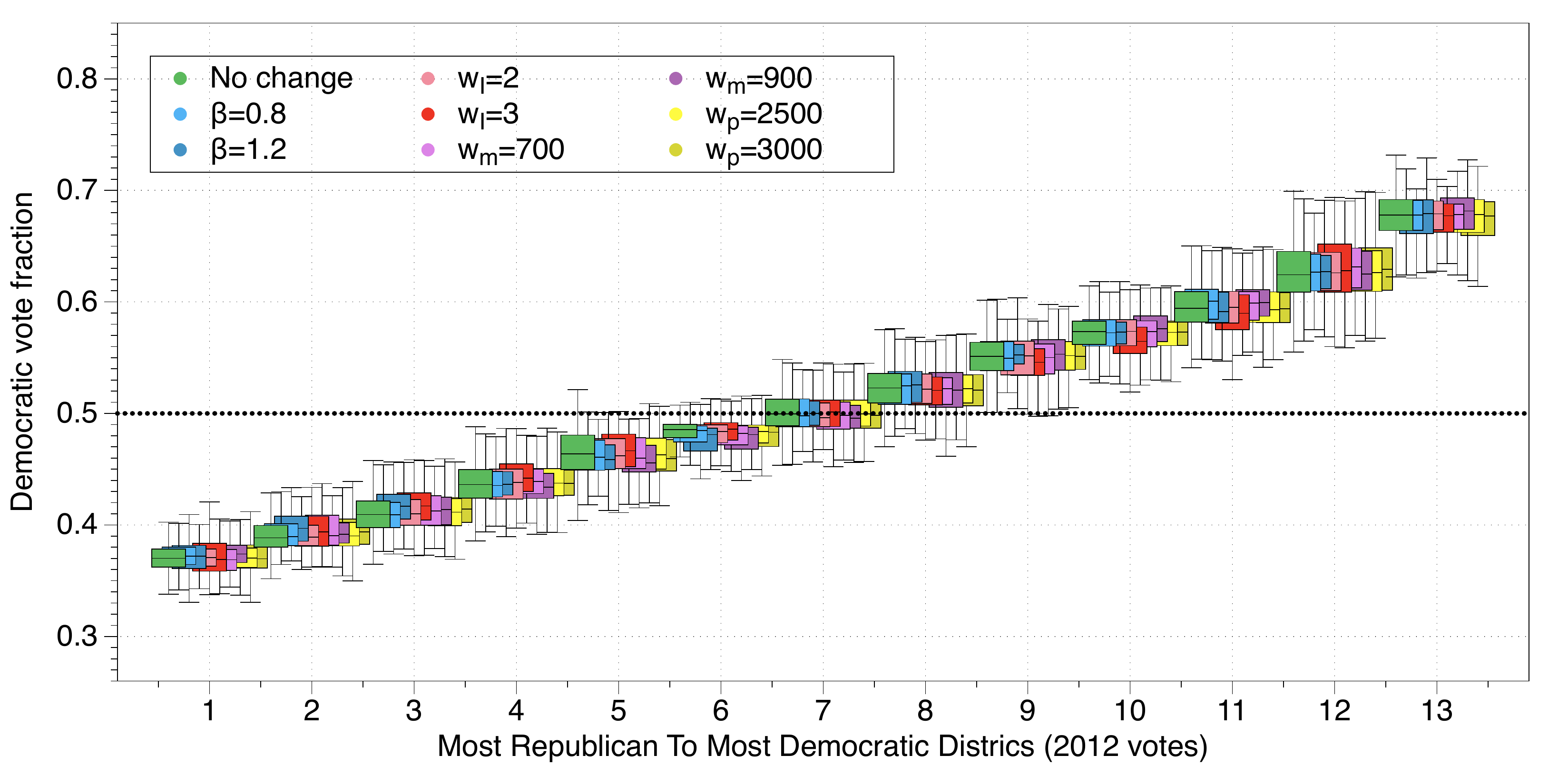}
   \caption{\capSize
   We display standard box-plots and demonstrate how the election results change with respect to changing the values of the weights.
   }
  \label{fig:varyWeights}
\end{figure}

We present our results in Figure~\ref{fig:varyWeights}, and find that the results are very robust in all examined directions of changing parameters.  We note, however, that the percentage of redistrictings that falls below our compactness acceptance threshold does change with varying parameters.  Based on our result that election results are robust with respect to large changes in the compactness threshold, we conclude that significant changes in the parameters will have little effect on the statistical results of the election data.

\subsection{Different weights for lower county splits}
\label{sec:lowcountysplit}
\begin{figure}[ht]
  \centering
\includegraphics[width=8cm]{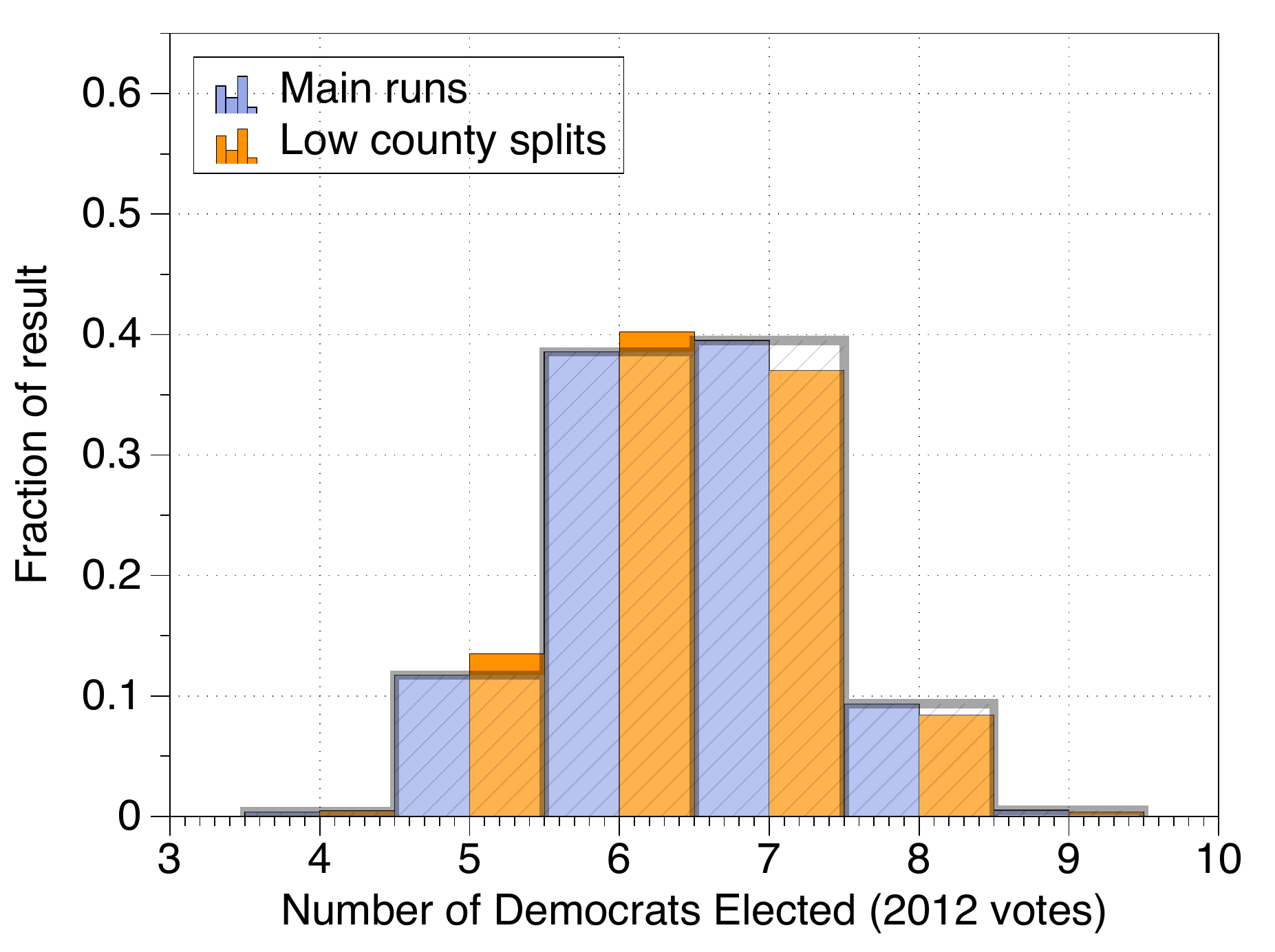}
\includegraphics[width=8cm]{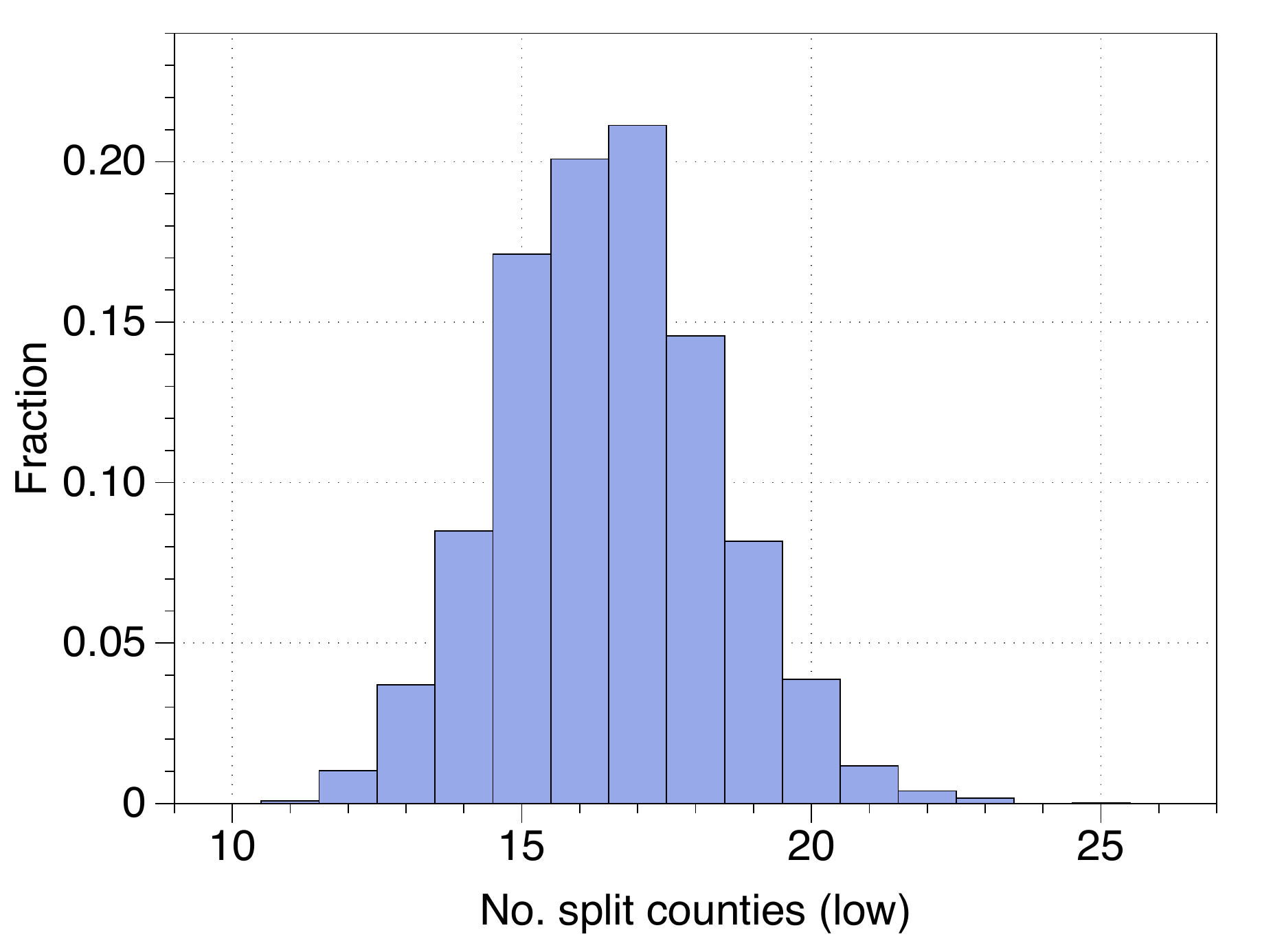}\\
\includegraphics[width=8cm]{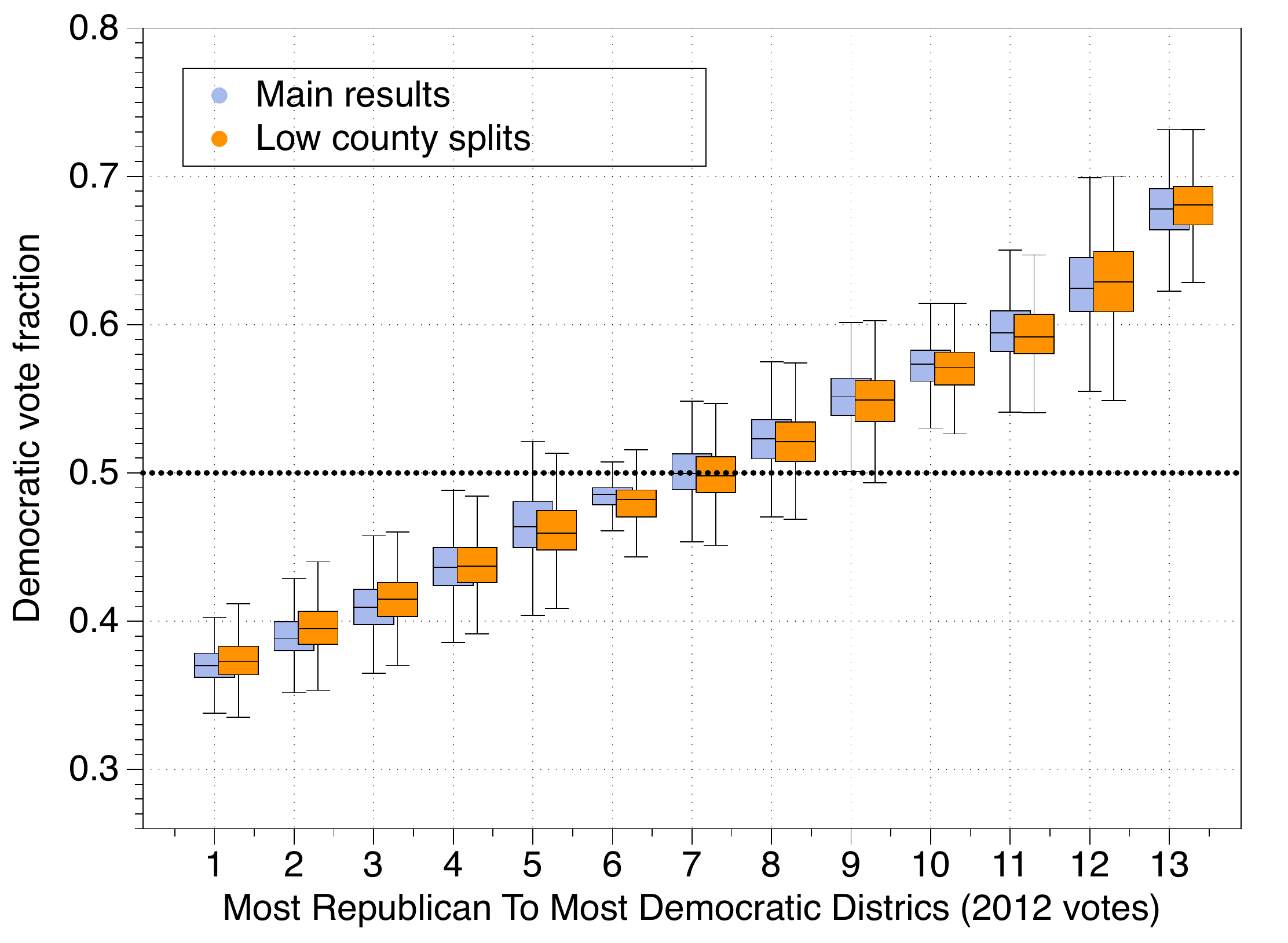}
\includegraphics[width=8cm]{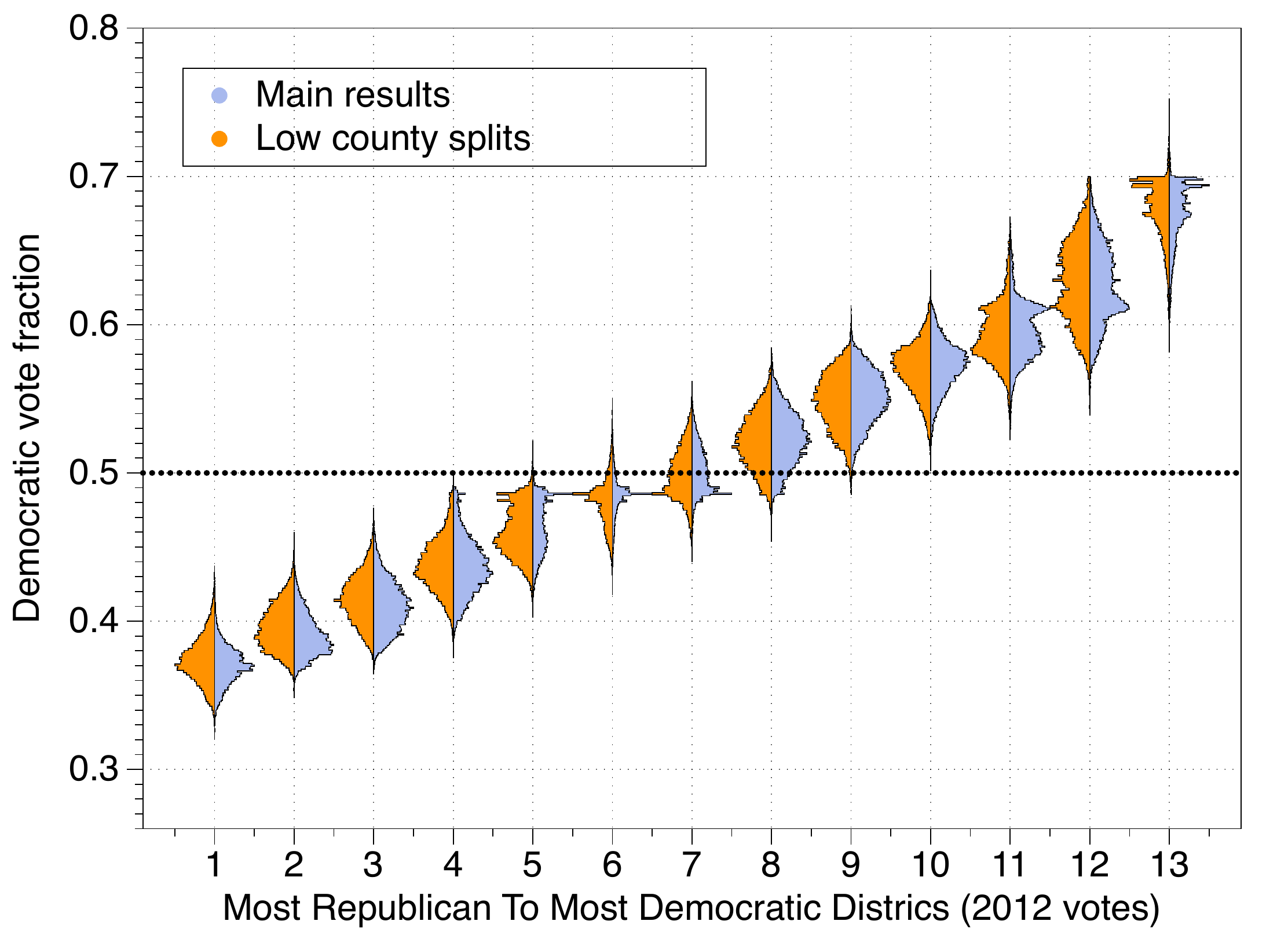}
   \caption{\capSize By changing the weights on the energy function we alter the distribution of two county splits (top right).  Despite these changes, the over all election results (top left) and box and box histogram plots by district (bottom) remain stable.}
  \label{fig:lowcs}
\end{figure}

In the above analysis we have prioritized compact districts over those with low two county splits.  The result of this is presented in Figure~\ref{fig:minoCountyStats} and Section~\ref{sec:characteristics} above, and we note again that the number of two county splits in our samples is far less than the NC2012 plan, but generally greater than that of NC2016 or the Judges.  In this section we determine the sensitivity of our results when we prioritize keeping a low number of county splits.  To make this examination, we double the county weight ($w_c=0.8$) and reduce the compactness weight ($w_I=2$).  By resetting the compactness threshold to be 80, we obtain just under 15 thousand redistrictings and note that all of them have a worst district better than the worst NC2016 district.  Keeping the threshold at 60 only yields a couple thousand samples, thus in order to obtain more samples and because we have found that compactness does not have a large effect on the results, we select the higher threshold.  We find that despite changing the weights in this severe way, the over all election results, in addition to the distribution of results per ordered district remains remarkably stable (see Figure~\ref{fig:lowcs}).  We also remark that we now have a median of 16 two county splits with a mean of 16.5, in contrast with a median of 21 and mean of 21.6 from the main results presented above.

\subsection{Using a different compactness energy}
We have used the isoparametric ratio for the compactness energy however there are other possible choices.  Dispersion, mentioned in Section~\ref{sec:sampl-rand-redistr}, measures how spread a district is.  Typically it is thought of as the ratio between the area of the minimal bounding circle and the districts area.  Although useful as a metric to compare two districtings, the dispersion score does not minimized for jagged perimeters and cannot be used as a sufficient criteria to draw reasonable districts.  Never-the-less, we examine the space of districts in which we replace the isoparametric ratios with the dispersion ratio.  Given \S120-4.52(g)(1) of HB92, which specifies district length and width, we chose to measure dispersion as the ratio between the area of the minimal bounding rectangle and the district area.  The redistricting plans we arrive at would never be used due to the jagged perimeters, however if such a drastic change in the compactness criteria gives similar results to those which we have found before, we would have stronger evidence still for the robustness of our analysis.  

\begin{figure}[ht]
  \centering
\includegraphics[width=5cm]{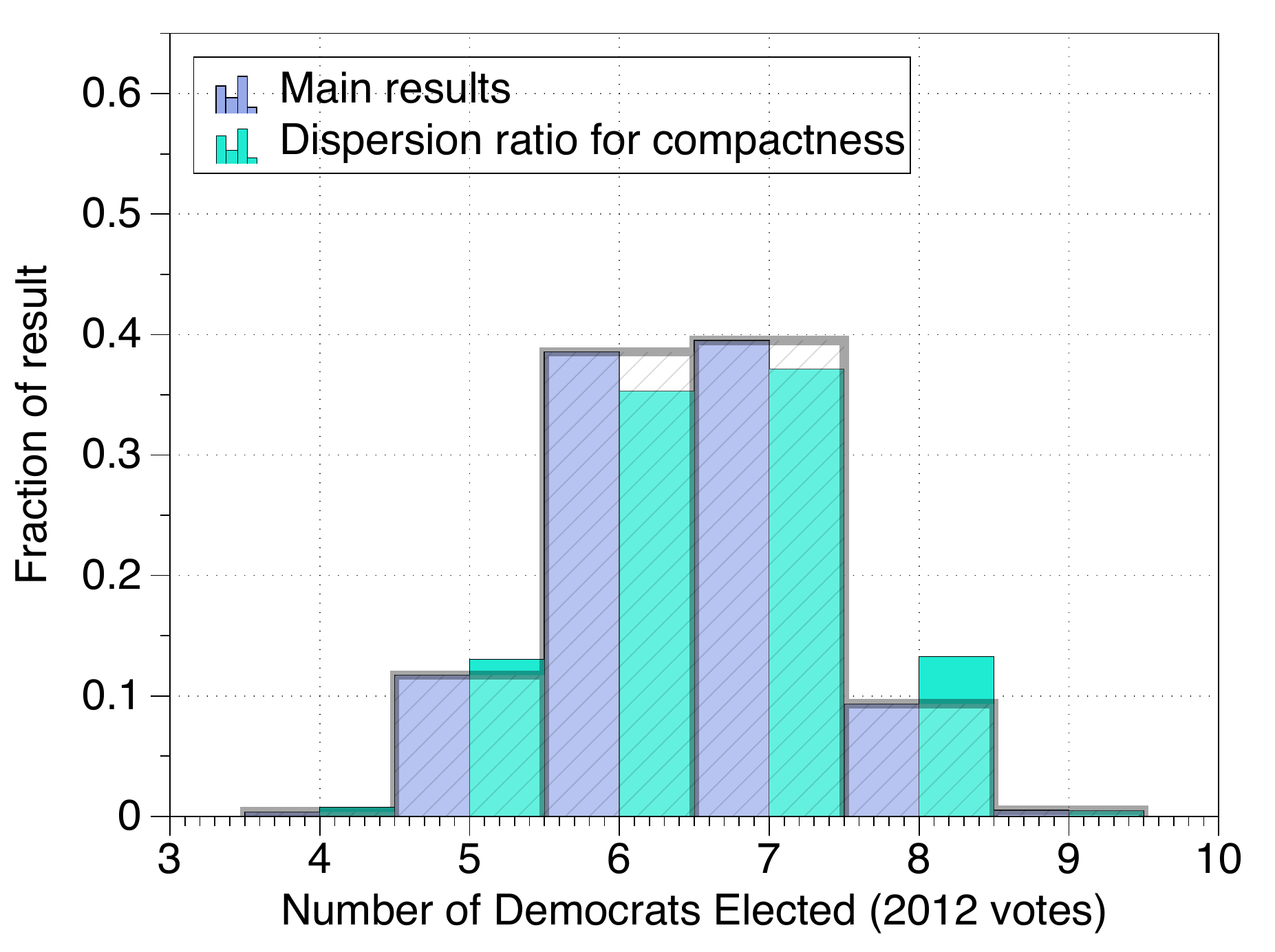}
\includegraphics[width=5cm]{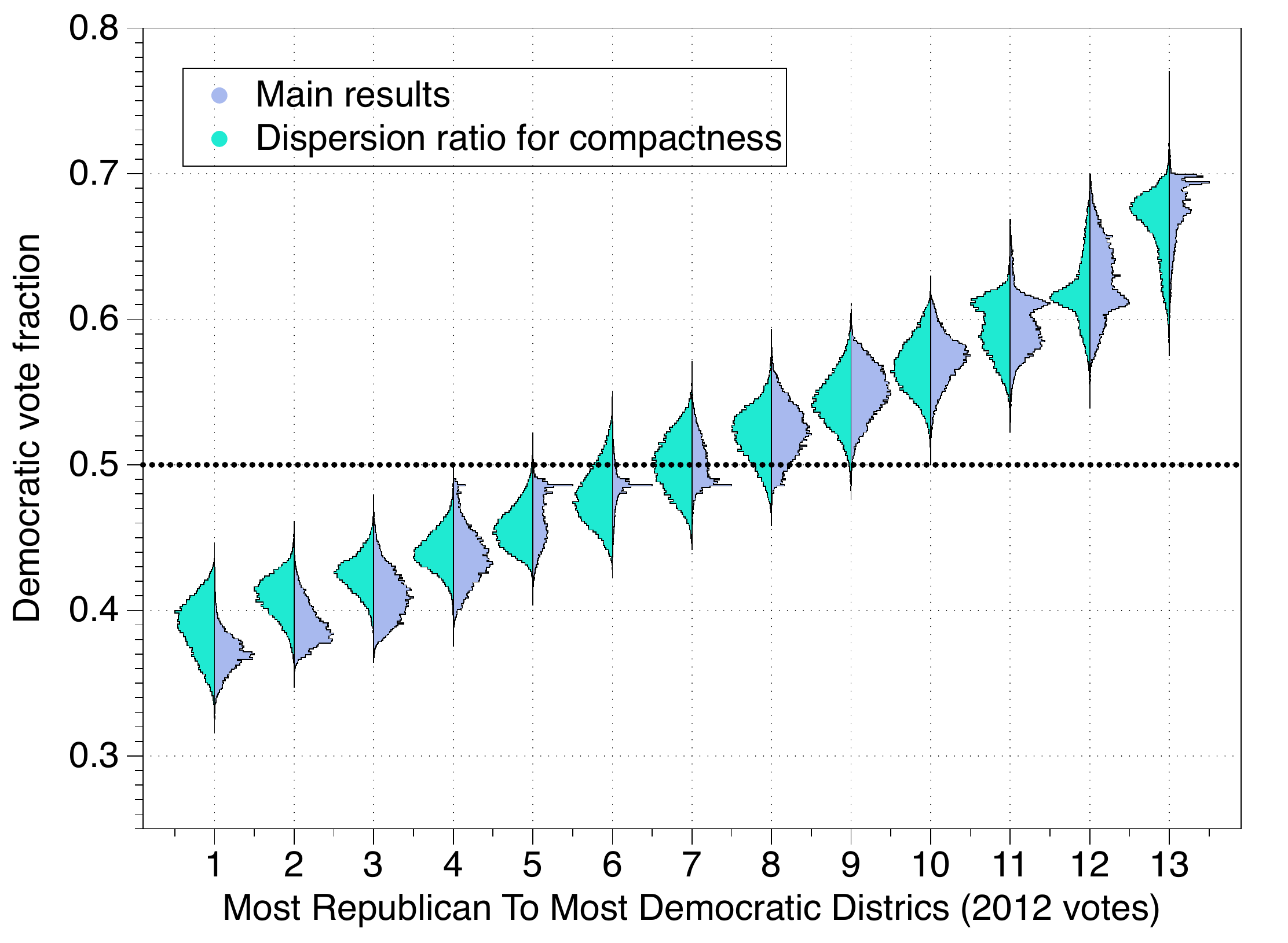}
\includegraphics[width=5cm]{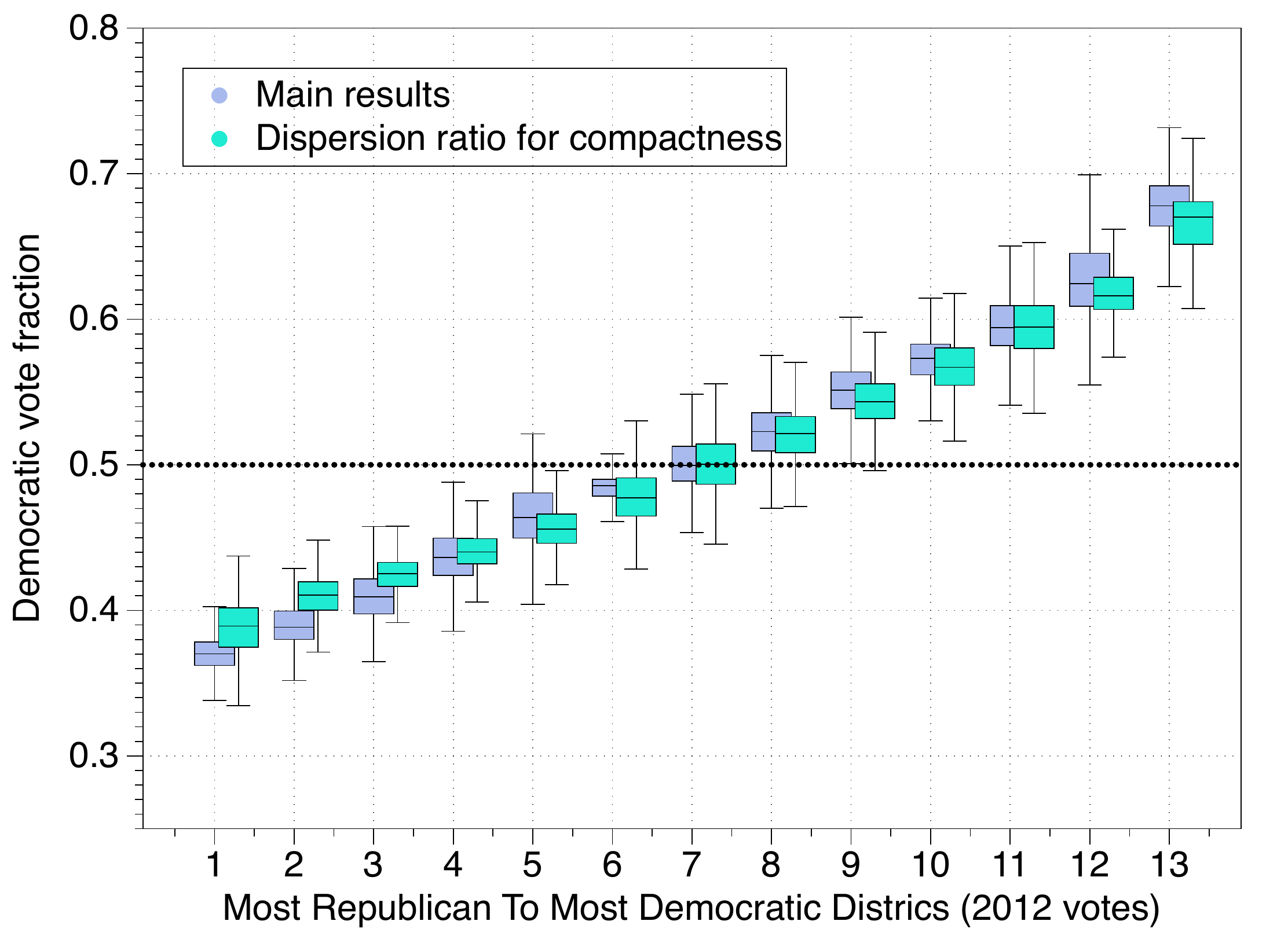}
   \caption{\capSize We change the compactness energy from the isoparametric ratio to a type of dispersion score.  Despite this drastic change in energy definition, we find the results to be remarkably similar.}
  \label{fig:dispersionComp}
\end{figure}

We threshold on everything but compactness as the isoparametric ratios become very high.  We keep the weights the same as in the main results.  With the resulting redistricting plans We display the histogram of the election results and district results and compare them with our main results in Figure~\ref{fig:dispersionComp}. Despite this drastic change in energy definition, we find the results to be remarkably similar to the main results.

\section{Technical Discussions}
\label{sec:more-details-results}

\subsection{Data sources and extraction}
The VTD geographic data were taken from the NCGA website (see \cite{NCGeo} from references)
and
the United States Census Bureau website (see \cite{shapeFilesCongress} from references), which
provide for each VTD its area, population count of the 2010 census,
the county in which the VTD lies, its shape and location.
Perimeter lengths shared by VTDs were extracted in ArcMap from this
data.  Minority voting age population was found on the NCGA website
using 2010 census data (see \cite{NCMino} from references).
Data for the vote counts in each VTD for the 2012 House elections was
taken from Harvard's Election Data Archive Dataverse (see
\cite{NC2012} from references).  Vote count data for the 2016 House elections was
provided by NCSBE Public Data (See \cite{VTD2016} from references). We note that for the 2016 election,
VTD data was not reported for all VTDs, but rather for each precinct;
2447 of the precincts are VTDs, meaning that we have data for the
majority of the 2692 VTDs.  However 172 precincts contain multiple
VTDs, 66 VTDs were reported with split data, and 7 VTDs were reported
with complex relationships.  To extrapolate VTD data on those
contained in the 172 precincts containing multiple VTDs, we split the
votes for a precinct among the VTDs it contained proportional to the
population of each VTD.  For the split VTDs, those containing multiple
precincts, we simply added up the votes among the precincts it
contained.  There was no extrapolation for these VTDs. For the VTDs
with complex relationships, we divided up the votes using estimates
based on the geography and population of the VTDs.  We note that
roughly 10\% of the population lies in the VTDs with imperfect data, and that we
do not expect significant deviation in our results based on the above
approximations.

 In using 2012 and 2016 data we 
have only used presidential election year data. Unfortunately, the 
2014 U.S. congressional election in North Carolina  contained an 
unopposed race which prevents the support for both parties being 
expressed in the VTDs contained in that district. In reference 
\cite{QuantifyingGerrymandering}, the missing votes were replaced with 
votes from the Senate race. However, since we had two full elections, 
namely 2012 and 2016, 
which needed little to no alterations, we chose not to include the 
2014 votes in our study.

\subsection{Examining nearby redistrictings within a distance}
The random sampling of the nearby districts is accomplished by running
the same MCMC algorithm described in Section~\ref{Sampling} with the
small modification that if a proposed step ever tries to increase the
deviation between any of the districts from the original redistricting
in question (either NC2012, NC2016, or the Judges) above 40 VTDs, then
the step is rejected and the chain does not move on that
round. Alternatively, one can think of $J(\xi)=\infty$ for any
$\xi \in \dist$ which has a district that differs from the original
redistricting by more than 40 VTDs. As before, we then threshold the
results for NC2016 and the Judges on the Population Score, the County
Score, and the Minority Score as described in
Section~\ref{subsec:threshold}. We do not threshold on the
Isoperimetric Score as (i) keeping the redistricting near the original is
likely sufficient and (ii) would be too severe for the NC2016 redistricting. We do not threshold the NC2012 at all since most of
the redistrictings close to NC2012 would fail the Population threshold since the compactness energy is so large in this region that it overwhelms population considerations.

We examine the difference in the local complementary cumulative
distribution function (thresholded and not) to get a sense of how
accurate the NC2012 local complementary cumulative distribution function is without thresholding.  We find that there is only a modest difference between the thresheld and non-thresheld results from the Judges which provides evidence that using the non-thresheld results for NC2012 is unimportant for obtaining a representative space of nearby districts.
\begin{figure}[ht]
\centering
\includegraphics[width=8cm]{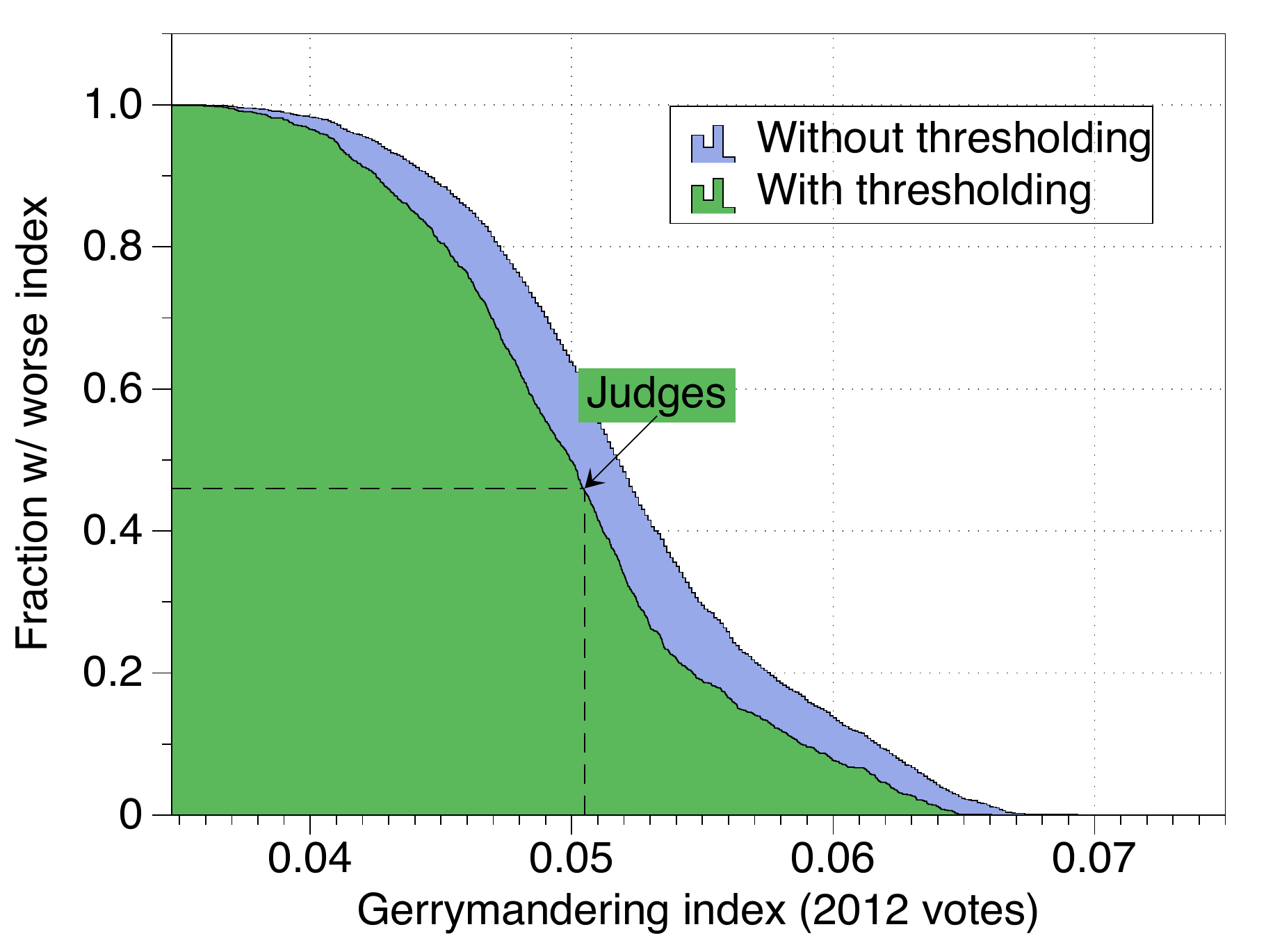}
\caption{\capSize There is not a large difference between the thresholded
and unthresholded results.}
\end{figure}

\section{Discussion}
\label{sec:conclusions}

We have provided a prototype probability distribution on the space of
congressional redistrictings of North Carolina. This distribution is
non-partisan in that it considers no information beyond the total
population, shape of the districts, and relevant demographic information for the VRA. The probability model was then
calibrated to produce redistrictings which are comparable to the
current district to the extent they partition the population equally
and produce compact districts. Then, effectively independent draws were
made from this probability distribution using the Metropolis-Hastings
variant of Markov chain Monte Carlo. For each redistricting drawn, the
2012 and 2016 U.S. House
of Representatives election was retabulated using the actual vote
counts to determine the party affiliation of the winner in each
district. The statistics of the number of Democratic winners gave a
portrait of the range of outcomes possible for the given set of votes
cast. This distribution could be viewed as the true will of the
people.

Redistricting's reach is beyond simply the election outcome. By packing 
and cracking groups, districts can be made safe, which arguably can 
shift the ideological center of the candidate elected away from the 
most representative positions. Our work also helps to identify when 
Gerrymandering has produced districts with unusually large 
concentrations of one party.

Redistrictings producing outcomes which are significantly
different than the typical results obtained from randomly sampled
redistrictings are arguably at odds with the will of the people
expressed in the record of their votes. The fact that the election outcomes are so
dependent on the choice of redistrictings demonstrates the need for
checks and balances to ensure that democracy is served when
redistrictings are drawn and the election outcome is representative of
the votes cast.

It seems unreasonable to expect that politics would not enter into the
process of redistricting. Since the legislators represent the people
and presumably express their will, restricting their ability to
express that will seems contrary to the very idea of democracy. Yet the work in this note could likely be developed into a
criteria to decide when a redistricting fails to be sufficiently
representative of the will of the people. It would perhaps be reasonable to only allow
redistrictings which yield the more typical results, eschewing the most
atypical as a subversion of the people's will. This would still leave
plenty of room for politics, but add a counter-weight to balance that role of
partisanship when it acts against the democratic ideals of a
republic governed by the people.

The most basic critiques of this work is that we have assumed that the candidate 
does not matter, that a vote for the Democrat or Republican will 
not change, even after the districts are rearranged. Furthermore, as 
districts become more polarized and many elections result in a forgone 
conclusion, voter turnout is likely suppressed. While we could try to 
correct for these effects, we find the simplicity and power of using 
the actual votes very compelling. The results are both striking and illuminating.

\section*{Acknowledgments}
We would like to thank the Duke Math Department, the Information
Initiative at Duke (iID), the PRUV and Data+ undergraduate research 
 programs for financial and material support. Bridget Dou and Sophie
 Guo were integral members of the Quantifying Gerrymandering
 Team. Their work was central in the first Data+ summer team which
 greatly influenced and informed the world of the second Data+ team on
 which this note is largely based. In particular, Bridget Dou wrote
 much of the code used that summer and  Sophie
 Guo headed the GIS efforts.
We would also like to thank
Mark Thomas, and the rest of the Duke Library GIS staff for
help with extracting the needed data from the congressional maps. We
are also indebted to Robert Calderbank, Galen Reeves, Henry
Pfister, Scott de Marchi, and Sayan Mukherjee for help during the many phases of this
project. John O'Hale helped in procuring the 2016 election Data. We are also extremely grateful to Tom Ross, Fritz
Mayer, Land Douglas Elliott, and  B.J. Rudell for letting us observe
the Beyond Gerrymandering Project. It was very educational and inspirational.
We also appreciated all of their help,
encouragement, and insightful discussions.

\bibliographystyle{plain}
\bibliography{gerrymandering}

\appendix

\begin{table}
\begin{tabular}{l || llllll}& &  & \multicolumn{2}{c}{\underline{Largest AA}} &
                                                                      \#
                                                                      Split
                                                                      \\
&\multicolumn{1}{c}{$J_p$} & \multicolumn{1}{c}{$J_I$} & \multicolumn{1}{c}{1st} & \multicolumn{1}{c}{2nd} & Counties\\
\hline
\hline
NC2012 & 1.59$\times 10^{-2}$ & 2068.53& 52.54 &50.01 & 40\\
NC2016& 7.93$\times 10^{-3}$ & 699.82 &  45.10& 36.28 & 13\\
Judges& 4.47$\times 10^{-3}$ & 527.14 & 41.68  & 32.96 & 12 \\
\hline
First sample& 1.38$\times 10^{-2}$ & 471.8 & 44.50 & 36.24 & 25\\
Second sample& 8.03$\times 10^{-3}$ & 489.44   & 44.48 & 36.31 &23\\
Third sample& 4.51$\times 10^{-3}$& 454.37 & 44.49 & 36.20 &23\\
Fourth sample& 8.98$\times 10^{-3}$& 465.03 &   41.65 &
                                                                  36.38 & 24\\
Fifth sample& 4.51$\times 10^{-3}$ & 463.86&  42.01 & 33,06 & 21 \\
Sixth sample&5.07$\times 10^{-3}$ & 460.12 &  44.95 & 36.26 & 21
\end{tabular}

 \centering
\vspace{1em}
\caption{\capSize
  We display the various energies for each of the districtings that we
  present in the appendix.  Note that reported numbers for
  districtings are before VTD splits. 
}
  \label{tab:varyWeights}
\end{table}

\begin{figure}[ht]\centering 
\includegraphics[scale=0.62,angle=90]{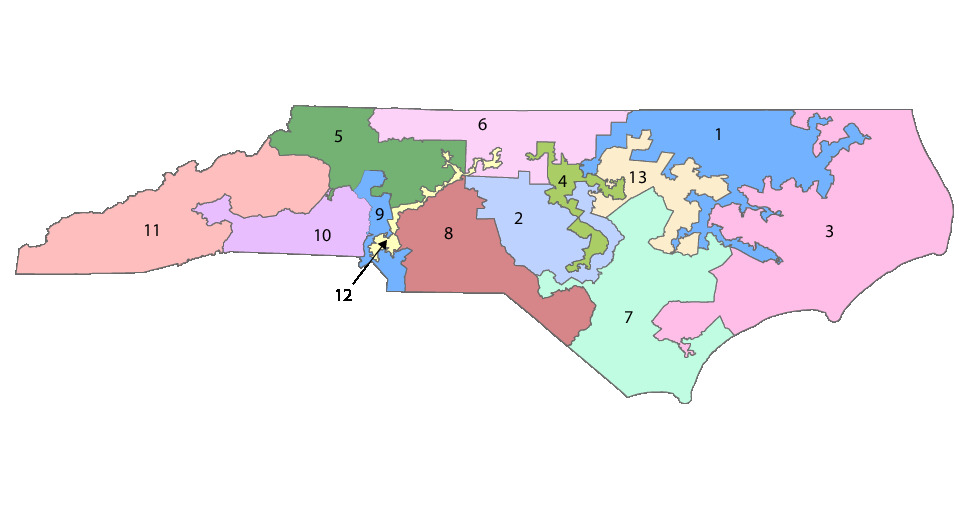} 
\caption{Map for NC2012. Numbers corespond to labels in Table~\ref{tab:table1}.} 
\label{fig:Districting2012}
\end{figure}

\begin{figure}[ht]\centering
\includegraphics[scale=0.62,angle=90]{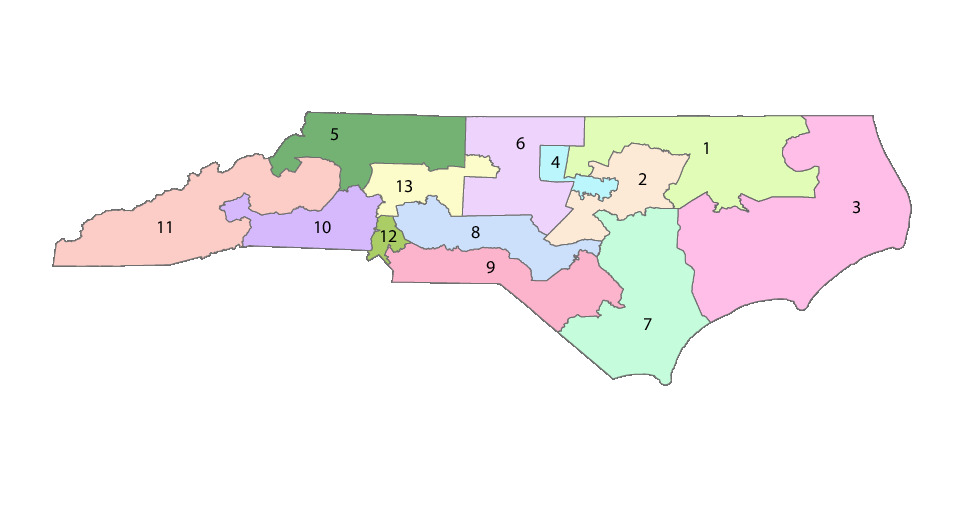} 
\caption{Map for m,NC2016. Numbers corespond to labels in Table~\ref{tab:table1}.} 
 \label{fig:Districting2016}
\end{figure}

\begin{figure}[ht]\centering 
\includegraphics[scale=0.62,angle=90]{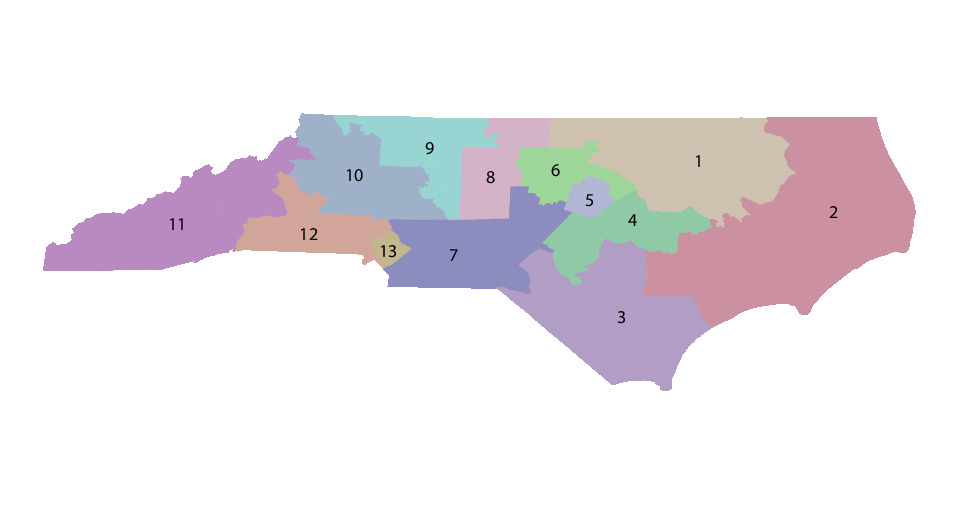} 
\caption{Map produced by bipartisan redistricting commission of
  retired Judges from Beyond Gerrymandering project. Numbers corespond to labels in Table~\ref{tab:table1}.} 
\label{fig:DistrictingJudges}
\end{figure}

\begin{figure}[ht]\centering 
\includegraphics[scale=0.62,angle=90]{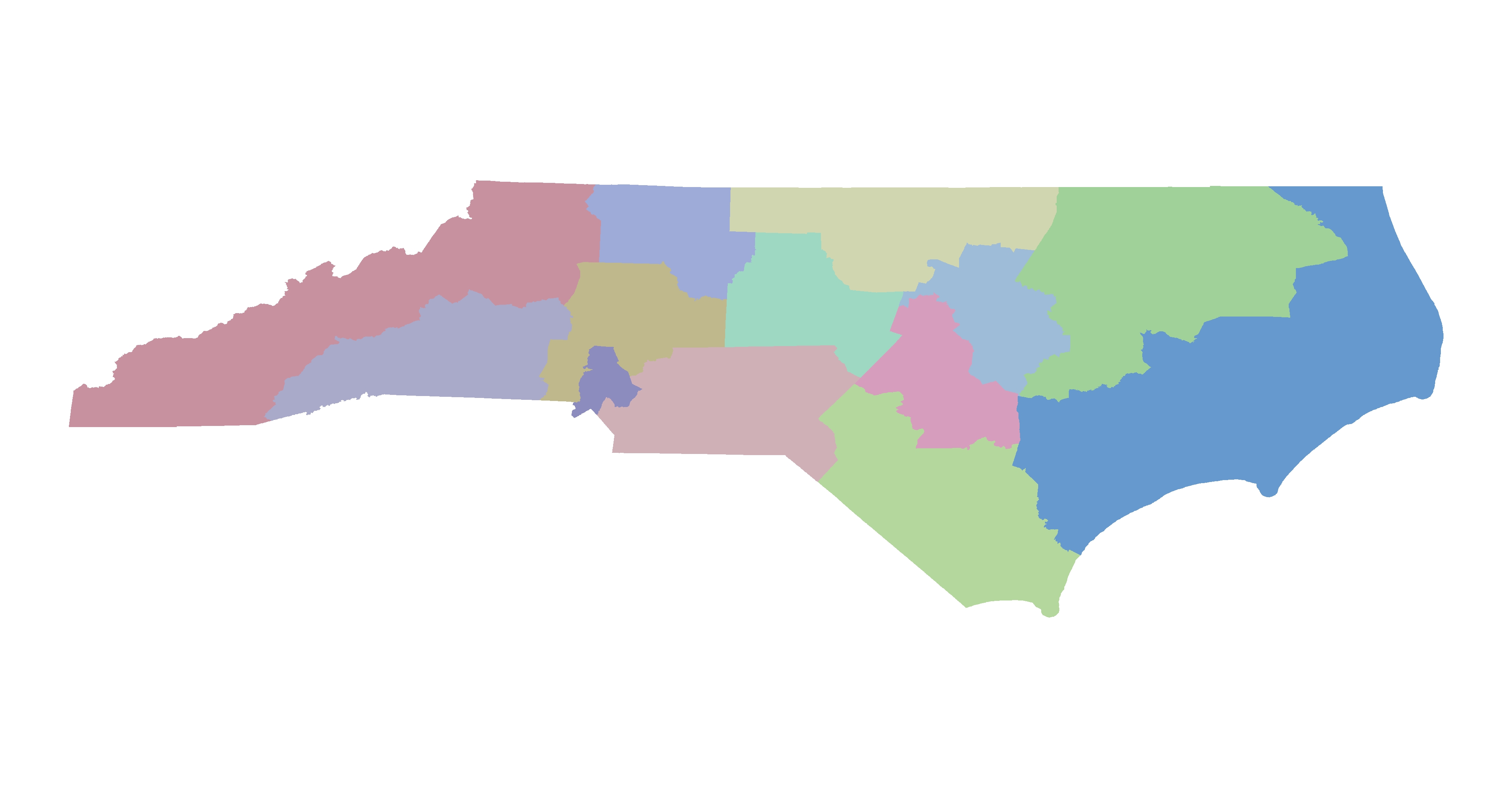} 
\caption{First sample redistricting generated by MCMC}
\label{fig:Districting1} 
\end{figure}
\begin{figure}[ht]\centering \label{fig:Districting2}
\includegraphics[scale=0.62,angle=90]{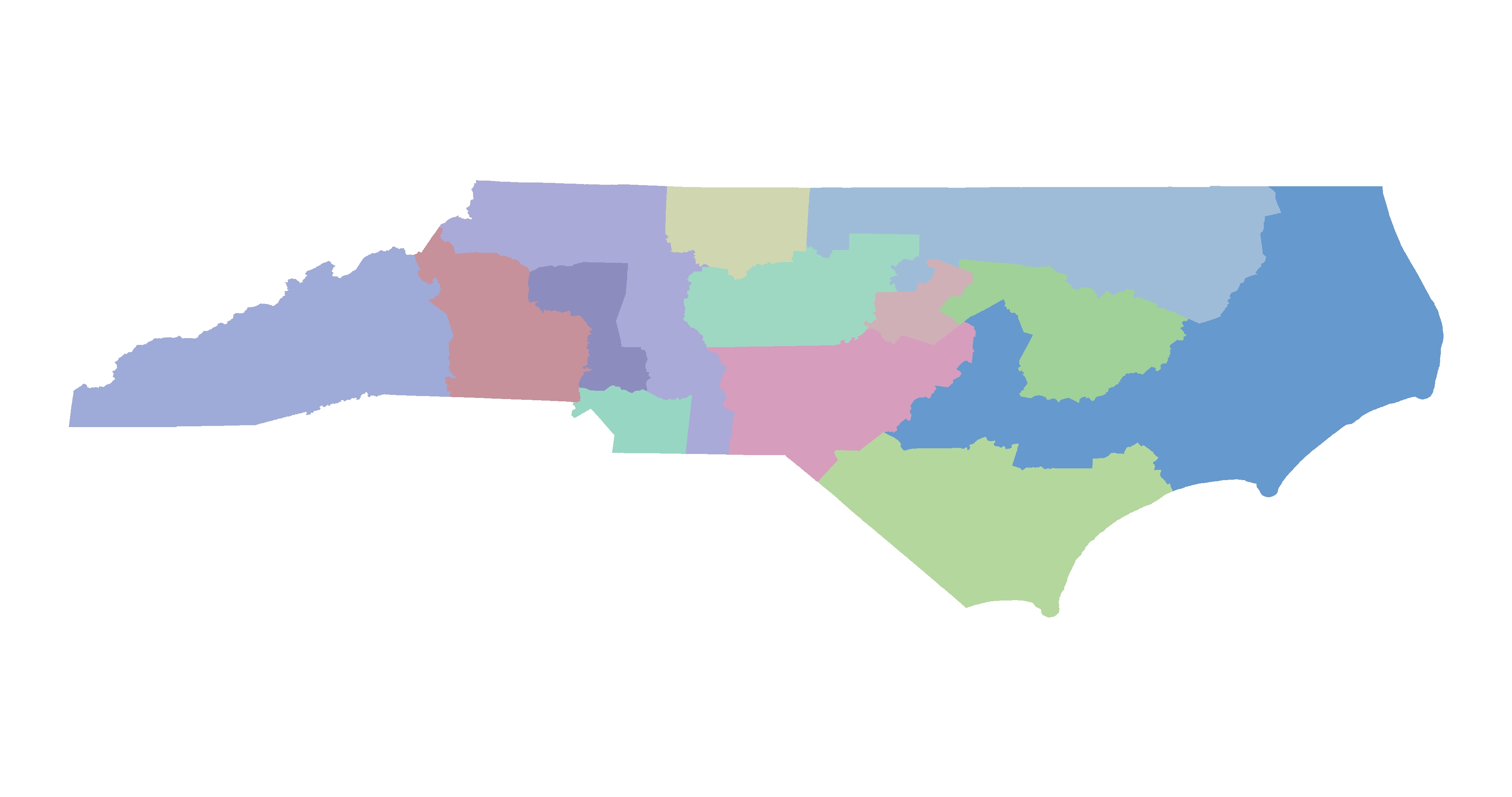} 
\caption{Second sample redistricting generated by MCMC} 
\end{figure}
\begin{figure}[ht]\centering \label{fig:Districting3}
\includegraphics[scale=0.62,angle=90]{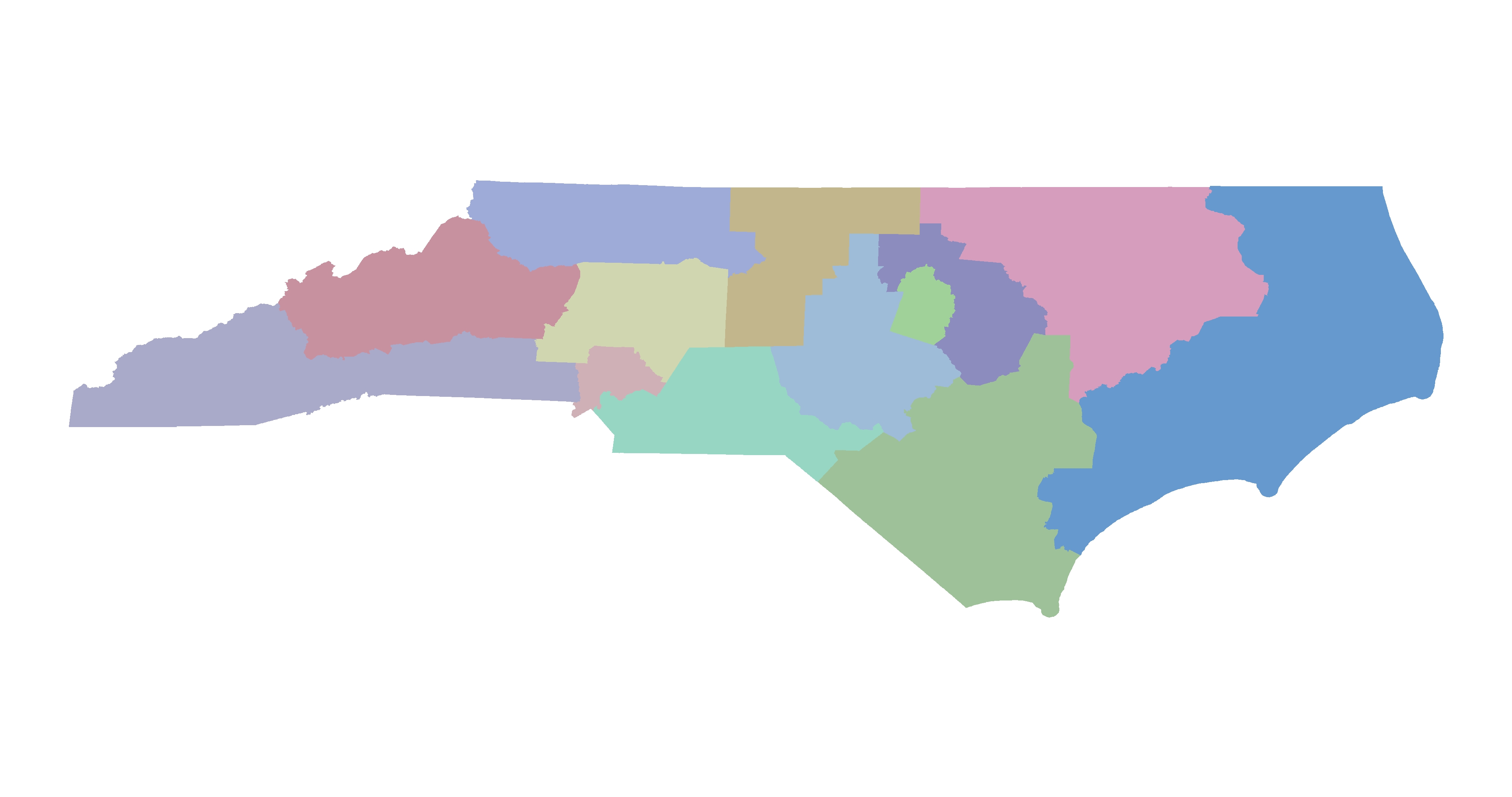} 
\caption{Third sample redistricting generated by MCMC} 
\end{figure}
\begin{figure}[ht]\centering \label{fig:Districting4}
\includegraphics[scale=0.62,angle=90]{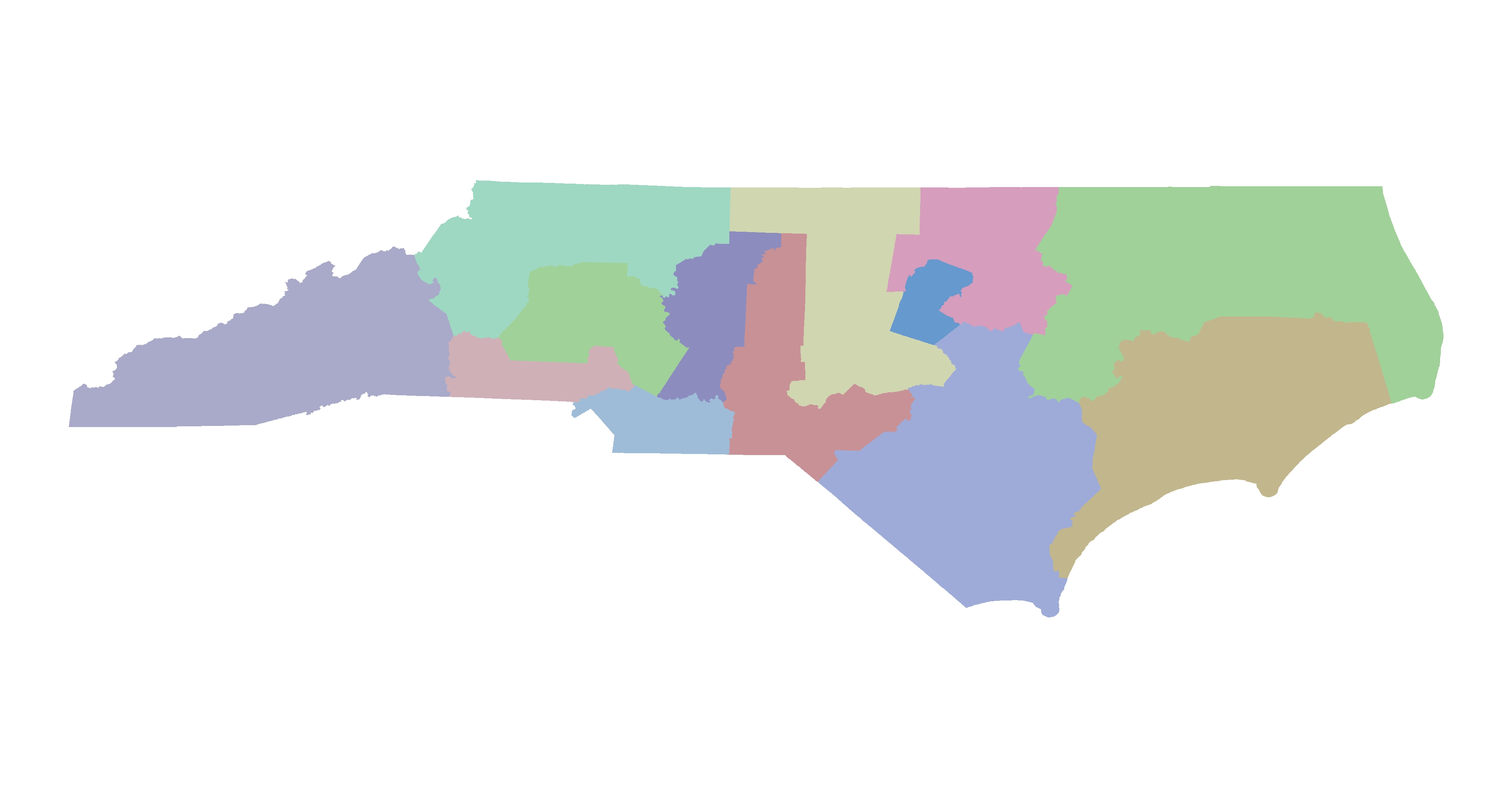} 
\caption{Fourth sample redistricting generated by MCMC} 
\end{figure}

\begin{figure}[ht]\centering \label{fig:Districting5}
\includegraphics[scale=0.62,angle=90]{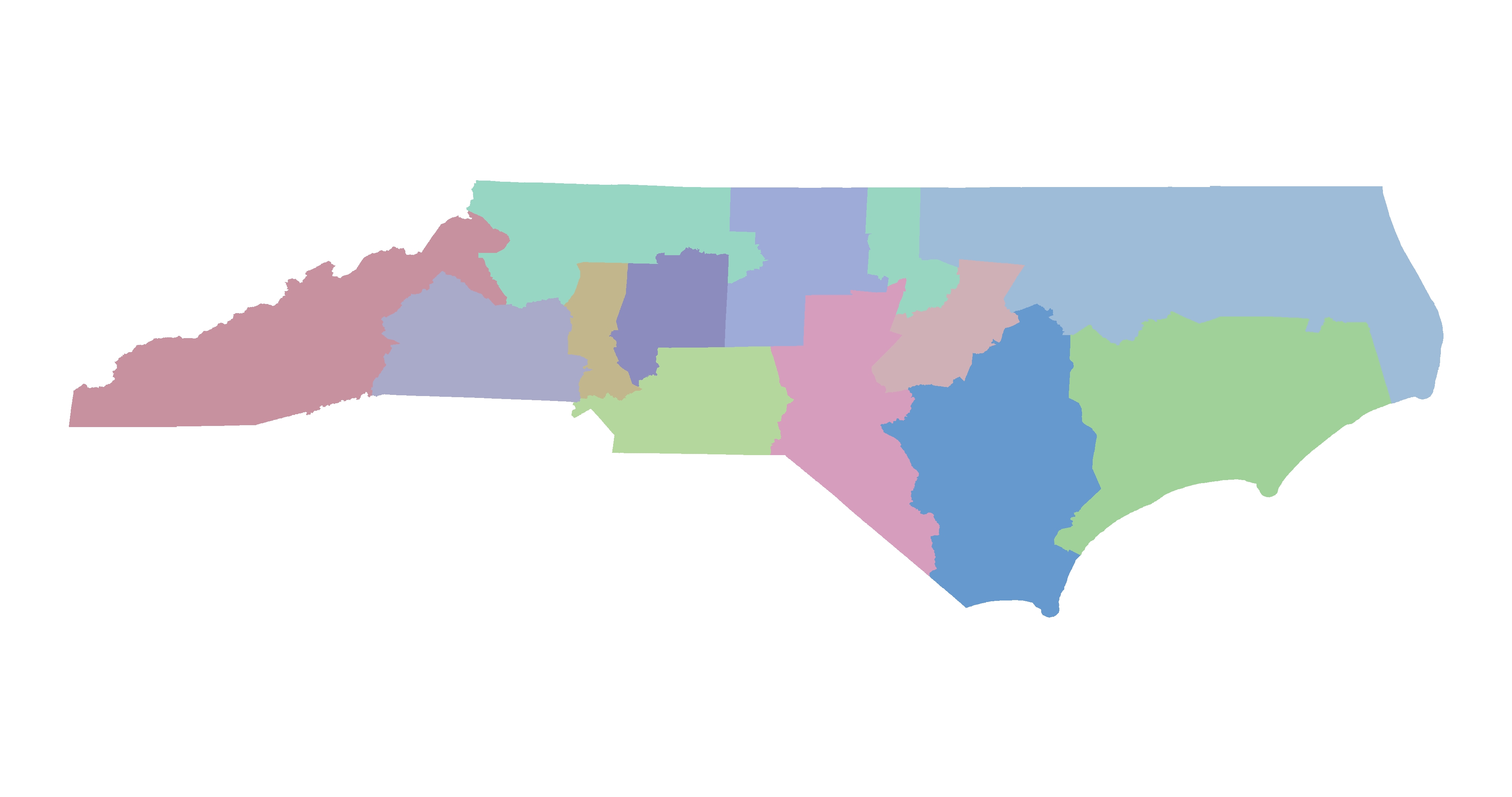} 
\caption{Fifth sample redistricting generated by MCMC} 
\end{figure}

\begin{figure}[ht]\centering \label{fig:Districting6}
\includegraphics[scale=0.62,angle=90]{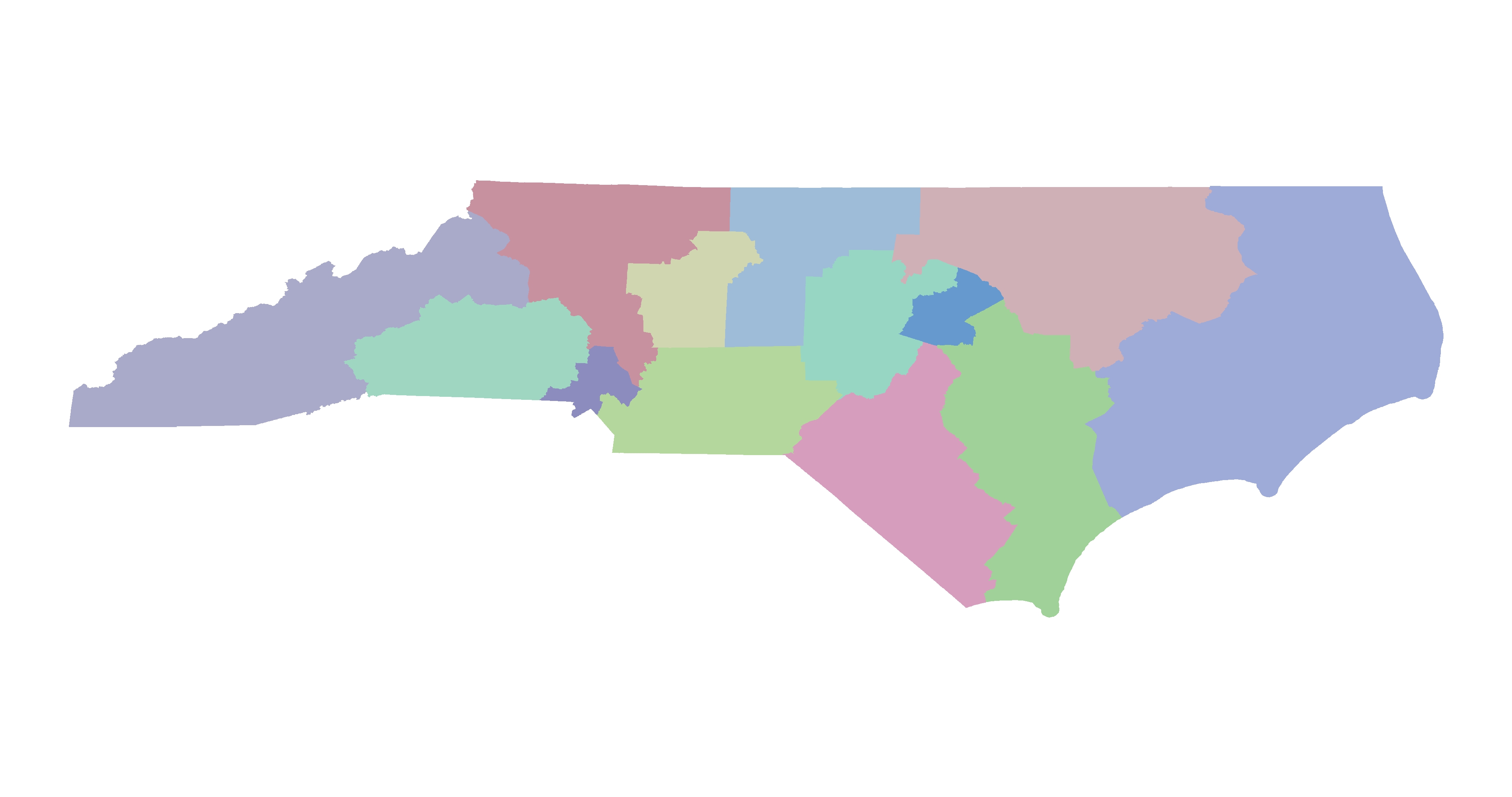} 
\caption{Sixth sample redistricting generated by MCMC} 
\end{figure}

\end{document}